\documentclass[preprint,
showkeys,
preprintnumbers,amsmath,amssymb,nofootinbib,eqsecnum]{revtex4}
\usepackage{hyperref}
\usepackage[mathscr]{eucal}
\usepackage{graphicx,wrapfig}

\begin{document}

\setlength{\baselineskip}{5.6mm}


\title{Twistor formulation of a massive particle with rigidity}

%
%
%
%
%

\author{Shinichi Deguchi} \email{ deguchi@phys.cst.nihon-u.ac.jp}
\affiliation{ Institute of Quantum Science, College of Science and Technology, 
Nihon University, Chiyoda-ku, Tokyo 101-8308, Japan}
\author{Takafumi Suzuki
} \email { takafumi@gaea.jcn.nihon-u.ac.jp} 
\affiliation { Junior College Funabashi Campus, 
Nihon University, Narashinodai, Funabashi, Chiba 274-8501, Japan  }

\begin{abstract}
A massive rigid particle model in $(3+1)$ dimensions is reformulated in terms of twistors. 
Beginning with a first-order Lagrangian,  
we establish a twistor representation of the Lagrangian for a massive particle with rigidity. 
The twistorial Lagrangian derived in this way remains invariant under a local $U(1)\times U(1)$ 
transformation of the twistor and other relevant variables. 
Considering this fact, we carry out a partial gauge-fixing so as to make our analysis simple and clear. 
We develop the canonical Hamiltonian formalism based on the gauge-fixed Lagrangian and 
perform the canonical quantization procedure of the Hamiltonian system. 
Also, we obtain an arbitrary-rank massive spinor field in $(3+1)$ dimensions via the Penrose transform of 
a twistor function defined in the quantization procedure. 
Then we prove, in a twistorial fashion, that the spin quantum number of a massive particle with rigidity 
can take only non-negative integer values, which result is in agreement with the one shown earlier by Plyushchay.  
Interestingly, the mass of the spinor field is determined depending on the spin quantum number.


\end{abstract}

%

\maketitle




\section{Introduction}

A model of a relativistic point particle with rigidity (or simply a rigid particle model) 
has first been proposed by Pisarski \cite{Pisarski} 
about 30 years ago as a 1-dimensional analog of the rigid string model presented by Polyakov \cite{Polyakov}. 
The action of the rigid particle model contains the extrinsic curvature, $K$, of a world-line traced out by a particle, 
in addition to the ordinary term that is identified as the arc-length of the world-line. 
By choosing the arc-length, $l$, to be a world-line parameter along the world-line, 
the action takes the following form:
\begin{align}
\mathcal{S}=\int_{l_0}^{\;\! l_1} dl \left[\;\! -m-kK(l\:\!) \:\! \right] ,
\label{1.1}
\end{align}
where $m$ is a mass parameter and $k$ is a dimensionless real constant 
(in units such that $\hbar=c=1$). 
Pisarski demonstrated, using the back-ground field method in sufficiently large Euclidean dimensions, that 
the renormalized version of $k^{-1}$ behaves asymptotically free. 
Not long after that, 
Plyushchay investigated the $(3+1)$-dimensional model governed by the action $\mathcal{S}$  
in both cases of $m=0$ and $m\neq0$ \cite{Ply01, Ply02, Ply03, Ply04}.\footnote{~This and similar models 
have also been investigated in, e.g., Refs. \cite{BGPR, Ram01, Ban01, Ban02, Pav01, Pav02, Der01, Ply05, Nes01, Pav03}.} 
He studied the canonical Hamiltonian formalism based on $\mathcal{S}$ and the subsequent 
canonical quantization of the system. 
Then it was clarified that the massless rigid particle model, specified by $m=0$,  
describes a massless spinning particle of helicity $k$ \cite{Ply02, Ply03},  
which can take both integer and half-integer values \cite{Ply03}. 
Also, it was shown that the massive rigid particle model, specified by $m\neq0$,  
describes a massive spinning particle whose spin quantum number 
can take only non-negative integer values \cite{Ply01, Ply04}. 
On the other hand, 
Deriglazov and Nersessian have recently claimed that the massive rigid particle model 
can yield the Dirac equation and hence 
can describe a massive spinning particle of spin one-half \cite{DerNer}. 
This statement is inconsistent with that of Plyushchay. 
It is therefore necessary to clarify which statement is correct.

Recently, the massless rigid particle model has been reformulated in terms of twistors \cite{DegSuz}. 
It was demonstrated in Ref. \cite{DegSuz} that the action $\mathcal{S}$ with $m=0$ 
is equivalent to the {\em gauged} Shirafuji action \cite{BarPic, DEN, DNOS} 
(rather than the original Shirafuji action \cite{Shirafuji}) 
that governs a twistor model of a massless spinning particle of helicity $k$ propagating in 
4-dimensional Minkowski space, $\mathbf{M}$. The gauged Shirafuji action can thus be regarded as 
a twistor representation of the action for a massless particle with rigidity. 
Upon canonical quantization of the twistor model, 
the allowed values of $k$ are restricted to either integer or half-integer values, 
which are in agreement with the allowed values obtained in Ref. \cite{Ply03}.  
At the same time, an arbitrary-rank spinor field in complexified Minkowski space, $\Bbb{C}\mathbf{M}$,  
can be elegantly derived via the Penrose transform \cite{PenMac, PenRin2, HugTod}. 
It can also be shown that this field satisfies generalized Weyl equations.

Since the twistor formulation of a massless particle with rigidity has been well established, 
it is quite natural to next consider the twistor formulation of a massive particle with rigidity.  
Twistor approaches to massive particle systems were investigated independently by 
Penrose, Perj\'{e}s, and Hughston about 40 years ago 
\cite{Penrose, Perjes1, Perjes2, Perjes3, Perjes4, Hughston}. 
After a long while, Lagrangian mechanics of a massive spinning particle has been formulated until recently 
in terms of two twistors 
\cite{FedZim, BALE, AFLM, FFLM, AIL, MRT, FedLuk, AFIL, MRT2, RouTow, DegOka}.\footnote{~Recently, it has been shown that 
the genuine $n$-twistor description of a massive particle in $\mathbf{M}$ fails for the case $n \geq 3$  
\cite{RouTow, OkaDeg}. 
Therefore it turns out that in the twistor formulation, 
the use of two twistors is the only possible choice to describe a massive particle in $\mathbf{M}$.}
In fact, the Shirafuji action for a massless spinning particle has been generalized in various ways   
to describe a massive spinning particle in $\mathbf{M}$ 
\cite{FedZim, AFLM, FFLM, MRT, FedLuk, AFIL, MRT2, DegOka}. 
Among the generalized Shirafuji actions, the gauged generalized Shirafuji (GGS) action presented in Ref. \cite{DegOka} 
is one of the most desirable actions, 
because it yields just sufficient constraints among the twistor variables 
in a systematic and consistent manner. 
All the constraints, excluding the mass-shell condition of Fedoruk-Lukierski type \cite{FedLuk, AFIL}, are indeed derived  
on the basis of the fact that the GGS action remains invariant under the reparametrization of the world-line parameter 
and under the local $U(1)\times \mathit{SU}(2)$ transformation of the twistor and other relevant variables.

In light of the equivalence between the action $\mathcal{S}$ with $m=0$ and the gauged Shirafuji action, 
we can expect that the action $\mathcal{S}$ with $m\neq0$ is equivalent to the GGS action or its analog. 
One of the purposes of the present paper is to confirm this expectation by reformulating 
the massive rigid particle model in terms of twistors. 
To this end, following the procedure developed in Ref. \cite{DegSuz}, 
we first provide an appropriate first-order Lagrangian and demonstrate that it is equivalent 
to the Lagrangian $L_{K}(l):=-m-|k|K$ found in Eq. (\ref{1.1}).  
(The constant $k$ contained in $\mathcal{S}$ is replaced by $|k|$ for the sake of consistency, 
as will be seen in Sec. 2.)  
The first-order Lagrangian eventually gives five constraints among dynamical variables. 
We simultaneously solve two of the five constraints  
by using two (commutative) 2-component spinors \cite{PenRin1} 
without spoiling compatibility with the other three constraints. 
Substituting the solution obtained there into the first-order Lagrangian, 
we have a Lagrangian written in terms of the spacetime coordinate variables and the spinor variables. 
Then, using the two (novel) twistors defined from these variables, we express the Lagrangian in a twistorial form. 
Each of these twistors satisfies the so-called null twistor condition \cite{PenMac, PenRin2, HugTod}. 
We incorporate the null twistor conditions for the two twistors 
into the Lagrangian with the aid of Lagrange multipliers 
so that all the twistor components can be treated as independent fundamental variables. 
In addition, we slightly modify the Lagrangian
so that it can immediately give the mass-shell condition of Fedoruk-Lukierski type. 
The modified Lagrangian is actually equivalent to the one before modification and hence to $L_{K}(l)$. 
In this way, we can elaborate an appropriate twistor representation of $L_{K}(l)$. 
The action defined with the modified Lagrangian is not the GGS action itself but its analog. 
It is thus confirmed that the action 
$\mathcal{S}=\int_{l_0}^{l_1} dl^{\:\!} L_{K}(l)$ with $m\neq0$ is equivalent to an analog of the GGS action.

Since the modified Lagrangian governs a novel twistor model that has not been studied yet, 
we need to investigate classical and quantum mechanical properties of this model in detail. 
For this purpose, we carry out a partial gauge-fixing for the local $U(1)\times U(1)$ symmetry of the twistor model 
by adding a gauge-fixing term and its associated term to the modified Lagrangian. 
(The modified Lagrangian remains invariant under the local $U(1)\times U(1)$ transformation 
rather than the local $U(1)\times \mathit{SU}(2)$ transformation.) 
The gauge-fixing condition and its associated condition are chosen so as to make our analysis simple and clear. 
With the gauge-fixed Lagrangian, 
we study the canonical Hamiltonian formalism of the twistor model  
by completely following the Dirac algorithm for Hamiltonian systems with constraints \cite{Dirac2, HRT, HenTei}. 
We see that by virtue of the appropriate gauge-fixing procedure, 
the Dirac brackets between the twistor variables take the form of canonical bracket relations. 
Also, we can obtain manageable first-class constraints. 
The subsequent canonical quantization of the twistor model is performed with the aid of the commutation relations 
between the operator versions of the twistor and other canonical variables. 
In the quantization procedure, 
the first-class constraints turn into the conditions imposed on a physical state vector. 
It is pointed out that the physical state vector can be chosen so as to be an eigenvector of the spin Casimir operator. 
Among a total of six physical state conditions, 
five are represented as simultaneous differential equations for a function of the canonical coordinate variables,  
while the remainder turns into the (algebraic) mass-shell condition. 
The five differential equations eventually reduce to two differential equations for a twistor function, 
a holomorphic function only of the twistor variables (excluding the dual twistor variables). 
We solve one of these equations by applying the method of separation of variables, 
finding a certain twistor function as its general solution. 
The other differential equation is used in some other stage.

After completing the quantization procedure, 
we consider the Penrose transform of the above-mentioned twistor function   
to obtain an arbitrary-rank massive spinor field in $\Bbb{C}\mathbf{M}$. 
The spinor field obtained has extra upper and lower indices 
in addition to dotted and undotted spinor indices.  
Because of the structure of the Penrose transform, the number of upper (lower) 
extra indices is equal to the number of undotted (dotted) spinor indices. 
We can find the allowed values of the spin quantum number of a massive particle with rigidity 
by evaluating the total number of extra indices of the spinor field in $\Bbb{C}\mathbf{M}$. 
Using a useful relation proven in Appendix B, 
we indeed prove that the spin quantum number can take only non-negative integer values. 
This result is in agreement with the one shown earlier by Plyushchay \cite{Ply01,Ply04} and 
contradicts the recent statement of Deriglazov and Nersessian \cite{DerNer}.

We also demonstrate, by using the mass-shell condition,  
that the spinor field satisfies generalized Dirac-Fierz-Pauli (DFP) equations with extra indices. 
In addition, we show that the spinor field symmetrized totally with respect to the extra indices 
satisfies the (ordinary) DFP equations \cite{Dirac1, Fierz, FiePau, IsaPod}. 
A physical mass parameter included in both the generalized and ordinary DFP equations 
is identical to the one found by Plyushchay \cite{Ply01,Ply04} 
and turns out to be dependent on the spin quantum number. 
We thus see that the physical mass of the spinor field is determined depending on its rank.

This paper is organized as follows: 
In section 2, we provide a first-order Lagrangian for a massive particle with rigidity.  
In section 3, beginning with the first-order Lagrangian, we elaborate an appropriate twistor representation of 
the Lagrangian for a massive particle with rigidity. 
A partial gauge-fixing is carried out for this twistor representation. 
The canonical Hamiltonian formalism based on the gauge-fixed Lagrangian is investigated in section 4, 
and the subsequent canonical quantization procedure is performed in section 5. 
In section 6, we derive an arbitrary-rank spinor field in $\Bbb{C}\mathbf{M}$ 
via the Penrose transform of a twistor function and demonstrate that this spinor field satisfies 
generalized DFP equations with extra indices. 
We also show that the allowed values of the spin quantum number 
are restricted to arbitrary non-negative integers. 
Section 7 is devoted to a summary and discussion. 
In Appendix A, we focus on the Pauli-Lubanski pseudovector and 
find a specific form of the physical mass parameter. 
In Appendix B, we prove a useful relation. 
In Appendix C, we extract some part of Plyushchay's noncovariant formulation 
from the twistor formulation developed in this paper.

\section{A first-order Lagrangian for a massive particle with rigidity}
\label{sec:intro}

In this section, we present a first-order Lagrangian and demonstrate that it is equivalent to 
the Lagrangian for a massive particle with rigidity.

Let $x^{\mu}=x^{\mu}(\tau)$ ($\mu=0,1,2,3$) be spacetime coordinates of a point 
particle propagating in 4-dimensional Minkowski space $\mathbf{M}$  
with the metric tensor $\eta_{\mu\nu}=\mathrm{diag}(1,-1,-1,-1)$.  
Here, $\tau$ ($\tau_{0} \le\tau\le \tau_{1}$) is an arbitrary world-line parameter 
along the particle's world-line, being chosen in such a manner that $dx^{0}/d\tau >0$. 
Under the proper reparametrization $\tau \rightarrow \tau^{\prime}=\tau^{\prime}(\tau)$  
($d\tau^{\prime}/d\tau >0$),  
the coordinate variables $x^{\mu}$ behave as real scalar fields on the 1-dimensional parameter space  
$\mathcal{T}:=\{^{\:\!}  \tau^{\;\!} |^{\,} \tau_{0} \le \tau \le \tau_{1} \}$, 
\begin{align}
x^{\mu}(\tau) \rightarrow x^{\prime\mu}(\tau^{\prime})=x^{\mu}(\tau) \,. 
\label{2.1}
\end{align}

Let us consider the action 
\begin{align}
{S}=\int_{\tau_0}^{\;\!\tau_1} d\tau \:\! {L}  
\label{2.2}
\end{align}
with the Lagrangian 
\begin{align}
L=-m\sqrt{q^2} +\left(q^{\mu}-\dot{x}^{\mu} \right) p_{\mu} 
-\left(\dot{q}^{\mu} -\beta q^{\mu} \right)  r_{\mu} +2e \left( \pm\sqrt{-q^{2} r^{2}}-k\:\! \right) ,   
\label{2.3}
\end{align}
supplemented by the conditions 
$q^{2}:=q_{\mu}q^{\mu} > 0$ and $r^{2}:=r_{\mu} r^{\mu} < 0$.  
Here, $m$ is a positive constant with dimensions of mass and 
$k$ is a dimensionless real constant (in units such that $\hbar=c=1$). 
A dot over a variable denotes its derivative with respect to $\tau$. 
The variables $q^{\mu}=q^{\mu}(\tau)$, $p_{\mu}=p_{\mu}(\tau)$, $r_{\mu}=r_{\mu}(\tau)$, 
$e=e(\tau)$, and $\beta=\beta(\tau)$ are understood as real fields on $\mathcal{T}$. 
These fields are assumed to transform under the proper reparametrization as follows:   
\begin{subequations}
\label{2.4}
\begin{align}
q^{\mu}(\tau)& \rightarrow q^{\prime\mu} (\tau^{\prime})
=\frac{d\tau}{d\tau^{\prime}}q^{\mu}(\tau) \,,
\label{2.4a}
\\
p_{\mu}(\tau)& \rightarrow p_{\mu}^{\prime}(\tau^{\prime})= p_{\mu}(\tau)\,, 
\label{2.4b}
\\
r_{\mu}(\tau)& \rightarrow r_{\mu}^{\prime}(\tau^{\prime})
=\frac{d\tau^{\prime}}{d\tau}r_{\mu}(\tau) \,,
\label{2.4c}
\\
e(\tau)& \rightarrow e^{\prime}(\tau^{\prime})
=\frac{d\tau}{d\tau^{\prime}} e(\tau) \,,
\label{2.4d}
\\
\beta(\tau)& \rightarrow \beta^{\prime}(\tau^{\prime})
= \frac{d\tau}{d\tau^{\prime}} \:\!\beta(\tau)
+\frac{d\tau^{\prime}}{d\tau}
\frac{d^2\tau}{d\tau^{\prime 2}} \,.
\label{2.4e}
\end{align}
\end{subequations}
Using Eqs. (\ref{2.1}) and (\ref{2.4}), 
we can verify that the action ${S}$ is reparametrization invariant.  
Note that the Lagrangian (\ref{2.3}) is first order in $\dot{x}^{\mu}$ and $\dot{q}^{\mu}$. 
In this Lagrangian, $p_{\mu}$, $r_{\mu}$, $e$, and $\beta$ are treated as independent auxiliary fields.

From the Lagrangian (\ref{2.3}),  
the Euler-Lagrange equations for 
$x^{\mu}$, $q^{\mu}$, $p_{\mu}$, $r_{\mu}$, $e$, and $\beta$ are derived, respectively, as 
\begin{subequations}
\label{2.5}
\begin{align}
\dot{p}_{\mu}=0 \,, 
\label{2.5a}
\\
\dot{r}_{\mu}+p_{\mu}+\beta r_{\mu} 
-\left( \frac{m}{\sqrt{q^{2}}}  
\pm\frac{2er^{2}}{\sqrt{-q^{2} r^{2}}} \right) q_{\mu}=0 \,,
\label{2.5b}
\\
\dot{x}{}^{\mu}-q^{\mu}=0 \,, 
\label{2.5c}
\\
\dot{q}{}^{\mu}-\beta q^{\mu} \pm \frac{2eq^{2}}{\sqrt{-q^{2} r^{2}}} r^{\mu}=0 \,,
\label{2.5d}
\\
\pm \sqrt{-q^{2} r^{2}}-k=0 \,, 
\label{2.5e}
\\
q^{\mu} r_{\mu}=0 \,. 
\label{2.5f}
\end{align}
\end{subequations}
Each of Eqs. (\ref{2.5a})--(\ref{2.5d}) includes a derivative term,  
while Eqs. (\ref{2.5e}) and (\ref{2.5f}) include no derivative terms. 
For this reason, Eqs. (\ref{2.5e}) and (\ref{2.5f}) are treated as (algebraic) constraints. 
It can be seen from Eq. (\ref{2.5e}) that the sign of $k$ is determined depending on 
which sign is chosen in $\pm\sqrt{-q^{2} r^{2}}$. 
Taking the derivative of Eq. (\ref{2.5e}) with respect to $\tau$ 
and using Eqs. (\ref{2.5b}), (\ref{2.5d}), and (\ref{2.5f}), we have $q^{2} r^{\mu} p_{\mu}=0$. 
Since $q^{2}>0$ has been postulated, it follows that 
\begin{align}
r^{\mu} p_{\mu}=0 \,.  
\label{2.6}
\end{align}
Taking the derivative of Eq. (\ref{2.5f}) with respect to $\tau$ 
and using Eqs. (\ref{2.5b}) and (\ref{2.5d}), we have 
\begin{align}
q^{\mu} p_{\mu}=m\sqrt{q^2} \,.
\label{2.7}
\end{align}
The derivative of Eq. (\ref{2.7}) with respect to $\tau$ is identically satisfied  
with the use of Eqs. (\ref{2.5a}), (\ref{2.5d}), (\ref{2.5f}), (\ref{2.6}), and (\ref{2.7}); 
hence no new conditions are derived from Eq. (\ref{2.7}). 
The derivative of Eq. (\ref{2.6}) with respect to $\tau$, 
together with Eqs. (\ref{2.5a}), (\ref{2.5b}), (\ref{2.6}), and (\ref{2.7}), yields
\begin{align}
p^2 \left(:=p^{\mu} p_{\mu} \right) =m^2 \mp 2me \sqrt{-r^2} \,. 
\label{2.8}
\end{align}
Taking the derivative of (\ref{2.8}) with respect to $\tau$ and using 
Eqs. (\ref{2.5a}), (\ref{2.5b}), (\ref{2.5f}), and (\ref{2.6}), we obtain $(\dot{e}-\beta e) \sqrt{-r^2} =0$. 
Since $r^{2}<0$ has been postulated, $\dot{e}-\beta e=0$ must be satisfied and $\beta$ is determined to be 
$\beta=\dot{e}/e =(d/d\tau) \ln e$. 
This is not an algebraic equation, and hence we do not need to consider it a constraint. 
In addition to Eqs. (\ref{2.5e}) and (\ref{2.5f}), Eqs (\ref{2.6})--(\ref{2.8}) are also regarded as (algebraic) constraints.

Using Eq. (\ref{2.5d}), we can eliminate the auxiliary field $r_{\mu}$ from the Lagrangian (\ref{2.3}) 
to obtain 
\begin{align}
{L}=-m\sqrt{q^2} +\left( q^{\mu}-\dot{x}^{\mu} \right) p_{\mu} -2ke \,. 
\label{2.9}
\end{align}
Here, $e$ is no longer an independent auxiliary field and is determined from Eq. (\ref{2.5d}) as follows: 
Contracting Eq. (\ref{2.5d}) with $q_{\mu}$ and using Eq. (\ref{2.5f}), we have 
\begin{align} 
\beta=\frac{q \dot{q}}{q^2} \,, 
\label{2.10}
\end{align}
where $q \dot{q}:=q_{\mu} \dot{q}^{\mu}$. 
Then Eq. (\ref{2.5d}) leads to 
\begin{align}
e^2 &=-\frac{1}{4q^2}\left(\dot{q}_{\mu}-\beta q_{\mu} \right) \left(\dot{q}^{\mu}-\beta q^{\mu} \right) 
=-\frac{\dot{q}_{\perp}^{2}}{4q^2} \,, 
\label{2.11}
\end{align}
where $\dot{q}_{\perp}^{2}$ is defined by $\dot{q}_{\perp}^{2}:=\dot{q}_{\perp \mu\:\!} \dot{q}_{\perp}^{\mu}$  
with $\dot{q}_{\perp}^{\mu}:=\dot{q}^{\mu}-q^{\mu}(q\dot{q}) /q^2$. 
In this way, $e$ is determined to be $e=\pm\frac{1}{2} \sqrt{-\dot{q}_{\perp}^{2}/q^2}$. 
Because $q^2 >0$, the inequality 
$\dot{q}_{\perp}^{2} =\big( q^{2} \dot{q}{}^{2} -(q \dot{q}){}^2 \big) / q^{2} \le 0$ holds,\footnote{~Let $u^{\mu}$ 
be an arbitrary timelike vector in ${\bf M}$ and $v^{\mu}$ an arbitrary 
vector in ${\bf M}$. Since $u^{\mu}$ is timelike, we can choose the rest frame 
such that $u^{i}=0$ ($i=1,2,3$). In this frame, 
$u^2=u_{\mu} u^{\mu}$ and $uv=u_{\mu} v^{\mu}$ reduce to 
$u^2=(u^0)^2$ and $uv=u^0 v^0$, respectively. 
Then it can readily be shown that 
$u^2 v^2 -(uv)^2 =-(u^0)^2 \sum_{i=1}^3 v^{i}v^{i} \leq 0$. 
Because $u^2$, $v^2$, and $uv$ are Lorentz scalars, $u^2 v^2 \leq (uv)^2$ 
holds true in arbitrary reference frames.}  
and hence $e$ still remains purely real. 
The Lagrangian (\ref{2.9}) can be written explicitly as 
\begin{align}
{L}=-m\sqrt{q^2} +\left(q^{\mu}-\dot{x}^{\mu} \right) p_{\mu} \mp k \sqrt{-\frac{\dot{q}_{\perp}^{2}}{q^2}} \,,  
\label{2.12}
\end{align}
which becomes  
\begin{align}
{L} &=-m\sqrt{\dot{x}{}^2} \mp k \sqrt{-\frac{\ddot{x}_{\perp}^{2}}{\dot{x}^2}} 
\notag
\\
&=-m\sqrt{\dot{x}{}^2} \mp k \frac{ \sqrt{(\dot{x} \ddot{x}){}^{2} -\dot{x}^2 \ddot{x}^{2}}}{\dot{x}^2}  
\label{2.13}
\end{align}
by eliminating $q^{\mu}$ and $p_{\mu}$ with the use of Eq. (\ref{2.5c}). 
In Eqs. (\ref{2.12}) and (\ref{2.13}), 
one of the signs in the symbol $\mp$ is chosen so that the Lagrangian 
can be negative definite even in the limit $m \downarrow 0$, whether $k$ is positive or negative. 
As a result, $\mp k=-|k|$ is realized and the Lagrangian (\ref{2.13}) reads 
\begin{align}
{L}=-m\sqrt{\dot{x}{}^2} -|k| \sqrt{-\frac{\ddot{x}_{\perp}^{2}}{\dot{x}^2}} \,.
\label{2.14}
\end{align}
This is exactly the Lagrangian for a massive particle with rigidity represented as a function of $\tau^{\:\!}$: 
$L(\tau)=\sqrt{\dot{x}^{2}} L_{K}=\sqrt{\dot{x}^{2}} (-m-|k|K)$ \cite{Pisarski,Ply01,Ply04}. 
(Note that $\mp k=-|k|$ is compatible with Eq. (\ref{2.5e});  
in fact, it leads to a consistent result $\sqrt{-q^{2} r^{2}}=|k|$.) 
The condition $\dot{x}^{2}>0$ implies that the particle moves at a speed less than the speed of light. 
As expected, the reparametrization invariance of the action 
${S}=\int_{\;\!\tau_0}^{\;\!\tau_1} d\tau^{\,\!} L$ is maintained with the Lagrangian (\ref{2.14}). 
In our present approach, the Lagrangian (\ref{2.14}) has been found from the Lagrangian (\ref{2.3}) 
by eliminating the auxiliary fields $p_{\mu}$, $r_{\mu}$, $e$, and $\beta$, and furthermore the field $q^{\mu}$. 
The Lagrangian (\ref{2.3}) is thus established as a first-order Lagrangian 
for a massive particle with rigidity,  
being considered the fact that the Lagrangian (\ref{2.3}) is first order in $\dot{x}^{\mu}$ and $\dot{q}^{\mu}$.

\section{Twistor representation of the Lagrangian}

In this section, we derive a twistor representation of the Lagrangian (\ref{2.14}) by following 
the method developed in Refs. \cite{DegSuz, Shirafuji}.  
A partial gauge-fixing for a local gauge symmetry of the twistor representation is also considered.

We first express the first-order Lagrangian (\ref{2.3}) using bispinor notation\footnote{~The 
bispinor notation $x^{\alpha \dot{\alpha}}$ ($p_{\alpha \dot{\alpha}}$) and 
the 4-vector notation $x^{\mu}$ ($p_{\mu}$) are related as follows \cite{PenRin1}: 
\begin{align}
\begin{pmatrix}
\: x^{0\dot{0}} & x^{0\dot{1}} \: \\
\: x^{1\dot{0}} & x^{1\dot{1}} \:
\end{pmatrix}
= \dfrac{1}{\sqrt{2}} 
\begin{pmatrix}
\: x^{0} + x^{3} &  x^{1} + i x^{2} \: \\
\: x^{1} - i x^{2} &  x^{0} - x^{3} \:
\end{pmatrix}, 
\qquad
\begin{pmatrix}
\: p_{0\dot{0}} &  p_{0\dot{1}} \, \\
\: p_{1\dot{0}} &  p_{1\dot{1}} \,
\end{pmatrix}
= \dfrac{1}{\sqrt{2}} 
\begin{pmatrix}
\: p_{0} + p_{3} &  p_{1} - i p_{2} \, \\
\: p_{1} + i p_{2} & p_{0} - p_{3} \,
\end{pmatrix}.
\notag
\end{align}
It is readily seen that $x^{\alpha \dot{\alpha}}$ and $p_{\alpha \dot{\alpha}}$ are Hermitian, 
because $x^{\mu}$ and $p_{\mu}$ are real.} as 
\begin{align}
L=-m\sqrt{q^2} +\left(q^{\alpha\dot{\alpha}}-\dot{x}{}^{\alpha\dot{\alpha}} \right) p_{\alpha\dot{\alpha}} 
-\left(\dot{q}^{\alpha\dot{\alpha}} -\beta q^{\alpha\dot{\alpha}} \right) r_{\alpha\dot{\alpha}} 
+2e \left( \sqrt{-q^{2} r^{2}}-|k| \:\! \right), 
\label{3.1}
\end{align}
where $q^{2}=q_{\alpha\dot{\alpha}}q^{\alpha\dot{\alpha}}$ and 
$r^{2}=r_{\alpha\dot{\alpha}} r^{\alpha\dot{\alpha}}$ 
$\:\big( \alpha=0,1;\, \dot{\alpha}=\dot{0}, \dot{1} \big)$. 
In this expression, $\pm k=|k|$ has been taken into account together with replacing $e$ by $\pm e$. 
The constraints (\ref{2.6}) and (\ref{2.5f}) can be written, respectively, as 
\begin{subequations}
\label{3.2}
\begin{align}
r^{\alpha\dot{\alpha}} p_{\alpha\dot{\alpha}} &=0 \,, 
\label{3.2a}
\\
q^{\alpha\dot{\alpha}} r_{\alpha\dot{\alpha}} &=0 \,. 
\label{3.2b}
\end{align}
\end{subequations}
In terms of two (commutative) 2-component spinors   
$\pi_{i\dot{\alpha}}=\pi_{i\dot{\alpha}}(\tau)$ $(i=1,2)$ and  
their complex conjugates 
$\bar{\pi}^{i}_{\alpha}=\bar{\pi}^{i}_{\alpha}(\tau)$,  
specified by $\bar{\pi}^{i}_{\alpha} :=\overline{\pi_{i\dot{\alpha}}}^{\:\!}$, 
we can simultaneously solve the constraint equations (\ref{3.2a}) and (\ref{3.2b}) as 
\begin{subequations}
\label{3.3}
\begin{align}
p_{\alpha\dot{\alpha}} &=\bar{\pi}^{1}_{\alpha} \pi_{1\dot{\alpha}} 
+\bar{\pi}^{2}_{\alpha} \pi_{2\dot{\alpha}} 
\equiv \bar{\pi}{}^{i}_{\alpha} \pi_{i\dot{\alpha}} \,,
\label{3.3a}
\\
q^{\alpha\dot{\alpha}} &=f \left(\bar{\pi}^{1\alpha} \pi_{1}^{\dot{\alpha}}   
+\gamma\bar{\pi}^{2\alpha} \pi_{2}^{\dot{\alpha}} \right) \,,
\label{3.3b}
\\
r_{\alpha\dot{\alpha}} &=i\left( g\bar{\pi}^{1}_{\alpha} \pi_{2\dot{\alpha}} 
-\bar{g} \bar{\pi}^{2}_{\alpha} \pi_{1\dot{\alpha}} \right) \,, 
\label{3.3c}
\end{align}
\end{subequations}
without spoiling compatibility with the other (algebraic) constraints (\ref{2.5e}), (\ref{2.7}), and (\ref{2.8}). 
Here, $\gamma$ is a positive constant, 
$f=f(\tau)$ is a positive-valued field on $\mathcal{T}$, 
and $g=g(\tau)$ is a complex field on $\mathcal{T}$. 
(The allowed values of $\gamma$ are found later in a quantization procedure.) 
Using the conventional formula  
$\iota^{\alpha}\kappa_{\alpha}=\epsilon^{\alpha\beta} \iota_{\beta}\kappa_{\alpha} 
=\iota^{\alpha}\kappa^{\beta} \epsilon_{\beta\alpha}$ valid for arbitrary commutative spinors 
$\iota_{\alpha}$ and $\kappa_{\alpha}$, together with 
its complex conjugate formula valid for 
$\bar{\iota}_{\dot{\alpha}}$ and $\bar{\kappa}_{\dot{\alpha}}$, 
we can verify that Eq (\ref{3.3}) satisfies Eq. (\ref{3.2}).  
(Here, $\epsilon^{\alpha\beta}$ and $\epsilon_{\alpha\beta}$ denote 
Levi-Civita symbols specified by $\epsilon^{01}=\epsilon_{01}=1$.)  
We assume that under the proper reparametrization, $\pi_{i\dot{\alpha}}$ and $\bar{\pi}^{i}_{\alpha}$  
behave as complex scalar fields on $\mathcal{T}$, precisely as 
$\pi_{i\dot{\alpha}}(\tau) \rightarrow \pi^{\prime}_{i\dot{\alpha}}(\tau^\prime)=\pi_{i\dot{\alpha}}(\tau)$ 
and $\bar{\pi}^{i}_{\alpha}(\tau) \rightarrow \bar{\pi}{}^{\prime\:\! i}_{\alpha} (\tau^\prime)=
\bar{\pi}^{i}_{\alpha}(\tau)$, 
while $f$ and $g$ transform  
 as 
\begin{subequations}
\label{3.4}
\begin{align}
f(\tau)& \rightarrow f^{\prime} (\tau^{\prime})
=\frac{d\tau}{d\tau^{\prime}} f(\tau) \,,
\label{3.4a}
\\
g(\tau)& \rightarrow g^{\prime}(\tau^{\prime})
=\frac{d\tau^{\prime}}{d\tau} g(\tau) \,. 
\label{3.4b}
\end{align}
\end{subequations}
Then it follows that Eqs. (\ref{3.3a}), (\ref{3.3b}), and (\ref{3.3c}) are compatible with 
the transformation rules (\ref{2.4b}), (\ref{2.4a}), and (\ref{2.4c}), respectively. 
For maintaining the reparametrization symmetry in the solution (\ref{3.3}), 
it is necessary to introduce scalar-density fields such as $f$ and $g$. 
We see that $p_{\alpha\dot{\alpha}}$, $q^{\alpha\dot{\alpha}}$, and $r_{\alpha\dot{\alpha}}$ 
in Eq. (\ref{3.3}) are invariant under the local $U(1)\times U(1)$ transformation 
\begin{subequations}
\label{3.5}
\begin{alignat}{3} 
\pi_{i \dot{\alpha}} &\rightarrow \pi_{i \dot{\alpha}}^{\prime}=e^{i\theta_i} \pi_{i \dot{\alpha}} \,, 
&\quad \;\;
\bar{\pi}{}^{i}_{\alpha} &\rightarrow \bar{\pi}{}^{\prime\:\! i}_{\alpha} =e^{-i\theta_i} \bar{\pi}{}^{i}_{\alpha} \,, 
\label{3.5a}
\\
e &\rightarrow e^{\prime}=e \,, 
&\quad \;\; 
f &\rightarrow f^{\prime}=f \,,
\label{3.5b}
\\
g &\rightarrow g^{\prime}=e^{i(\theta_1 -\theta_2)} g \,, 
&\quad \;\;
\bar{g} &\rightarrow \bar{g}^{\prime}=e^{-i(\theta_1 -\theta_2)} \bar{g}  
\label{3.5c}
\end{alignat}
\end{subequations}
with real gauge functions $\theta_{i}=\theta_{i}(\tau)$ $(i=1,2)$. 
Hereafter, we refer to the local $U(1)$ transformation with $\theta_{i}$ as the $U(1)_{i}$ transformation. 
Its corresponding gauge group is simply denoted as $U(1)_{i}$.

From Eqs. (\ref{3.3a})--(\ref{3.3c}), we can obtain 
\begin{subequations}
\label{3.6}
\begin{align}
p^{2} &=2 |\varPi|^{2} \:\!,
\label{3.6a}
\\
q^{2} &=2\gamma f^{2} |\varPi|^{2} \:\!,
\label{3.6b}
\\
r^{2} &=-2 \tilde{g}^{2} |\varPi|^{2} \:\!, 
\label{3.6c}
\\
q^{\alpha\dot{\alpha}} p_{\alpha\dot{\alpha}} &
=(\gamma+1) f \:\!|\varPi|^{2} \:\!,  
\label{3.6d}
\end{align}
\end{subequations}
where $\tilde{g}:=|g|$, and $\varPi$ is a complex scalar field defined by 
$\varPi:=\frac{1}{2} \epsilon^{ij} \pi_{i\dot{\alpha}} \pi{}_{j}^{\dot{\alpha}} 
=\pi_{1\dot{\alpha}} \pi{}_{2}^{\dot{\alpha}}$. 
(In this paper, $\epsilon^{ij}$ and $\epsilon_{ij}$ denote Levi-Civita symbols specified by $\epsilon^{12}=\epsilon_{12}=1$ 
and follow the rules $\overline{\epsilon^{ij}}=\epsilon_{ij}$ and $\overline{\epsilon_{ij}}=\epsilon^{ij}$.)  
From Eq. (\ref{3.5a}), we see that $\varPi$ transforms as 
$\varPi \rightarrow \varPi^{\prime}=e^{i(\theta_{1}+\theta_{2})} \varPi$ under the $U(1)_{1}\times U(1)_{2}$ transformation. 
In order that Eqs. (\ref{3.6b}) and (\ref{3.6c}) can be consistent with the conditions $q^{2}>0$ and $r^{2}<0$, 
it is necessary to postulate that $\pi_{1\dot{\alpha}}$ and $\pi_{2\dot{\alpha}}$ are not proportional to each other, as 
$\pi_{1\dot{\alpha}} \neq l^{\:\!} \pi_{2\dot{\alpha}}$ (${}^{\,\!} l \in \Bbb{C}$). 
Clearly, Eqs. (\ref{3.6b}) and (\ref{3.6c}) are compatible with the constraint $\sqrt{-q^{2} r^{2}}=|k|$.   
Also, Eqs. (\ref{3.6b}) and (\ref{3.6d}) are compatible with Eq. (\ref{2.7}), and 
Eqs. (\ref{3.6a}) and (\ref{3.6c}) are compatible with Eq. (\ref{2.8}), provided $e$ is determined by Eq. (\ref{2.8}).

Substituting Eqs. (\ref{3.3}), (\ref{3.6b}), and (\ref{3.6c}) into (\ref{3.1}), we obtain 
\begin{align}
L = & -\dot{x}{}^{\alpha\dot{\alpha}} \left(\bar{\pi}^{1}_{\alpha} \pi_{1\dot{\alpha}} 
+\bar{\pi}^{2}_{\alpha} \pi_{2\dot{\alpha}} \right) 
\notag
\\
& -if \left( g \varPi \Dot{\Bar{\pi}}{}^{1}_{\alpha} \bar{\pi}{}^{1\alpha} 
-\bar{g} \bar{\varPi} \Dot{\pi}{}_{1\dot{\alpha}} \pi{}_{1}^{\dot{\alpha}}  
+\gamma \bar{g} \varPi \Dot{\Bar{\pi}}{}^{2}_{\alpha} \bar{\pi}{}^{2\alpha} 
-\gamma g \bar{\varPi} \Dot{\pi}{}_{2\dot{\alpha}} \pi{}_{2}^{\dot{\alpha}} \right) 
\notag
\\
& +\tilde{f} \left( \sqrt{2} \:\! |\varPi| -M \right) 
+2e \left( \frac{2\sqrt{2\gamma}}{\gamma+1} \:\! \tilde{f} \tilde{g} |\varPi| -|k| \right) ,
\label{3.7}
\end{align}
where $\tilde{f}:=(\gamma+1)\:\! |\varPi|f /\sqrt{2}^{\,}(>0)$ and 
\begin{align}
M:=\frac{2\sqrt{\gamma}}{\gamma+1} m \,.
\label{3.8}
\end{align}
The constant $M$ is identified as a physical mass parameter. At this place, 
$\big( x^{\alpha\dot{\alpha}}, \pi{}_{i \dot{\alpha}}, \bar{\pi}{}^{i}_{\alpha}, e, f, $ $g, \bar{g} \big)$ 
are considered to be independent coordinate variables. 
The Lagrangian (\ref{3.7}) is, of course, invariant under the $U(1)_{1}\times U(1)_{2}$ transformation. 
Now we define the new 2-component spinors $\omega_{i}^{\alpha}=\omega_{i}^{\alpha}(\tau)$ by 
\begin{subequations}
\label{3.9}
\begin{align}
\omega_{1}^{\alpha} &:= ix^{\alpha\dot{\alpha}} \pi_{1\dot{\alpha}} 
+fg \varPi \bar{\pi}^{1\alpha} ,
\label{3.9a}
\\
\omega_{2}^{\alpha} &:= ix^{\alpha\dot{\alpha}} \pi_{2\dot{\alpha}} 
+\gamma f\bar{g} \varPi \bar{\pi}^{2\alpha} . 
\label{3.9b}
\end{align}
\end{subequations}
Their complex conjugates, $\bar{\omega}^{i \dot{\alpha}}=\bar{\omega}^{i \dot{\alpha}}(\tau)$, are given by$^{\;\!}$\footnote{~A similar, 
but essentially different expression 
$\bar{\omega}^{i \dot{\alpha}} = -ix^{\alpha\dot{\alpha}} \bar{\pi}{}^{i}_{\alpha} 
+\rho^{-1} \epsilon^{ij} \pi_{j}^{\dot{\alpha}}$ 
has been presented in Ref. \cite{CRZ}. 
(Here, $\rho$ denotes the radial coordinate of $\mathrm{AdS}_{5}$ space.)  
This expression is covariant under the $U(1) \times \mathit{SU}(2)$ transformation, whereas 
Eqs. (\ref{3.9}) and (\ref{3.10}) are covariant only under the $U(1)_{1}\times U(1)_{2}$ transformation.} 
\begin{subequations}
\label{3.10}
\begin{align}
\bar{\omega}^{1 \dot{\alpha}} &:= -ix^{\alpha\dot{\alpha}} \bar{\pi}{}^{1}_{\alpha} 
+f\bar{g} \bar{\varPi} \pi_{1}^{\dot{\alpha}} ,
\label{3.10a}
\\
\bar{\omega}^{2 \dot{\alpha}} &:= -ix^{\alpha\dot{\alpha}} \bar{\pi}{}^{2}_{\alpha} 
+\gamma fg \bar{\varPi} \pi_{2}^{\dot{\alpha}} . 
\label{3.10b}
\end{align}
\end{subequations}
Here we note that $x^{\alpha\dot{\alpha}}$ satisfy the Hermiticity condition $\overline{x^{\alpha\dot{\beta}}}=x^{\beta\dot{\alpha}}$, 
because $x^{\mu}$ are real. 
It is evident that $\omega_{i}^{\alpha}$ and $\bar{\omega}^{i \dot{\alpha}}$ behave as complex scalar fields on $\mathcal{T}$. 
Under the $U(1)_{1} \times U(1)_{2}$ transformation, 
$\omega_{i}^{\alpha}$ and $\bar{\omega}^{i \dot{\alpha}}$ transform as 
\begin{align} 
\omega_{i}^{\alpha} \rightarrow \omega_{i}^{\prime \alpha}=e^{i\theta_i} \omega_{i}^{\alpha} \,, 
\qquad 
\bar{\omega}{}^{i \dot{\alpha}} \rightarrow \bar{\omega}{}^{\prime\:\! i \dot{\alpha}} 
=e^{-i\theta_i} \bar{\omega}{}^{i \dot{\alpha}} \,.  
\label{3.11}
\end{align}
We can easily show that 
\begin{align}
\bar{\pi}{}^{1}_{\alpha} \omega_{1}^{\alpha} 
+\bar{\omega}{}^{1 \dot{\alpha}} \pi_{1 \dot{\alpha}} =0 \,, 
\qquad 
\bar{\pi}{}^{2}_{\alpha} \omega_{2}^{\alpha} 
+\bar{\omega}{}^{2 \dot{\alpha}} \pi_{2 \dot{\alpha}} =0 \,, 
\label{3.12}
\end{align} 
which, respectively, remain invariant under the $U(1)_{1}$ and $U(1)_{2}$ transformations.

With Eqs. (\ref{3.9}) and (\ref{3.10}), the Lagrangian (\ref{3.7}) can be written as  
\begin{align}
L &= \frac{i}{2} \left( \bar{\pi}{}^{i}_{\alpha} \dot{\omega}{}_{i}^{\alpha} 
+\bar{\omega}{}^{i \dot{\alpha}} \dot{\pi}_{i \dot{\alpha}} 
-\omega_{i}^{\alpha} \Dot{\bar{\pi}}{}^{i}_{\alpha}  
-\pi_{i \dot{\alpha}} \Dot{\bar{\omega}}{}^{i \dot{\alpha}} \right)
\notag
\\
&\quad\, +\tilde{f} \left( \sqrt{2} \:\! |\varPi| -M \right) 
+2e \left( \frac{2\sqrt{2\gamma}}{\gamma+1} \:\! \tilde{f} \tilde{g} |\varPi| -|k| \:\! \right) ,
\label{3.13}
\end{align}
which can be expressed more concisely as 
\begin{align}
L& =\frac{i}{2} \left( \bar{Z}{}^{i}_{A} \dot{Z}{}_{i}^{A} -Z_{i}^{A} \Dot{\bar{Z}}{}^{i}_{A} \right)
+\tilde{f} \left( \sqrt{2} \:\! |\varPi| -M \right) 
+2e \left( \frac{2\sqrt{2\gamma}}{\gamma+1} \:\! \tilde{f} \tilde{g} |\varPi| -|k| \:\! \right) 
\label{3.14}
\end{align}
in terms of the (novel) twistors defined by $Z_{i}^{A}:=\big(\omega_{i}^{\alpha}, \pi_{i \dot{\alpha}} \big)$ 
and their dual twistors $\bar{Z}{}^{i}_{A}:=\big(\bar{\pi}{}^{i}_{\alpha}, \bar{\omega}{}^{i \dot{\alpha}} \big)$ 
($A=0,1,2,3$). 
Equation (\ref{3.12}) can now be expressed as 
\begin{align}
\bar{Z}{}^{1}_{A} Z{}_{1}^{A}=0 \,, \qquad  \bar{Z}{}^{2}_{A} Z{}_{2}^{A}=0 \,. 
\label{3.15}
\end{align} 
These are precisely the null twistor conditions, and hence $Z_{i}^{A}$ turn out to be null twistors \cite{PenMac, PenRin2, HugTod}. 
The transformation rules (\ref{3.5a}) and (\ref{3.11}) can be combined into 
\begin{align} 
Z_{i}^{A} \rightarrow Z_{i}^{\prime A}=e^{i\theta_i} Z_{i}^{A} \,, 
\qquad 
\bar{Z}{}^{i}_{A} \rightarrow \bar{Z}{}^{\prime\:\! i}_{A}=e^{-i\theta_i} \bar{Z}{}^{i}_{A}\,.  
\label{3.16}
\end{align}

If we regard $\omega_{i}^{\alpha}$ and $\bar{\omega}{}^{i \dot{\alpha}}$ as primary independent variables, 
without referring to Eqs. (\ref{3.9}) and (\ref{3.10}), 
then the Lagrangian (\ref{3.14}) itself does not remain invariant under the $U(1)_{1} \times U(1)_{2}$ transformation 
owing to the existence of derivatives with respect to $\tau$. This Lagrangian becomes invariant only if 
the null twistor conditions in Eq. (\ref{3.15}) are used. 
Now we incorporate the null twistor conditions into the Lagrangian (\ref{3.14}) with the aid of 
real Lagrange multiplier fields, $a_{i}=a_{i}(\tau)$, on $\mathcal{T}$,  
so that all the twistor components can be treated as independent variables at the Lagrangian level. 
The Lagrangian (\ref{3.14}) is thus improved as follows: 
\begin{align}
L& =\frac{i}{2} \left( \bar{Z}{}^{i}_{A} \dot{Z}{}_{i}^{A} -Z_{i}^{A} \Dot{\bar{Z}}{}^{i}_{A} \right) 
+\sum_{i=1,2} a_{i} \bar{Z}{}^{i}_{A} Z{}_{i}^{A}
\notag
\\
&\quad\, +\tilde{f} \left( \sqrt{2} \:\! |\varPi| -M \right) 
+2e \left( \frac{2\sqrt{2\gamma}}{\gamma+1} \:\! \tilde{f} \tilde{g} |\varPi| -|k| \:\! \right) ,   
\label{3.17}
\end{align}
From this Lagrangian, 
the null twistor conditions can be derived as the Euler-Lagrange equations for $a_{1}$ and $a_{2}$. 
The field $a_{i}$ is assumed to transform, under the proper reparametrization, as 
\begin{align}
a_{i}(\tau)& \rightarrow a_{i}^{\prime} (\tau^{\prime})
=\frac{d\tau}{d\tau^{\prime}} a_{i}(\tau) 
\label{3.18}
\end{align}
so as to maintain the reparametrization invariance of the action  
${S}=\int_{\;\!\tau_0}^{\;\!\tau_1} d\tau^{\,\!} L$ with Eq. (\ref{3.17}).  
Also, we assume that under the $U(1)_{i}$ transformation, $a_{i}$ transforms  as 
\begin{align}
a_{i} \rightarrow a_{i}^{\prime}=a_{i} +\dot{\theta}_{i} \,.
\label{3.19}
\end{align}
Thereby, the $U(1)_{1} \times U(1)_{2}$ invariance of the Lagrangian (\ref{3.7}) is recovered in 
the Lagrangian (\ref{3.17}). 
If we substitute Eqs. (\ref{3.9}) and (\ref{3.10}) into Eq. (\ref{3.17}), it reduces to Eq. (\ref{3.7}). 
The Lagrangian (\ref{3.17}) can be written as 
\begin{align}
L& =\frac{i}{2} \sum_{i=1,2} \left( \bar{Z}{}^{i}_{A} D_{i} Z_{i}^{A} -Z_{i}^{A} \bar{D}_{i} \bar{Z}{}^{i}_{A} \right) 
+\tilde{f} \left( \sqrt{2} \:\! |\varPi| -M \right) 
+2e \left( \frac{2\sqrt{2\gamma}}{\gamma+1} \:\! \tilde{f} \tilde{g} |\varPi| -|k| \:\! \right) ,
\label{3.20}
\end{align}
with 
\begin{align}
& D_{i} Z_{i}^{A}:= \dot{Z}{}_{i}^{A}-ia_{i} Z_{i}^{A} , \quad\;\;
\bar{D}_{i} \bar{Z}{}^{i}_{A}:= \Dot{\bar{Z}}{}^{i}_{A}+ia_{i} \bar{Z}{}^{i}_{A} \,. 
\label{3.21}
\end{align}
(In Eq. (\ref{3.21}),  no summation is taken over $i$.) 
Using Eqs. (\ref{3.16}) and (\ref{3.19}), we can show that $D_{i} Z_{i}^{A}$ transforms like $Z_{i}^{A}$ under the $U(1)_{i}$ 
transformation. For this reason, $a_{i}$ and $D_{i}$ are identified as a $U(1)$ gauge field on $\mathcal{T}$ and  
its associated covariant derivative operator, respectively. 
At this stage, $\big( Z_{i}^{A}, \bar{Z}{}^{i}_{A}, a_{i}, e, \tilde{f}, \tilde{g} \big)$ 
are treated as independent coordinate variables.

From the Lagrangian (\ref{3.20}), the Euler-Lagrange equations for $e$, $\tilde{f}$, and $\tilde{g}$ 
are respectively found to be 
\begin{subequations}
\label{3.22}
\begin{align}
\frac{2\sqrt{2\gamma}}{\gamma+1} \:\! \tilde{f} \tilde{g} |\varPi| -|k| &=0 \,, 
\label{3.22a}
\\
\sqrt{2} \:\! |\varPi| -M +\frac{4\sqrt{2\gamma}}{\gamma+1} \:\! e \tilde{g} |\varPi| &=0 \,,
\label{3.22b}
\\
\frac{4\sqrt{2\gamma}}{\gamma+1} \:\! e \tilde{f} \:\! |\varPi| &=0 \,, 
\label{3.22c}
\end{align}
\end{subequations}
Since $\gamma$, $\tilde{f}$, and $|\varPi|$ are positive quantities, Eq. (\ref{3.22c}) reduces to $e=0$. 
Hence, Eq. (\ref{3.22b}) reads
\begin{align}
\sqrt{2}\:\!|\varPi|=M .
\label{3.23}
\end{align}
(This condition can also be derived from the Lagrangian (\ref{3.7}).) 
Equation (\ref{3.23}) is equivalent to$^{\;\!}$\footnote{~Equation (\ref{3.24}) can be written in the form of the 
mass-shell condition of Fedoruk-Lukierski type, as  
$\sqrt{2}^{\:\!} \varPi=Me^{i\varphi}$ and $\sqrt{2}^{\:\!} \bar{\varPi}=Me^{-i\varphi}$ \cite{FedLuk, AFIL}.}  
\begin{align}
\sqrt{2}\:\! e^{-i\varphi} \varPi=M , \qquad  \sqrt{2}\:\! e^{i\varphi} \bar{\varPi}=M . 
\label{3.24}
\end{align}
Here it is assumed that $\varphi=\varphi(\tau)$ is a real scalar field on $\mathcal{T}$ and transforms,   
under the $U(1)_{1}\times U(1)_{2}$ transformation, as 
\begin{align}
\varphi \rightarrow \varphi^{\prime}=\varphi +\theta_{1} +\theta_{2} \,.  
\label{3.25}
\end{align}
Accordingly, it follows that Eq. (\ref{3.24}) remains unchanged 
under both of the reparametrization and the $U(1)_{1}\times U(1)_{2}$ transformation. 
In order to obtain Eq. (\ref{3.24}) immediately, 
without referring to Eq. (\ref{3.23}), we modify the Lagrangian (\ref{3.20}) as 
\begin{align}
L& =\frac{i}{2} \sum_{i=1,2} \left( \bar{Z}{}^{i}_{A} D_{i} Z_{i}^{A} -Z_{i}^{A} \bar{D}_{i} \bar{Z}{}^{i}_{A} \right) 
+h\left( \sqrt{2}\:\! e^{-i\varphi} \varPi-M \right)
+\bar{h} \left( \sqrt{2}\:\! e^{i\varphi} \bar{\varPi}-M \right) 
\notag
\\
& \quad\, +2e \left\{ \frac{2\sqrt{2\gamma}}{\gamma+1} \big(h+\bar{h} \big)\tilde{g} |\varPi|-|k| \:\! \right\} ,
\label{3.26}
\end{align}
where $h$ is a $U(1)_{1} \times U(1)_{2}$ invariant complex field on $\mathcal{T}$ such that  
\begin{align}
h+\bar{h}=\tilde{f}\,(>0) \,, 
\label{3.27}
\end{align}
and obeys the transformation rule 
\begin{align}
h(\tau)& \rightarrow h^{\prime} (\tau^{\prime})
=\frac{d\tau}{d\tau^{\prime}} h(\tau) \,.
\label{3.28}
\end{align}
The action ${S}=\int_{\;\!\tau_0}^{\;\!\tau_1} d\tau^{\,\!} L$ with the modified Lagrangian (\ref{3.26}) has a form 
very similar to the GGS action found in Ref. \cite{DegOka}. 
Clearly, the action $S$ with Eq. (\ref{3.26}) is invariant under the reparametrization  
and the $U(1)_{1} \times U(1)_{2}$ transformation. 
At this place, $\big(Z_{i}^{A}, \bar{Z}{}^{i}_{A}, a_{i}, e, h, \bar{h}, \varphi, \tilde{g} \big)$   
are treated as independent coordinate variables. 
From the Lagrangian (\ref{3.26}), the Euler-Lagrange equations for $e$, $h$, $\bar{h}$, $\varphi$, and $\tilde{g}$  
are respectively found to be 
\begin{subequations}
\label{3.29}
\begin{align}
\frac{2\sqrt{2\gamma}}{\gamma+1} \big(h+\bar{h} \big) \tilde{g} |\varPi|-|k|=0 \,, 
\label{3.29a}
\\
\sqrt{2}\:\! e^{-i\varphi} \varPi -M+\frac{4\sqrt{2\gamma}}{\gamma+1} e \tilde{g} |\varPi|=0 \,, 
\label{3.29b}
\\
\sqrt{2}\:\! e^{i\varphi} \bar{\varPi} -M+\frac{4\sqrt{2\gamma}}{\gamma+1} e \tilde{g} |\varPi|=0 \,, 
\label{3.29c}
\\
\sqrt{2}\:\! e^{-i\varphi} \varPi h -\sqrt{2}\:\! e^{i\varphi} \bar{\varPi} \bar{h}=0 \,,
\label{3.29d}
\\
\frac{4\sqrt{2\gamma}}{\gamma+1} e\big(h+\bar{h} \big)|\varPi|=0 \,. 
\label{3.29e}
\end{align}
\end{subequations}
Equations (\ref{3.29a}) and (\ref{3.29e}) are identical to (\ref{3.22a}) and (\ref{3.22c}), respectively.  
Equation (\ref{3.29e}) reduces to $e=0$, and accordingly Eqs. (\ref{3.29b}) and (\ref{3.29c}) reduce to the conditions in Eq. (\ref{3.24}). 
Applying these conditions to Eq. (\ref{3.29d}), we have $h=\bar{h}$, and it turns out that $h=\tilde{f}/2$. 
We thus see that the Lagrangian (\ref{3.26}) is equivalent to the Lagrangian (\ref{3.20}) 
and is established as a twistor representation of the Lagrangian for a massive particle with rigidity.

Now, we carry out a partial gauge-fixing for later convenience by imposing the condition 
\begin{align}
a_{1} +a_{2}=0 \,.
\label{3.30}
\end{align} 
This condition partially breaks the $U(1)_{1} \times U(1)_{2}$ invariance of the Lagrangian (\ref{3.26})  
so as to maintain its invariance under the restricted gauge transformation with the gauge functions $\theta_{i}$  
such that $\theta_{1}+\theta_{2}=0$. (The condition (\ref{3.30}) is equivalent to the Lorenz-type gauge condition 
$\dot{a}_{1} +\dot{a}_{2}=0$, provided that this condition is reparametrization-covariant.) 
We also impose the additional condition 
\begin{align} 
\varphi=0 \,, 
\label{3.31}
\end{align} 
which is compatible with the condition (\ref{3.30}), as can be seen from the transformation rule (\ref{3.25}). 
The condition (\ref{3.31}) itself can be regarded as a gauge-fixing condition. 
In fact, if the condition (\ref{3.31}) alone is primarily imposed, 
then Eq. (\ref{3.30}) is derived as a secondary constraint in the canonical Hamiltonian formalism. 
However,  it eventually turns out that imposing both the conditions (\ref{3.30}) and (\ref{3.31}) makes 
our analysis procedure simple and clear. For this reason, we now set the two conditions simultaneously. 
The conditions (\ref{3.30}) and (\ref{3.31}) can be incorporated into the Lagrangian (\ref{3.26}) by adding 
the gauge-fixing term $b(a_{1}+a_{2})$ and the associated term $M\zeta \varphi$ to it as follows: 
\begin{align}
L& =\frac{i}{2} \sum_{i=1,2} \left( \bar{Z}{}^{i}_{A} D_{i} Z_{i}^{A} -Z_{i}^{A} \bar{D}_{i} \bar{Z}{}^{i}_{A} \right) 
+h\left( \sqrt{2}\:\! e^{-i\varphi} \varPi-M \right)
+\bar{h} \left( \sqrt{2}\:\! e^{i\varphi} \bar{\varPi}-M \right) 
\notag
\\
& \quad\, +2e \left\{ \frac{2\sqrt{2\gamma}}{\gamma+1} \big(h+\bar{h} \big)\tilde{g} |\varPi|-|k| \:\! \right\} 
+b(a_{1}+a_{2})+M\zeta \varphi \,. 
\label{3.32}
\end{align}
Here, $b=b(\tau)$ is a Nakanishi-Lautrup real scalar field on $\mathcal{T}$, 
and $\zeta=\zeta(\tau)$ is a real field on $\mathcal{T}$ obeying 
the transformation rule 
\begin{align}
\zeta(\tau)& \rightarrow \zeta^{\prime} (\tau^{\prime})
=\frac{d\tau}{d\tau^{\prime}} \zeta(\tau) \,.
\label{3.33}
\end{align}
The reparametrization invariance of the action $S$ with Eq. (\ref{3.32}) is ensured. 
The conditions (\ref{3.30}) and (\ref{3.31}) can indeed be derived from the Lagrangian (\ref{3.32}) 
as the Euler-Lagrange equations for $b$ and $\zeta$, respectively. 
After adding the new term 
$L_{\rm N}:=b(a_{1}+a_{2})+M\zeta \varphi$, 
Eqs. (\ref{3.15}) and (\ref{3.29d}) are modified as follows: 
\begin{align}
\bar{Z}{}^{1}_{A} Z{}_{1}^{A}+b=0 \,, \qquad   \bar{Z}{}^{2}_{A} Z{}_{2}^{A}+b=0 \,, 
\label{3.34}
\\[5pt]
\sqrt{2}\:\! e^{-i\varphi} \varPi h -\sqrt{2}\:\! e^{i\varphi} \bar{\varPi} \bar{h}+iM\zeta=0 \,.   
\label{3.35}
\end{align}
By contrast,  the Euler-Lagrange equations (\ref{3.29a}), (\ref{3.29b}), (\ref{3.29c}), and (\ref{3.29e}) 
do not change after adding $L_{\rm N}$. 
Equation (\ref{3.35}) ultimately becomes $h-\bar{h}+i\zeta=0$. 
Combining this with Eq. (\ref{3.27}) gives $h=(\tilde{f} -i\zeta)/2$. 
We thus see that $-\zeta /2$ can be identified as the imaginary part of $h$.

\section{Canonical formalism}

In this section, we study the canonical Hamiltonian formalism of the twistor model governed by  
the Lagrangian (\ref{3.32}).  
The canonical momenta conjugate to the canonical coordinates  
$\big(Z_{i}^{A}, \bar{Z}{}^{i}_{A}, a_{i}, b, e, h, \bar{h}, \varphi, \zeta, \tilde{g} \big)$ are found from Eq. (\ref{3.32}) to be 
\begin{subequations}
\label{4.1}
\begin{align}
P_{A}^{i}
&:= \frac{\partial L}{\partial \dot{Z}^{A}_{i}}
=\frac{i}{2} \bar{{Z}}_{A}^{i} \,, 
\label{4.1a}
\\
\bar{P}^{A}_{i}
&:= \frac{\partial L}{\partial \dot{\bar{Z}}_{A}^{i}}
=-\frac{i}{2} {Z}^{A}_{i} , 
\label{4.1b}
\\
P^{(a)i}
&:= \frac{\partial L}{\partial \dot{a}_{i}}
= 0 \,, 
\label{4.1c}
\\
P^{(b)}
&:= \frac{\partial L}{\partial \dot{b}} 
= 0 \,, 
\label{4.1d}
\\
P^{(e)}
&:= \frac{\partial L}{\partial \dot{e}}
= 0 \,, 
\label{4.1e}
\\
P^{(h)}
&:= \frac{\partial L}{\partial \dot{h}}
= 0 \,, 
\label{4.1f}
\\
P^{(\bar{h})}
&:= \frac{\partial L}{\partial \dot{\bar{h}}}
= 0 \,, 
\label{4.1g} 
\\
P^{(\varphi)}
&:= \frac{\partial L}{\partial \dot{\varphi}}
= 0 \,, 
\label{4.1h}
\\
P^{(\zeta)}
&:= \frac{\partial L}{\partial \dot{\zeta}}
= 0 \,, 
\label{4.1i}
\\
P^{(\tilde{g})}
&:= \frac{\partial L}{\partial \Dot{\Tilde{g}}}
= 0 \,. 
\label{4.1j}
\end{align}
\end{subequations}
The canonical Hamiltonian corresponding to $L$ is defined by the Legendre transform of $L$, 
\begin{align}
H_{\rm{C}} &:=\dot{Z}^{A}_{i} P_{A}^{i} 
+\Dot{\Bar{Z}}_{A}^{i} \bar{P}^{A}_{i} 
+\dot{a}_{i} P^{(a)i} +\dot{b} P^{(b)} +\dot{e} P^{(e)} 
\notag
\\
& \quad\;\;\!
+\dot{h} P^{(h)} +\dot{\bar{h}} P^{(\bar{h})} 
+\dot{\varphi} P^{(\varphi)} +\dot{\zeta} P^{(\zeta)} 
+\dot{\tilde{g}} P^{(\tilde{g})} -L 
\notag 
\\
&\;=-\sum_{i=1,2} a_{i} \left(\bar{Z}_{A}^{i} Z^{A}_{i} +b \right) 
-h\left( \sqrt{2}\:\! e^{-i\varphi} \varPi-M \right) 
-\bar{h} \left( \sqrt{2}\:\! e^{i\varphi} \bar{\varPi}-M \right) 
\notag
\\
& \quad\;\;\! 
-2e \left\{ \frac{2\sqrt{2\gamma}}{\gamma+1} \big(h+\bar{h} \big)\tilde{g} |\varPi|-|k| \:\! \right\} 
-M\zeta \varphi \,. 
\label{4.2}
\end{align}
The equal-time Poisson brackets between the canonical variables are given by 
\begin{subequations}
\label{4.3}
\begin{alignat}{3}
\left\{ Z^{A}_{i} , P_{B}^{j} \right\} &= \delta_{i}^{j} \delta_{B}^{A}  \,, 
&\qquad \;
\left\{ \bar{Z}_{A}^{i} , \bar{P}^{B}_{j} \right\} &= \delta^{i}_{j} \delta_{A}^{B} \,, 
\label{4.3a}
\\
\left\{ a_{i} \:\!, P^{(a)j} \right\} &= \delta_{i}^{j}  \,, 
&\qquad \;
\left\{ b \:\!, P^{(b)} \right\} &= 1 \,, 
\label{4.3b}
\\
\left\{ e \:\!, P^{(e)} \right\} &= 1 \,, 
\label{4.3c}
\\
\left\{ h \:\!, P^{(h)} \right\} &= 1 \,, 
&\qquad \;
\left\{ \bar{h} \:\!, P^{(\bar{h})} \right\} &= 1 \,, 
\label{4.3d}
\\
\left\{ \varphi \:\!, P^{(\varphi)} \right\} &= 1 \,, 
&\qquad \;
\left\{ \zeta \:\!, P^{(\zeta)} \right\} &= 1 \,, 
\label{4.3e}
\\
\left\{ \tilde{g} \:\!, P^{(\tilde{g})} \right\} &= 1 \,, 
&\qquad \;
\mbox{all others} &=0 \,, 
\label{4.3f}
\end{alignat}
\end{subequations}
which can be used for calculating the Poisson bracket between two arbitrary analytic functions 
of the canonical variables.

Equations (\ref{4.1a})--(\ref{4.1j}) are read as the primary constraints 
\begin{subequations}
\label{4.4}
\begin{align}
\phi_{A}^{i} &:=P_{A}^{i} -\frac{i}{2} \bar{Z}_{A}^{i} \approx 0 \,, 
\label{4.4a}
\\
\bar{\phi}^{A}_{i} &:=\bar{P}^{A}_{i} +\frac{i}{2} {Z}^{A}_{i} \approx 0 \,, 
\label{4.4b}
\\
\phi^{(a)i}	&:=P^{(a)i} \approx 0 \,, 
\label{4.4c}
\\
\phi^{(b)} &:=P^{(b)} \approx 0 \,, 
\label{4.4d}
\\
\phi^{(e)} &:=P^{(e)} \approx 0 \,, 
\label{4.4e}
\\
\phi^{(h)}	&:=P^{(h)} \approx 0 \,, 
\label{4.4f}
\\
\phi^{(\bar{h})}	&:=P^{(\bar{h})} \approx 0 \,, 
\label{4.4g}
\\
\phi^{(\varphi)} &:=P^{(\varphi)} \approx 0 \,, 
\label{4.4h}
\\
\phi^{(\zeta)} &:=P^{(\zeta)} \approx 0 \,, 
\label{4.4i}
\\
\phi^{(\tilde{g})} &:=P^{(\tilde{g})} \approx 0 \,, 
\label{4.4j}
\end{align}
\end{subequations}
where the symbol ^^ ^^ $\approx$" denotes the weak equality. 
Now, we apply the Dirac algorithm for constrained Hamiltonian systems \cite{Dirac2, HRT, HenTei} 
to develop the canonical formalism of the present model. 
Using Eq. (\ref{4.3}), the Poisson brackets between the primary constraint functions $\phi$'s 
are found to be  
\begin{align}
\left\{ \phi^{i}_{A} , \bar{\phi}{}_{j}^{B} \right\} =-i\delta^{i}_{j} \delta_{A}^{B} \,,  
\quad \;\:
\mbox{all others} =0 \,. 
\label{4.5}
\end{align}
The Poisson brackets between $H_{\rm{C}}$ and the primary constraint functions can be calculated to obtain 
\begin{subequations}
\label{4.6}
\begin{align}
\left\{ \phi^{i}_{A} , H_{\rm{C}} \right\}
&= a_{i} \bar{Z}{}^{i}_{A} 
+\left\{ \sqrt{2}\:\! e^{-i\varphi} h+\frac{2\sqrt{2\gamma}}{\gamma+1} e \left(h+\bar{h} \right)\tilde{g} 
\frac{\bar{\varPi}}{|\varPi|} \right\} \:\! \epsilon^{ij} I_{AB} Z_{j}^{B}, 
\label{4.6a}
\\
\left\{ \bar{\phi}_{i}^{A} , H_{\rm{C}} \right\}
&= a_{i} Z_{i}^{A} 
+\left\{ \sqrt{2}\:\! e^{i\varphi} \bar{h} +\frac{2\sqrt{2\gamma}}{\gamma+1} e \left(h+\bar{h} \right)\tilde{g} 
\frac{\varPi}{|\varPi|} \right\} \:\! \epsilon_{ij} I^{AB} \bar{Z}^{j}_{B} \,, 
\label{4.6b}
\\[2pt]
\left\{ \phi^{(a)i} , H_{\rm{C}} \right\}
&=\bar{Z}_{A}^{i} Z^{A}_{i} +b \,, 
\label{4.6c}
\\
\left\{ \phi^{(b)} , H_{\rm{C}} \right\}
&=a_{1}+a_{2} \,, 
\label{4.6d}
\\[2pt]
\left\{ \phi^{(e)} , H_{\rm{C}} \right\}
&=2 \left\{ \frac{2\sqrt{2\gamma}}{\gamma+1} \big(h+\bar{h} \big)\tilde{g} |\varPi|-|k| \:\! \right\} \,, 
\label{4.6e}
\\
\left\{ \phi^{(h)} , H_{\rm{C}} \right\}
&=\sqrt{2}\:\! e^{-i\varphi} \varPi-M 
+\frac{4\sqrt{2\gamma}}{\gamma+1} e \tilde{g} |\varPi|  \,, 
\label{4.6f}
\\[2pt]
\left\{ \phi^{(\bar{h})} , H_{\rm{C}} \right\}
&=\sqrt{2}\:\! e^{i\varphi} \bar{\varPi}-M 
+\frac{4\sqrt{2\gamma}}{\gamma+1} e \tilde{g} |\varPi|  \,, 
\label{4.6g}
\\[2pt]
\left\{ \phi^{(\varphi)} , H_{\rm{C}} \right\}
&=-i \sqrt{2} \left( e^{-i\varphi} \varPi h -e^{i\varphi} \bar{\varPi} \bar{h} \right) +M\zeta \,, 
\label{4.6h}
\\
\left\{ \phi^{(\zeta)} , H_{\rm{C}} \right\}
&=M\varphi \,, 
\label{4.6i}
\\
\left\{ \phi^{(\tilde{g})} , H_{\rm{C}} \right\}
&= \frac{4\sqrt{2\gamma}}{\gamma+1} e \left(h+\bar{h} \right) |\varPi| \,,
\label{4.6j}
\end{align}
\end{subequations}
where $I_{AB}$ and $I^{AB}$ are the so-called infinity twistors \cite{PenMac, PenRin2}, 
defined by 
\begin{align}
I_{AB}:= \left(\,
\begin{array}{cc}
0 & \;\! 0 \\
0 & \;\, \epsilon^{\dot{\alpha} \dot{\beta}} 
\end{array}
\! \right) , 
\qquad 
I^{AB}:= \left(
\begin{array}{cc}
\epsilon^{\alpha\beta} &\:\:\! 0 \\
0 & \:\:\! 0 
\end{array}
\:\! \right) .
\label{4.7}
\end{align}
(In Eqs. (\ref{4.6a}), (\ref{4.6b}), and (\ref{4.6c}),  no summation is taken over $i$.) 
With $H_{\rm{C}}$ and the primary constraint functions, we define the total Hamiltonian 
\begin{align}
H_{\rm{T}} &:=H_{\rm{C}} +u_{i}^{A} \phi^{i}_{A} +\bar{u}^{i}_{A} \bar{\phi}_{i}^{A} 
+u_{(a)i} \phi^{(a)i} +u_{(b)} \phi^{(b)} +u_{(e)} \phi^{(e)} 
\notag
\\
&\quad\:\, 
+u_{(h)} \phi^{(h)} +u_{(\bar{h})} \phi^{(\bar{h})} +u_{(\varphi)} \phi^{(\varphi)} +u_{(\zeta)} \phi^{(\zeta)} 
+u_{(\tilde{g})} \phi^{(\tilde{g})} , 
\label{4.8}
\end{align}
where $u_{i}^{A}$, $\bar{u}^{i}_{A}$, $u_{(a)i}$, $u_{(b)}$, $u_{(e)}$, 
$u_{(h)}$, $u_{(\bar{h})}$, $u_{(\varphi)}$, $u_{(\zeta)}$, and $u_{(\tilde{g})}$ are Lagrange multipliers.  
The time evolution of a function $\mathcal{F}$ of the canonical variables is governed by 
the canonical equation 
\begin{align}
\dot{\mathcal{F}}=\{ \mathcal{F}, H_{\rm{T}} \}\,. 
\label{4.9}
\end{align}
Using this equation together with Eqs. (\ref{4.4}), (\ref{4.5}), (\ref{4.6}), and (\ref{4.8}),  
we can evaluate the time evolution of the primary constraint functions. 
Because the primary constraints (\ref{4.4a})--(\ref{4.4j}) are valid at any time, 
they must be preserved in time. This fact leads to the consistency conditions 
\begin{subequations}
\label{4.10}
\begin{align}
\dot{\phi}{}^{i}_{A} &
=\left\{ \phi^{i}_{A} , H_{\rm{T}} \right\}
\approx a_{i} \bar{Z}{}^{i}_{A} 
+\left\{ \sqrt{2}\:\! e^{-i\varphi} h +\frac{2\sqrt{2\gamma}}{\gamma+1} e \left(h+\bar{h} \right)\tilde{g} 
\frac{\bar{\varPi}}{|\varPi|} \right\} \:\! \epsilon^{ij} I_{AB} Z_{j}^{B} -i\bar{u}^{i}_{A} 
\approx 0 \,, 
\label{4.10a}
\\
\Dot{\Bar{\phi}}{}_{i}^{A} &
=\left\{ \bar{\phi}_{i}^{A} , H_{\rm{T}} \right\} 
\approx a_{i} Z_{i}^{A} 
+\left\{ \sqrt{2}\:\! e^{i\varphi} \bar{h} +\frac{2\sqrt{2\gamma}}{\gamma+1} e \left(h+\bar{h} \right)\tilde{g} 
\frac{\varPi}{|\varPi|} \right\} \:\! \epsilon_{ij} I^{AB} \bar{Z}^{j}_{B} +iu_{i}^{A} 
\approx 0 \,, 
\label{4.10b}
\\[2pt]
\dot{\phi}{}^{(a)i} &
=\left\{ \phi^{(a)i} , H_{\rm{T}} \right\}
\approx 
\bar{Z}_{A}^{i} Z^{A}_{i} +b
\approx 0 \,,
\label{4.10c}
\\
\dot{\phi}{}^{(b)} &
=\left\{ \phi^{(b)} , H_{\rm{T}} \right\}
\approx 
a_{1}+a_{2} 
\approx 0 \,,
\label{4.10d}
\\[2pt]
\dot{\phi}{}^{(e)} &
=\left\{ \phi^{(e)} , H_{\rm{T}} \right\} 
\approx 
2 \left\{ \frac{2\sqrt{2\gamma}}{\gamma+1} \left(h+\bar{h} \right)\tilde{g} |\varPi|-|k| \:\! \right\} 
\approx 0 \,, 
\label{4.10e}
\\
\dot{\phi}{}^{(h)} &
=\left\{ \phi^{(h)} , H_{\rm{T}} \right\}
\approx 
\sqrt{2}\:\! e^{-i\varphi} \varPi-M 
+\frac{4\sqrt{2\gamma}}{\gamma+1} e \tilde{g} |\varPi|  
\approx 0 \,, 
\label{4.10f}
\\[2pt]
\dot{\phi}{}^{(\bar{h})} &
=\left\{ \phi^{(\bar{h})} , H_{\rm{T}} \right\}
\approx 
\sqrt{2}\:\! e^{i\varphi} \bar{\varPi}-M 
+\frac{4\sqrt{2\gamma}}{\gamma+1} e \tilde{g} |\varPi|  
\approx 0 \,, 
\label{4.10g}
\\[2pt]
\dot{\phi}{}^{(\varphi)} &
=\left\{ \phi^{(\varphi)} , H_{\rm{T}} \right\}
\approx 
-i \sqrt{2} \left( e^{-i\varphi} \varPi h -e^{i\varphi} \bar{\varPi} \bar{h} \right) +M\zeta 
\approx 0 \,, 
\label{4.10h}
\\
\dot{\phi}{}^{(\zeta)} &
=\left\{ \phi^{(\zeta)} , H_{\rm{T}} \right\}
\approx 
M\varphi 
\approx 0 \,, 
\label{4.10i}
\\
\dot{\phi}{}^{(\tilde{g})} & 
=\left\{ \phi^{(\tilde{g})} , H_{\rm{T}} \right\}
\approx 
\frac{4\sqrt{2\gamma}}{\gamma+1} e \left(h+\bar{h} \right) |\varPi| 
\approx 0 \,.
\label{4.10j}
\end{align}
\end{subequations}
Equations (\ref{4.10a}) and (\ref{4.10b}) determine $\bar{u}^{i}_{A}$ and $u_{i}^{A}$, respectively, 
as follows: 
\begin{subequations}
\label{4.11}
\begin{align}
\bar{u}^{i}_{A} &=
-ia_{i} \bar{Z}{}^{i}_{A} 
-i\left\{ \sqrt{2}\:\! e^{-i\varphi} h +\frac{2\sqrt{2\gamma}}{\gamma+1} e \left(h+\bar{h} \right)\tilde{g} 
\frac{\bar{\varPi}}{|\varPi|} \right\} \:\! \epsilon^{ij} I_{AB} Z_{j}^{B} , 
\label{4.11a}
\\
u_{i}^{A} &= 
ia_{i} Z_{i}^{A} 
+i\left\{ \sqrt{2}\:\! e^{i\varphi} \bar{h} +\frac{2\sqrt{2\gamma}}{\gamma+1} e \left(h+\bar{h} \right)\tilde{g} 
\frac{\varPi}{|\varPi|} \right\} \:\! \epsilon_{ij} I^{AB} \bar{Z}^{j}_{B} \,. 
\label{4.11b}
\end{align}
\end{subequations}
Equations (\ref{4.10i}) and (\ref{4.10j}) reduce to $\varphi \approx 0$ and $e \approx 0$, respectively. 
Taking into account $\varphi \approx 0$ and $e \approx 0$, 
we see that Eqs. (\ref{4.10c})--(\ref{4.10j}) yield the secondary constraints 
\begin{subequations}
\label{4.12}
\begin{align}
\chi^{(a)i} &:=\bar{Z}_{A}^{i} Z^{A}_{i} +b 
\approx 0 \quad (\mbox{no sum with respect to $i\:\!$}), 
\label{4.12a}
\\
\chi^{(b)} &:=a_{1}+a_{2} 
\approx 0 \,, 
\label{4.12b}
\\
\chi^{(e)} &:=\left(h+\bar{h} \right)\tilde{g} -\frac{\left(\gamma+1 \right) |k|}{2\sqrt{\gamma} M} 
\approx 0 \,, 
\label{4.12c}
\\
\chi^{(h)} &:=\sqrt{2}\:\! \varPi-M 
\approx 0 \,, 
\label{4.12d}
\\
\chi^{(\bar{h})} &:=\sqrt{2}\:\! \bar{\varPi}-M 
\approx 0 \,, 
\label{4.12e}
\\
\chi^{(\varphi)} &:=i \left( h -\bar{h} \right) -\zeta 
\approx 0 \,, 
\label{4.12f}
\\
\chi^{(\zeta)} &:= \varphi \approx 0 \,.
\label{4.12g}
\\
\chi^{(\tilde{g})} &:= e \approx 0 \,.
\label{4.12h}
\end{align}
\end{subequations}
All the Poisson brackets between $H_{\rm C}$ and the secondary constraint functions $\chi$'s vanish. 
The Poisson brackets between the primary and secondary constraint functions  
are found to be   
\begin{subequations}
\label{4.13}
\begin{alignat}{3}
\left\{\:\! \chi^{(a)i}, \phi^{j}_{A} \right\} &=\delta_{i}^{j} \bar{Z}_{A}^{i} \,, 
& 
\left\{\:\! \chi^{(a)i}, \bar{\phi}_{j}^{A} \right\} &=\delta_{j}^{i} Z^{A}_{i} , 
\label{4.13a}
\\
\left\{\:\! \chi^{(a)i}, \phi^{(b)} \right\} &=1 \,, 
& 
\left\{\:\! \chi^{(b)}, \phi^{(a)i} \right\} &=1 \,, 
\label{4.13b}
\\
\left\{\:\!  \chi^{(e)}, \phi^{(h)} \right\} &=\tilde{g} \,, 
& 
\left\{\:\!  \chi^{(e)}, \phi^{(\bar{h})} \right\} &=\tilde{g} \,, 
\label{4.13c}
\\
\left\{\:\!  \chi^{(e)}, \phi^{(\tilde{g})} \right\} &=h+\bar{h} \,, 
& 
\left\{\:\!  \chi^{(\tilde{g})}, \phi^{(e)} \right\} & =1 \,, 
\label{4.13d}
\\
\left\{\:\!  \chi^{(h)}, \phi^{i}_{A} \right\} &=\sqrt{2} \:\! \epsilon^{ij} I_{AB} Z^{B}_{j} , 
& \qquad \;
\left\{\:\!  \chi^{(\bar{h})}, \bar{\phi}_{i}^{A} \right\} &=\sqrt{2} \:\! \epsilon_{ij} I^{AB} \bar{Z}_{B}^{j} \,, 
\label{4.13e} 
\\
\left\{\:\!  \chi^{(\varphi)}, \phi^{(h)} \right\} & = i \,, 
& 
\left\{\:\!  \chi^{(\varphi)}, \phi^{(\bar{h})} \right\} & = -i \,, 
\label{4.13f}
\\
\left\{\:\!  \chi^{(\varphi)}, \phi^{(\zeta)} \right\} & = -1 \,, 
& 
\left\{\:\!  \chi^{(\zeta)}, \phi^{(\varphi)} \right\} & = 1 \,, 
\label{4.13g}
\\
\mbox{all others} &=0 \,. 
\notag
\end{alignat}
\end{subequations}
All the Poisson brackets between the secondary constraint functions vanish.

Next we investigate the time evolution of the secondary constraint functions using Eqs. (\ref{4.9}) 
and (\ref{4.13}). The time evolution of $\chi^{(a)i}$ can be calculated as 
\begin{align}
\dot{\chi}{}^{(a)i} 
=\left\{ \:\! \chi^{(a)i}, H_{\rm{T}} \right\}
\approx
u_{i}^{A} \bar{Z}_{A}^{i} +\bar{u}^{i}_{A} Z^{A}_{i} +u_{(b)} 
\approx -M\zeta +u_{(b)} 
\label{4.14}
\\
(\mbox{no sum with respect to $i\:\!$}) 
\notag
\end{align}
by using   
Eqs. (\ref{4.11a}), (\ref{4.11b}), (\ref{4.12d}), (\ref{4.12e}), (\ref{4.12f}), (\ref{4.12g}), and (\ref{4.12h}), 
together with the formulas 
\begin{subequations}
\label{4.15}
\begin{align}
I_{AB} Z^{A}_{i} Z^{B}_{j}
&=\epsilon^{\dot{\alpha} \dot{\beta}} \pi{}_{i \dot{\alpha}} \pi{}_{j \dot{\beta}} =\varPi \epsilon_{ij} \,,  
\label{4.15a}
\\
I^{AB} \bar{Z}{}^{i}_{A} \bar{Z}{}^{j}_{B}
&=\epsilon^{\alpha \beta} \bar{\pi}{}^{i}_{\alpha} \bar{\pi}{}^{j}_{\beta} =\bar{\varPi} \epsilon^{ij} . 
\label{4.15b}
\end{align}
\end{subequations} 
Then the condition $\dot{\chi}{}^{(a)i}\approx 0$ determines $u_{(b)}$ to be $M\zeta$. 
The time evolution of $\chi^{(b)}$ and $\chi^{(e)}$ is found to be 
\begin{subequations}
\label{4.16}
\begin{align}
\dot{\chi}{}^{(b)}
&=\left\{\:\!  \chi^{(b)}, H_{\rm{T}} \right\} \approx u_{(a)1} +u_{(a)2} \,.
\label{4.16a}
\\
\dot{\chi}{}^{(e)}
&=\left\{\:\!  \chi^{(e)}, H_{\rm{T}} \right\}
\approx \left( u_{(h)}+u_{(\bar{h})} \right) \tilde{g} +\left(h+\bar{h} \right) u_{(\tilde{g})} \,. 
\label{4.16b}
\end{align}
\end{subequations}
The time evolution of $\chi^{(h)}$ is evaluated as 
\begin{align}
\dot{\chi}{}^{(h)} 
=\left\{\:\!  \chi^{(h)}, H_{\rm{T}} \right\}
& \approx 
\sqrt{2} \:\! \epsilon^{ij} I_{AB} u^{A}_{i} Z^{B}_{j}  
=i\sqrt{2} \:\! \varPi \sum_{i=1,2} \epsilon_{ij} \epsilon^{ij} a_{i}  
\notag
\\
& \approx iM (a_{1}+a_{2}) =iM\chi^{(b)} \approx 0 
\label{4.17}
\end{align}
by using $I_{AB} I^{AC}=0$ and Eqs. (\ref{4.11b}), (\ref{4.15a}), (\ref{4.12d}), and (\ref{4.12b}). 
Hence the condition $\dot{\chi}{}^{(h)} \approx 0$ is identically fulfilled. 
Similarly, it can be shown that $\dot{\chi}{}^{(\bar{h})}\approx 0$ is identically fulfilled.   
The time evolution of $\chi^{(\varphi)}$, $\chi^{(\zeta)}$, and $\chi^{(\tilde{g})}$  
is found to be 
\begin{subequations}
\label{4.18}
\begin{align}
\dot{\chi}{}^{(\varphi)} 
&=\left\{\:\!  \chi^{(\varphi)}, H_{\rm{T}} \right\}
\approx i\left( u_{(h)}-u_{(\bar{h})} \right) -u_{(\zeta)} \,, 
\label{4.18a}
\\
\dot{\chi}{}^{(\zeta)} 
&=\left\{\:\!  \chi^{(\zeta)}, H_{\rm{T}} \right\} \approx u_{(\varphi)} \,, 
\label{4.18b}
\\
\dot{\chi}{}^{(\tilde{g})} 
&=\left\{\:\!  \chi^{(\tilde{g})}, H_{\rm{T}} \right\} \approx u_{(e)} \,. 
\label{4.18c}
\end{align}
\end{subequations}
The conditions $\dot{\chi}{}^{(b)} \approx 0$, 
$\dot{\chi}{}^{(e)} \approx 0$, and $\dot{\chi}{}^{(\varphi)} \approx 0$, respectively, give 
\begin{subequations} 
\label{4.19}
\begin{align}
& u_{(a)1} +u_{(a)2}=0 \,, 
\label{4.19a}
\\
& u_{(\tilde{g})}=-\frac{\tilde{g}}{h+\bar{h}} \left( u_{(h)}+u_{(\bar{h})} \right) \,,  
\label{4.19b}
\\
& u_{(\zeta)}=i\left( u_{(h)}-u_{(\bar{h})} \right) \,.
\label{4.19c}
\end{align}
\end{subequations}
From the conditions $\dot{\chi}{}^{(\zeta)} \approx 0$ and $\dot{\chi}{}^{(\tilde{g})} \approx 0$, 
the Lagrange multipliers $u_{(\varphi)}$ and $u_{(e)}$ are determined to be zero. 
As can be seen from the above analysis, 
no any further constraints are derived, and hence the procedure for finding secondary constraints is now complete.  
We have seen that $u_{(a)1} +u_{(a)2}$, $u_{(\varphi)}$, and $u_{(e)}$ vanish and  
$u_{i}^{A}$, $\bar{u}^{i}_{A}$, $u_{(b)}$, $u_{(\zeta)}$, and $u_{(\tilde{g})}$ are determined to be what are written 
in terms of other variables such as the canonical coordinates. 
In contrast, $u_{(a)1} -u_{(a)2}$, $u_{(h)}$, and $u_{(\bar{h})}$ still remain as undetermined functions of $\tau$.

We have obtained all the Poisson brackets between 
the constraint functions, as in Eqs. (\ref{4.5}) and (\ref{4.13}). 
However, it is difficult to classify the constraints in Eqs. (\ref{4.4}) and (\ref{4.12}) 
into first and second classes on the basis of Eqs. (\ref{4.5}) and (\ref{4.13}) 
together with the vanishing Poisson brackets between the secondary constraint functions. 
To find simpler forms of the relevant Poisson brackets, we first define  
\begin{subequations}
\label{4.20}
\begin{align}
\tilde{\phi}{}^{(\tilde{g})}&:=\frac{1}{h+\bar{h}} \phi^{(\tilde{g})} , 
\label{4.20a}
\\
\tilde{\chi}^{(a)i}&:=\chi^{(a)i}+i\bar{Z}{}^{i}_{A} \bar{\phi}{}_{i}^{A} -i\phi^{i}_{A} Z_{i}^{A} 
\quad (\mbox{no sum with respect to $i\:\!$}),  
\label{4.20b}
\\
\tilde{\chi}{}^{(h)}&:=\chi^{(h)}+i\sqrt{2} \:\! \epsilon^{jk} I_{AB} \bar{\phi}{}_{j}^{A} Z^{B}_{k} -iM \phi^{(b)} , 
\label{4.20c}
\\
\tilde{\chi}{}^{(\bar{h})}&:=\chi^{(\bar{h})}-i\sqrt{2} \:\! \epsilon_{jk} I^{AB} \phi^{j}_{A} \bar{Z}{}^{k}_{B} +iM \phi^{(b)} ,    
\label{4.20d}
\end{align}
\end{subequations}
and furthermore define 
\begin{subequations}
\label{4.21}
\begin{align}
\phi^{(a)\pm}&:=\frac{1}{2} \left( \phi^{(a)1} \pm \phi^{(a)2} \right) \,, 
\label{4.21a}
\\
\tilde{\chi}^{(a)\pm}&:=\frac{1}{2} \left( \:\! \tilde{\chi}^{(a)1} \pm \tilde{\chi}^{(a)2} \right) \,, 
\label{4.21b}
\\
\phi^{(+)}&:=\frac{1}{2\tilde{g}}  \left( \phi^{(h)} +\phi^{(\bar{h})} \right) -\tilde{\phi}{}^{(\tilde{g})} , 
\label{4.21c}
\\
\phi^{(-)}&:=\frac{1}{2i} \left( \phi^{(h)} -\phi^{(\bar{h})} \right) +\phi^{(\zeta)} \,.  
\label{4.21d}
\end{align}
\end{subequations}
It is easy to see that the set of all the constraints given in Eqs. (\ref{4.4}) and (\ref{4.12}), i.e., 
\begin{align}
\Big( & \:\! \phi_{A}^{i}, \bar{\phi}{}^{A}_{i}, 
\phi^{(a)i}, \phi^{(b)}, \phi^{(e)}, \phi^{(h)}, \phi^{(\bar{h})}, \phi^{(\varphi)}, \phi^{(\zeta)}, \phi^{(\tilde{g})}, 
\notag
\\
& \chi^{(a)i}, \chi^{(b)}, \chi^{(e)}, \chi^{(h)}, \chi^{(\bar{h})},  \chi^{(\varphi)}, \chi^{(\zeta)}, \chi^{(\tilde{g})} \Big)  
\approx 0 \,, 
\label{4.22}
\end{align}
is equivalent to the new set of constraints 
\begin{align}
\Big( & \:\! \phi_{A}^{i}, \bar{\phi}{}^{A}_{i}, 
\phi^{(a)\pm}, \phi^{(b)}, \phi^{(e)}, \phi^{(+)}, \phi^{(-)}, \phi^{(\varphi)}, \phi^{(\zeta)}, \tilde{\phi}{}^{(\tilde{g})},  
\notag
\\
& \tilde{\chi}{}^{(a)\pm}, \chi^{(b)}, \chi^{(e)}, \tilde{\chi}{}^{(h)}, \tilde{\chi}{}^{(\bar{h})}, 
\chi^{(\varphi)}, \chi^{(\zeta)}, \chi^{(\tilde{g})} \Big)  
\approx 0 \,. 
\label{4.23}
\end{align}
We can show that except for 
\begin{subequations}
\label{4.24}
\begin{alignat}{3}
\left\{ \phi^{i}_{A} , \bar{\phi}{}_{j}^{B} \right\} &=-i\delta^{i}_{j} \delta_{A}^{B} \,,  
& ~ & 
\label{4.24a}
\\
\left\{\:\! \tilde{\chi}{}^{(a)+}, \phi^{(b)} \right\} &=1 \,, 
& \qquad
\left\{\:\! \chi^{(b)}, \phi^{(a)+} \right\} &=1 \,, 
\label{4.24b}
\\
\left\{\:\! \chi^{(e)}, \tilde{\phi}{}^{(\tilde{g})} \right\} &=1 \,,
& \qquad
\left\{\:\! \chi^{(\tilde{g})},  \phi^{(e)} \right\} &= 1 \,, 
\label{4.24c}
\\
\left\{\:\! \chi^{(\varphi)}, \phi^{(\zeta)} \right\} &= -1 \,, 
& \qquad
\left\{\:\! \chi^{(\zeta)}, \phi^{(\varphi)} \right\} &=1 \,,  
\label{4.24d}
\end{alignat}
\end{subequations}
all other Poisson brackets between the constraint functions in Eq. (\ref{4.23}) vanish. 
In this way, the relevant Poisson brackets are simplified in terms of 
the new constraint functions. 
Now we can readily see from the Poisson brackets (\ref{4.24}) that 
$\phi^{(a)-} \approx 0$, 
$\phi^{(+)} \approx 0$, 
$\phi^{(-)} \approx 0$, 
$\tilde{\chi}{}^{(a)-} \approx 0$, 
$\tilde{\chi}{}^{(h)} \approx 0$, 
and $\tilde{\chi}{}^{(\bar{h})} \approx 0$ 
are first-class constraints, while 
$\phi_{A}^{i} \approx 0$, 
$\bar{\phi}{}^{A}_{i} \approx 0$, 
$\phi^{(a)+} \approx 0$, 
$\phi^{(b)} \approx 0$, 
$\phi^{(e)} \approx 0$, 
$\phi^{(\varphi)} \approx 0$, 
$\phi^{(\zeta)} \approx 0$, 
$\tilde{\phi}{}^{(\tilde{g})} \approx 0$, 
$\tilde{\chi}{}^{(a)+} \approx 0$, 
$\chi^{(b)} \approx 0$, 
$\chi^{(e)} \approx 0$, 
$\chi^{(\varphi)} \approx 0$, 
$\chi^{(\zeta)} \approx 0$, 
and $\chi^{(\tilde{g})}  \approx 0$ 
are second-class constraints.

Following Dirac's approach to second-class constraints, we define the Dirac bracket between 
arbitrary smooth functions $\mathcal{F}$ and $\mathcal{G}$ of the canonical variables as follows: 
\begin{align}
\left\{ \mathcal{F} , \mathcal{G} \right\}_{\rm D} 
&:=\left\{ \mathcal{F} , \mathcal{G} \right\} 
+i \left\{ \mathcal{F} , \phi_{A}^{i} \right\}  \left\{ \bar{\phi}^{A}_{i} , \mathcal{G} \right\} 
-i \left\{ \mathcal{F} , \bar{\phi}^{A}_{i} \right\}  \left\{ \phi_{A}^{i} , \mathcal{G} \right\} 
\notag
\\
&\;\,\quad 
+\left\{ \mathcal{F} , \tilde{\chi}{}^{(a)+} \right\}  \left\{ \phi^{(b)} , \mathcal{G} \right\} 
-\left\{ \mathcal{F} , \phi^{(b)} \right\}  \left\{ \:\! \tilde{\chi}{}^{(a)+} , \mathcal{G} \right\} 
\notag
\\
&\;\,\quad 
+\left\{ \mathcal{F} , \chi^{(b)} \right\}  \left\{ \phi^{(a)+}, \mathcal{G} \right\} 
-\left\{ \mathcal{F} , \phi^{(a)+} \right\}  \left\{ \:\! \chi^{(b)}, \mathcal{G} \right\} 
\notag
\\
&\;\,\quad 
+\left\{ \mathcal{F} , \chi^{(e)} \right\}  \left\{ \tilde{\phi}{}^{(\tilde{g})}, \mathcal{G} \right\} 
-\left\{ \mathcal{F} , \tilde{\phi}{}^{(\tilde{g})} \right\}  \left\{ \:\! \chi^{(e)}, \mathcal{G} \right\} 
\notag
\\
&\;\,\quad 
+\left\{ \mathcal{F} , \chi^{(\tilde{g})} \right\}  \left\{ \phi^{(e)}, \mathcal{G} \right\} 
-\left\{ \mathcal{F} , \phi^{(e)} \right\}  \left\{ \:\! \chi^{(\tilde{g})}, \mathcal{G} \right\} 
\notag
\\
&\;\,\quad 
-\left\{ \mathcal{F} , \chi^{(\varphi)} \right\}  \left\{ \phi^{(\zeta)}, \mathcal{G} \right\} 
+\left\{ \mathcal{F} , \phi^{(\zeta)} \right\}  \left\{ \:\! \chi^{(\varphi)}, \mathcal{G} \right\} 
\notag
\\
&\;\,\quad 
+\left\{ \mathcal{F} , \chi^{(\zeta)} \right\}  \left\{ \phi^{(\varphi)}, \mathcal{G} \right\} 
-\left\{ \mathcal{F} , \phi^{(\varphi)} \right\}  \left\{ \:\! \chi^{(\zeta)}, \mathcal{G} \right\} . 
\label{4.25} 
\end{align}
As long as the Dirac bracket is adopted, the second-class constraints 
can be set strongly equal to zero and can be expressed as  
$\phi_{A}^{i} =0$, 
$\bar{\phi}{}^{A}_{i} =0$, 
$\phi^{(a)+} =0$, 
$\phi^{(b)} =0$, 
$\phi^{(e)} =0$, 
$\phi^{(\varphi)} =0$, 
$\phi^{(\zeta)} =0$, 
$\tilde{\phi}{}^{(\tilde{g})} =0$, 
$\tilde{\chi}{}^{(a)+} =0$, 
$\chi^{(b)} =0$, 
$\chi^{(e)} =0$, 
$\chi^{(\varphi)} =0$, 
$\chi^{(\zeta)} =0$, 
and $\chi^{(\tilde{g})} =0$. 
The second-class constraints lead to 
\begin{subequations}
\label{4.26}
\begin{alignat}{3}
P_{A}^{i} &=\frac{i}{2} \bar{Z}_{A}^{i} \,, 
&\qquad 
\bar{P}^{A}_{i} &=-\frac{i}{2} Z^{A}_{i} \,, 
\label{4.26a}
\\
& \!\!\! a_{1}+a_{2}=0 \,, 
&\qquad 
& \!\!\!\!\!\!\!\!\!\! P^{(a)1}+P^{(a)2}=0 \,,
\label{4.26b}
\\
b &=-\frac{1}{2} \left( \bar{Z}_{A}^{1} Z^{A}_{1} +\bar{Z}_{A}^{2} Z^{A}_{2} \right) \,, 
&\qquad \;\; 
P^{(b)} &=0 \,,
\label{4.26c}
\\
e &=0 \,, 
&\qquad 
P^{(e)} &=0 \,, 
\label{4.26d}
\\
\varphi &=0 \,, 
&\qquad 
P^{(\varphi)} &=0 \,, 
\label{4.26e}
\\
\zeta &=i(h-\bar{h}) \,,
&\qquad
P^{(\zeta)} &=0 \,, 
\label{4.26f}
\\
\tilde{g} &=\frac{\left(\gamma+1 \right) |k|}{2\sqrt{\gamma} M \left(h+\bar{h} \right)} 
&\qquad
P^{(\tilde{g})} &=0 \,. 
\label{4.26g}
\end{alignat}
\end{subequations}
Then we see that $P_{A}^{i}$, $\bar{P}^{A}_{i}$, 
$a_{+}:=a_1+a_2$, $P^{(a)+}:=\frac{1}{2} \big(P^{(a)1}+P^{(a)2}\big)$, 
$b$, $P^{(b)}$, $e$, $P^{(e)}$, $\varphi$, $P^{(\varphi)}$, $\zeta$, $P^{(\zeta)}$, 
$\tilde{g}$, and $P^{(\tilde{g})}$ are treated as dependent variables 
specified by Eq. (\ref{4.26}), while the remaining canonical variables 
$\bar{Z}_{A}^{i}$, $Z^{A}_{i}$, $a_{-}:=a_1-a_2$, $P^{(a)-}:=\frac{1}{2} \big(P^{(a)1}-P^{(a)2}\big)$, 
$h$, $P^{(h)}$, $\bar{h}$, and $P^{(\bar{h})}$ 
are treated as independent variables. 
By virtue of the strong equalities of the second-class constraints, 
the set of all the first-class constraints, i.e., 
\begin{align}
\Big(\:\! \phi^{(a)-}, \phi^{(+)}, \phi^{(-)}, \tilde{\chi}{}^{(a)-}, \tilde{\chi}{}^{(h)}, \tilde{\chi}{}^{(\bar{h})} \Big)  \approx 0 \,, 
\label{4.27}
\end{align}
turns out to be equivalent to the set consisting of 
\begin{subequations}
\label{4.28}
\begin{align}
\phi^{(a)-}	& \approx 0 \,, 
\label{4.28a}
\\
\phi^{(h)} & \approx 0 \,,
\label{4.28b}
\\
\phi^{(\bar{h})} & \approx 0 \,,
\label{4.28c}
\\
\chi^{(a)-} & \approx 0 \,, 
\label{4.28d}
\\
\chi^{(h)} & \approx 0 \,,
\label{4.28e}
\\
\chi^{(\bar{h})} & \approx 0 \,, 
\label{4.28f}
\end{align}
\end{subequations}
where 
\begin{align}
\chi^{(a)-}:=\frac{1}{2} \left( \:\! \chi^{(a)1} -\chi^{(a)2} \right) 
=\frac{1}{2} \left( \bar{Z}_{A}^{1} Z^{A}_{1} -\bar{Z}_{A}^{2} Z^{A}_{2} \right) \,.
\label{4.29}
\end{align}
The emergence of these first-class constraints is consistent with the fact that 
the Lagrange multipliers $u_{(a)1} -u_{(a)2}$, $u_{(h)}$, and $u_{(\bar{h})}$ remain as 
undetermined functions of $\tau$. 
The Dirac brackets between the independent canonical variables 
$\bar{Z}_{A}^{i}$, $Z^{A}_{i}$, $a_{-}$, $P^{(a)-}$, $h$, $P^{(h)}$, $\bar{h}$, and $P^{(\bar{h})}$ 
are found from Eq. (\ref{4.25}) to be 
\begin{subequations}
\label{4.30}
\begin{alignat}{3}
\left\{ Z^{A}_{i} , \bar{Z}_{B}^{j} \right\}_{\rm D} &= -i\delta_{i}^{j} \delta_{B}^{A}  \,, 
&\qquad 
~
\label{4.30a}
\\
\left\{ Z^{A}_{i} , Z^{B}_{j} \right\}_{\rm D} &=0  \,, 
&\qquad 
\left\{ \bar{Z}_{A}^{i} , \bar{Z}_{B}^{j} \right\}_{\rm D} &=0 \,, 
\label{4.30b}
\\
\left\{ a_{-} \:\!, P^{(a)-} \right\}_{\rm D} &= 1 \,, 
&\qquad 
~
\label{4.30c}
\\
\left\{ h \:\!, P^{(h)} \right\}_{\rm D} &=1 \,, 
&\qquad 
\left\{ \bar{h} \:\!, P^{(\bar{h})} \right\}_{\rm D} &=1 \,, 
\label{4.30d}
\\
\mbox{all others} &=0 \,. 
\notag
\end{alignat}
\end{subequations}
In this manner, we obtain the canonical Dirac bracket relations 
appropriate for the subsequent quantization procedure. 
In fact, the Dirac brackets in Eqs. (\ref{4.30a}) and (\ref{4.30b}) immediately lead 
to the twistor quantization procedure.

\section{Canonical quantization}

In this section, we perform the canonical quantization of the Hamiltonian system investigated in Sec. 4. 
To this end, in accordance with Dirac's method of quantization, we introduce the operators 
$\hat{\mathcal{F}}$ and $\hat{\mathcal{G}}$ corresponding to the functions $\mathcal{F}$ and $\mathcal{G}$, respectively, 
and impose the commutation relation 
\begin{align}
\left[\:\! \hat{\mathcal{F}}, \hat{\mathcal{G}} \:\! \right] \!
=i \;\! \widehat {\left\{ \mathcal{F} , \mathcal{G} \right\} }_{\rm D} 
\label{5.1}
\end{align}
in units such that $\hbar=1$.  
Here, $\widehat {\left\{ \mathcal{F} , \mathcal{G} \right\} }_{\rm D}$ denotes the operator corresponding to 
the Dirac bracket $\left\{ \mathcal{F} , \mathcal{G} \right\}_{\rm D}$. 
From Eqs. (\ref{4.30}) and (\ref{5.1}), we have the canonical commutation relations 
\begin{subequations}
\label{5.2}
\begin{alignat}{3}
\left[ \:\! \hat{Z}{}^{A}_{i} , \Hat{\Bar{Z}}{}_{B}^{j} \right] &=\delta_{i}^{j} \delta_{B}^{A}  \,, 
&\qquad 
~
\label{5.2a}
\\
\left[ \:\! \hat{Z}{}^{A}_{i} , \hat{Z}{}^{B}_{j} \right] &=0  \,, 
&\qquad 
\left[ \:\! \Hat{\Bar{Z}}{}_{A}^{i} , \Hat{\Bar{Z}}{}_{B}^{j} \right] &=0 \,, 
\label{5.2b}
\\
\left[ \;\! \hat{a}_{-} \:\!, \hat{P}^{(a)-} \right] &= i \,, 
&\qquad 
~
\label{5.2c}
\\
\left[ \;\! \hat{h} \:\!, \hat{P}{}^{(h)} \right] &=i \,, 
&\qquad 
\left[ \;\! \Hat{\Bar{h}} \:\!, \hat{P}{}^{(\bar{h})} \right] &=i \,, 
\label{5.2d}
\\
\mbox{all others} &=0 \,. 
\notag
\end{alignat}
\end{subequations}
The commutation relations in Eqs. (\ref{5.2a}) and (\ref{5.2b}) govern together so-called twistor quantization 
\cite{PenMac, PenRin2}. 
The operators $\hat{Z}{}^{A}_{i}$ and $\Hat{\Bar{Z}}{}_{A}^{i}$, referred to as the twistor operators, 
can be expressed in terms of their spinor components as 
$\hat{Z}{}_{i}^{A}=\big(\hat{\omega}{}_{i}^{\alpha}, \hat{\pi}{}_{i \dot{\alpha}} \big)$ 
and $\Hat{\Bar{Z}}{}_{A}^{i}=\big(\Hat{\Bar{\pi}}{}^{i}_{\alpha}, \Hat{\Bar{\omega}}{}^{i \dot{\alpha}} \big)$.  
Accordingly, Eq. (\ref{5.2a}) can be decomposed as follows: 
\begin{alignat}{3}
\begin{aligned}
\left[ \:\! \hat{\omega}{}_{i}^{\alpha}, \:\! \Hat{\Bar{\pi}}{}^{\:\! j}_{\beta} \:\! \right] 
&=\delta_{i}^{j} \delta_{\beta}^{\alpha} \,, 
&\qquad 
\left[ \:\! \hat{\pi}{}_{i \dot{\alpha}}, \:\! \Hat{\Bar{\omega}}{}^{\:\! j \dot{\beta}}  \:\! \right] 
&= \delta_{i}^{j} \delta_{\dot{\alpha}}^{\dot{\beta}} \,, 
\\
\left[ \:\! \hat{\omega}{}_{i}^{\alpha}, \:\! \Hat{\Bar{\omega}}{}^{\:\! j \dot{\beta}} \:\! \right] 
&=0 \,, 
&\qquad 
\left[ \:\! \hat{\pi}{}_{i \dot{\alpha}}, \:\! \Hat{\Bar{\pi}}{}^{\:\! j}_{\beta} \:\! \right] 
&=0 \,.
\end{aligned}
\label{5.3}
\end{alignat}

In the canonical quantization procedure, 
the first-class constraints are transformed into the conditions for specifying physical states,  
after the replacement of the first-class constraint functions by the corresponding operators. 
In the present model, the first-class constraints (\ref{4.28a})--(\ref{4.28f}) lead to 
the following physical state conditions imposed on 
a physical state vector $|F^{\:\!} \rangle$: 
\begin{subequations}
\label{5.4}
\begin{align}
\hat{\phi}{}^{(a)-} |F^{\:\!} \rangle & =\hat{P}{}^{(a)-} |F^{\:\!} \rangle =0 \,, 
\label{5.4a}
\\
\hat{\phi}{}^{(h)} |F^{\:\!} \rangle & =\hat{P}{}^{(h)} |F^{\:\!} \rangle =0 \,, 
\label{5.4b}
\\
\hat{\phi}{}^{(\bar{h})} |F^{\:\!} \rangle & =\hat{P}{}^{(\bar{h})} |F^{\:\!} \rangle =0 \,, 
\label{5.4c}
\\
\hat{\chi}{}^{(a)-} |F^{\:\!} \rangle 
& =\frac{1}{4} \left( \Hat{\Bar{Z}}{}^{1}_{A} \hat{Z}_{1}^{A} +\hat{Z}_{1}^{A} \Hat{\Bar{Z}}{}^{1}_{A} 
-\Hat{\Bar{Z}}{}^{2}_{A} \hat{Z}_{2}^{A} -\hat{Z}_{2}^{A} \Hat{\Bar{Z}}{}^{2}_{A} \right) |F^{\:\!} \rangle
\notag
\\
&=\frac{1}{2} \left( \hat{Z}_{1}^{A} \Hat{\Bar{Z}}{}^{1}_{A} 
-\hat{Z}_{2}^{A} \Hat{\Bar{Z}}{}^{2}_{A} \right) |F^{\:\!} \rangle =0 \,, 
\label{5.4d}
\\
\hat{\chi}{}^{(h)} |F^{\:\!} \rangle 
& =\left( \sqrt{2} \:\! \hat{\varPi} -M \right) |F^{\:\!} \rangle =0 \,, 
\label{5.4e}
\\
\hat{\chi}{}^{(\bar{h})} |F^{\:\!} \rangle 
& =\left( \sqrt{2} \:\! \Hat{\Bar{\varPi}}- M \right) |F^{\:\!} \rangle =0 \,, 
\label{5.4f}
\end{align}
\end{subequations}
where $\hat{\varPi}:=\frac{1}{2} \epsilon^{ij} \hat{\pi}_{i\dot{\alpha}} \hat{\pi}{}_{j}^{\dot{\alpha}}$ 
and $\Hat{\Bar{\varPi}}:=\frac{1}{2} \epsilon_{ij} \Hat{\Bar{\pi}}{}^{i}_{\alpha} \Hat{\Bar{\pi}}{}^{j\alpha}$. 
In defining the operator $\hat{\chi}{}^{(a)-}$, we have obeyed the Weyl ordering rule and have used 
the commutation relation (\ref{5.2a}) to simplify the Weyl ordered operator. 
As can be easily verified, the operators $\hat{\phi}{}^{(a)-}$, $\hat{\phi}{}^{(h)}$, $\hat{\phi}{}^{(\bar{h})}$, 
$\hat{\chi}{}^{(a)-}$, $\hat{\chi}{}^{(h)}$, and $\hat{\chi}{}^{(\bar{h})}$ commute with each other. 
For this reason, the conditions (\ref{5.4a})--(\ref{5.4f}) can be simultaneously imposed on $|F^{\:\!} \rangle$ 
without yielding further conditions.

We can also verify that the spin Casimir operator $\hat{W}{}^{2}$ and the $\mathit{SU}(2)$ Casimir operator $\hat{T}{}^{2}$, 
which are defined in Appendix A, commute with the operators 
$\hat{\phi}{}^{(a)-}$, $\hat{\phi}{}^{(h)}$, $\hat{\phi}{}^{(\bar{h})}$, 
$\hat{\chi}{}^{(a)-}$, $\hat{\chi}{}^{(h)}$, and $\hat{\chi}{}^{(\bar{h})}$. 
In addition, $\hat{W}{}^{2}$ and $\hat{T}{}^{2}$ commute with each other, as can be seen in Eq. (\ref{A19}). 
Hence, it is possible to choose $|F^{\:\!} \rangle$ such that it satisfies 
\begin{subequations}
\label{5.5}
\begin{alignat}{3}
\hat{W}^{2} |F^{\:\!} \rangle &=-J(J+1) |F^{\:\!} \rangle\,, & \qquad J &=0, \frac{1}{2}, 1, \frac{3}{2}, \ldots\,,  
\label{5.5a}  
\\
\hat{T}^{2} |F^{\:\!} \rangle &=I(I+1) |F^{\:\!} \rangle\,, & \qquad I &=0, \frac{1}{2}, 1, \frac{3}{2}, \ldots\,,  
\label{5.5b}
\end{alignat}
\end{subequations}
as well as Eq. (\ref{5.4}) [$^{\:\!}$see Eqs. (\ref{A24}) and (\ref{A31})]. 
Here, $J$ denotes the spin quantum number of a massive particle with rigidity. 
(In Sec. 6, it is clarified that in actuality, $J$ takes only non-negative integer values: $J=0, 1, 2, \ldots\,$.)
As shown in Appendix A, it eventually turns out that $I=J$ [$^{\:\!}$see Eq. (\ref{A32})]. 
Taking into account this fact and Eqs. (\ref{5.4e}) and (\ref{5.4f}), 
one may conclude that the physical state vector $|F^{\:\!} \rangle$ is specified by $J$ and $M$. 
However, after some analysis, 
we see that the physical mass parameter $M$ is actually determined depending on $J$ as follows: 
\begin{align}
M_{J} :=\frac{m}{\sqrt{1+\dfrac{J(J+1)}{k^{2}}}} 
\label{5.6}
\end{align}
[$^{\:\!}$see Eq. (\ref{A27})]. 
For this reason, $|F^{\:\!} \rangle$ can be specified only by $J$ 
and accordingly can be denoted as $|F_{J} \rangle$. 
[$^{\:\!}$Recall here that $m$ and $k$ are fixed constant parameters provided beforehand in the Lagrangian (\ref{2.3}).]  
From now on, the $|F^{\:\!} \rangle$ and $M$ contained in Eqs. (\ref{5.4}) and (\ref{5.5}) are, respectively,  
denoted as $|F_{J} \rangle$ and $M_{J}$.

Now we introduce the bra-vector 
\begin{align}
\big\langle Z, a_{-}, h, \bar{h} \;\! \big|
& :=\langle 0 | \exp \left( -Z^{A}_{i} \Hat{\Bar{Z}}{}^{i}_{A} 
+ia_{-} \hat{P}^{(a)-} +ih \hat{P}^{(h)} +i\bar{h} \hat{P}^{(\bar{h})} \right)
\label{5.7}
\end{align}
with a reference bra-vector $\langle 0 |$ satisfying 
\begin{align}
\langle 0 |\:\! \hat{Z}_{i}^{A} =\langle 0 |\:\! \hat{a}_{-}
=\langle 0 |\:\! \hat{h} =\langle 0 |\:\! \Hat{\Bar{h}} =0 \,. 
\label{5.8}
\end{align}
Using the commutation relations in Eq. (\ref{5.2}), we can show that 
\begin{subequations}
\label{5.9}
\begin{align}
\big\langle Z, a_{-}, h, \bar{h} \;\! \big| \;\! \hat{Z}{}^{A}_{i} 
&=Z^{A}_{i} \big\langle Z, a_{-}, h, \bar{h} \;\! \big| \,, 
\label{5.9a}
\\
\big\langle Z, a_{-}, h, \bar{h} \;\! \big| \;\! \hat{a}_{-} 
&=a_{-} \big\langle Z, a_{-}, h, \bar{h} \;\! \big| \,, 
\label{5.9b}
\\
\big\langle Z, a_{-}, h, \bar{h} \;\! \big| \;\! \hat{h} 
&=h\:\! \big\langle Z, a_{-}, h, \bar{h} \;\! \big| \,, 
\label{5.9c}
\\
\big\langle Z, a_{-}, h, \bar{h} \;\! \big| \;\! \Hat{\Bar{h}}  
&=\bar{h}\:\! \big\langle Z, a_{-}, h, \bar{h} \;\! \big| \,. 
\label{5.9d}
\end{align}
\end{subequations}
Equation (\ref{5.9a}) can be decomposed into two parts,  
\begin{subequations}
\label{5.10}
\begin{align}
\big\langle Z, a_{-}, h, \bar{h} \;\! \big| \;\! \hat{\omega}{}_{i}^{\alpha}
&=\omega_{i}^{\alpha} \big\langle Z, a_{-}, h, \bar{h} \;\! \big| \,, 
\label{5.10a}
\\
\big\langle Z, a_{-}, h, \bar{h} \;\! \big| \;\! \hat{\pi}_{i \dot{\alpha}}
&=\pi_{i \dot{\alpha}} \big\langle Z, a_{-}, h, \bar{h} \;\! \big| \,. 
\label{5.10b}
\end{align}
\end{subequations}
Also, it is easy to see that 
\begin{subequations}
\label{5.11}
\begin{align}
\big\langle Z, a_{-}, h, \bar{h} \;\! \big| \;\! \Hat{\Bar{Z}}{}^{i}_{A} 
&=-\frac{\partial}{\partial Z^{A}_{i}} \big\langle Z, a_{-}, h, \bar{h} \;\! \big| \,, 
\label{5.11a}
\\
\big\langle Z, a_{-}, h, \bar{h} \;\! \big| \:\! \hat{P}^{(a)-} 
&=-i\frac{\partial}{\partial a_{-}} \big\langle Z, a_{-}, h, \bar{h} \;\! \big| \,, 
\label{5.11b}
\\
\big\langle Z, a_{-}, h, \bar{h} \;\! \big| \:\! \hat{P}^{(h)} 
&=-i\frac{\partial}{\partial h} \big\langle Z, a_{-}, h, \bar{h} \;\! \big| \,, 
\label{5.11c}
\\
\big\langle Z, a_{-}, h, \bar{h} \;\! \big| \:\! \hat{P}^{(\bar{h})}  
&=-i\frac{\partial}{\partial \bar{h}} \big\langle Z, a_{-}, h, \bar{h} \;\! \big| \,. 
\label{5.11d}
\end{align}
\end{subequations}
Equation (\ref{5.11a}) can be decomposed into two parts, 
\begin{subequations}
\label{5.12}
\begin{align}
\big\langle Z, a_{-}, h, \bar{h} \;\! \big| \;\! \Hat{\Bar{\pi}}{}^{i}_{\alpha} 
&=-\frac{\partial}{\partial \omega_{i}^{\alpha}} \big\langle Z, a_{-}, h, \bar{h} \;\! \big| \,, 
\label{5.12a}
\\
\big\langle Z, a_{-}, h, \bar{h} \;\! \big| \;\! \Hat{\Bar{\omega}}{}^{i \dot{\alpha}} 
&=-\frac{\partial}{\partial \pi_{i \dot{\alpha}}} \big\langle Z, a_{-}, h, \bar{h} \;\! \big| \,.  
\label{5.12b}
\end{align}
\end{subequations}
Multiplying each of Eqs. (\ref{5.4a})--(\ref{5.4f}) by 
$\big\langle Z, a_{-}, h, \bar{h}{}^{\;\!} \big|$ on the left,  
and using some of Eqs. (\ref{5.9})--(\ref{5.12}) appropriately, we obtain the algebraic equation 
\begin{align}
\sqrt{2}\:\! \varPi=M_{J} 
\label{5.13}
\end{align}
and the following set of differential equations for  
$F_{J} \big(Z, a_{-}, h, \bar{h} \big):=\big\langle Z, a_{-}, h, \bar{h}{}^{\;\!} \big|F_{J} \big\rangle$: 
\begin{subequations}
\label{5.14}
\begin{align}
\frac{\partial}{\partial a_{-}} F_{J} &=0 \,, 
\label{5.14a}
\\
\frac{\partial}{\partial h} F_{J} &=0 \,, 
\label{5.14b}
\\
\frac{\partial}{\partial \bar{h}} F_{J} &=0 \,, 
\label{5.14c}
\\
\left( Z_{1}^{A} \frac{\partial}{\partial Z_{1}^{A}} -Z_{2}^{A} \frac{\partial}{\partial Z_{2}^{A}} \right) F_{J} &=0 \,, 
\label{5.14d}
\\
\epsilon_{ij} \epsilon^{\alpha\beta} 
\frac{\partial}{\partial \omega_{i}^{\alpha}} \frac{\partial}{\partial \omega_{j}^{\beta}} F_{J}
&=\sqrt{2} \:\! M_{J} F_{J} \,. 
\label{5.14e}
\end{align}
\end{subequations}
Here, Eq. (\ref{5.13}) is obtained under the natural condition $F_{J} \neq0$. 
Equations (\ref{5.14a})--(\ref{5.14c}) imply that $F_{J}$ does not depend on $a_{-}$, $h$, and $\bar{h}$. 
Hence it follows that $F_{J}$ is a function only of the twistors $Z^{A}_{i}$ $(i=1,2)$. 
Such a holomorphic function of $Z^{A}_{i}$ is often called a twistor function.   
As can be seen immediately, Eqs. (\ref{5.13}) and (\ref{5.14e}) are, respectively, equivalent to
\begin{subequations}
\label{5.15}
\begin{align}
\pi_{i \dot{\alpha}} \pi_{j}^{\dot{\alpha}}  
&=\frac{M_{J}}{\sqrt{2}}\:\! \epsilon_{ij} \,, 
\label{5.15a}
\\
\epsilon^{\alpha \beta}
\frac{\partial}{\partial \omega_{i}^{\alpha}} \frac{\partial}{\partial \omega_{j}^{\beta}} F_{J}  
&=\frac{M_{J}}{\sqrt{2}}\:\! \epsilon^{ij} F_{J} \,.    
\label{5.15b}
\end{align}
\end{subequations}

We now apply the method of separation of variables to Eq. (\ref{5.14d}) to find its general solution. 
Substituting the factorized function 
$F_{J} \big(Z^{A}_{1}, Z^{A}_{2} \big)=F_{J}^{(1)}\big(Z^{A}_{1} \big) F_{J}^{(2)}\big(Z^{A}_{2} \big)$ 
into Eq. (\ref{5.14d}), we can separate it into the two equations 
\begin{subequations}
\label{5.16}
\begin{align}
Z_{1}^{A} \frac{\partial}{\partial Z_{1}^{A}} F_{J}^{(1)} &=-2(s_{\ast}+1) F_{J}^{(1)} , 
\label{5.16a}
\\
Z_{2}^{A} \frac{\partial}{\partial Z_{2}^{A}} F_{J}^{(2)} &=-2(s_{\ast}+1) F_{J}^{(2)} , 
\label{5.16b}
\end{align}
\end{subequations}
where $s_{\ast}$ is a constant and $F_{J}^{(i)}$ is a function only of $Z^{A}_{i}$. 
Obviously, Eqs. (\ref{5.16a}) and (\ref{5.16b}) are satisfied by homogeneous twistor functions of degree $-2s_{\ast}-2$.  
This degree must be an integer so that $F_{J}^{(i)}$ can be a single-valued function of $Z^{A}_{i}$. 
As a result, the allowed values of $s_{\ast}$ are restricted to arbitrary integer or half-integer values. 
Hereafter, the twistor functions of the homogeneity degree $-2s_{\ast}-2$ are denoted as  
$F_{Js_{\ast}}^{(1)}$ and $F_{Js_{\ast}}^{(2)}$. 
From these functions, the general solution of Eq. (\ref{5.14d}) is constructed as follows: 
\begin{align}
F_{J}(Z)=\sum_{s_{\ast} \in {\textstyle{\frac{1}{2}}}\Bbb{Z}} C_{Js_{\ast}} F_{Js_{\ast}}(Z) 
\label{5.17}
\end{align}
with $F_{Js_{\ast}}(Z):=F_{Js_{\ast}}^{(1)} \big(Z^{A}_{1} \big) F_{Js_{\ast}}^{(2)} \big(Z^{A}_{2} \big)$ 
being particular solutions of Eq. (\ref{5.14d}). 
Here, $C_{Js_{\ast}}$ are complex coefficients.

\section{Penrose transform}

In this section, 
we obtain an arbitrary-rank spinor field in complexified Minkowski space 
$\Bbb{C}\mathbf{M}$ 
via the Penrose transform of $F_{J}(Z)$. 
Then it is shown that the allowed values of the spin quantum number $J$ 
are restricted to arbitrary non-negative integers. 
We also demonstrate that the spinor field satisfies generalized DFP equations with extra indices. 
Furthermore we mention the total symmetrization of the spinor field and a bra-ket formalism of the Penrose transform.

Let us consider the Penrose transform of $F_{J}(Z)$ specified by 
\begin{align}
& \varPsi^{\:\! i_1\ldots i_p}_{\alpha_1 \ldots \alpha_p ; \;\! 
 j_1\ldots j_q, \:\! \dot{\alpha}_1 \ldots \dot{\alpha}_q} (z)
=\frac{1}{(2\pi i)^{4}} \oint_{\varSigma} 
\pi_{j_{1} \dot{\alpha}_{1}} \cdots \pi_{j_{q} \dot{\alpha}_{q}} 
\frac{\partial}{\partial \omega_{i_{1}}^{\alpha_{1}}} \cdots  \frac{\partial}{\partial \omega_{i_{p}}^{\alpha_{p}}} 
F_{J}(Z) 
\:\! d^4 \pi 
\label{6.1}
\end{align} 
with $d^4 \pi :=d\pi_{1\dot{0}} \wedge d\pi_{1\dot{1}} \wedge d\pi_{2\dot{0}} \wedge d\pi_{2\dot{1}}$ 
to define a rank-$(p+q)$ spinor field 
$\varPsi^{\:\! i_1\ldots i_{p}}_{\alpha_1 \ldots \alpha_{p} ; \;\!  j_1\ldots j_q, \:\! \dot{\alpha}_1 \ldots \dot{\alpha}_q}$ 
(sometimes abbreviated as $\varPsi$). 
Here, $\varSigma$ denotes a 4-dimensional contour, 
and the twistors ${Z}^{A}_{i}=\big({\omega}{}_{i}^{\alpha}, \pi_{i \dot{\alpha}} \big)$ 
satisfy the incidence relations 
\begin{align}
\omega_{i}^{\alpha}= iz^{\alpha\dot{\alpha}} \pi_{i\dot{\alpha}} \,, 
\label{6.2}
\end{align}
with $z^{\alpha \dot{\alpha}}\big(=x^{\alpha \dot{\alpha}}-iy^{\alpha \dot{\alpha}}\big)$ being 
coordinates of a point in $\Bbb{C}\mathbf{M}$. 
The Penrose transform (\ref{6.1}) corresponds to the one treated in Ref. \cite{DegOka}. 
It should be noted that the spinor field $\varPsi$ 
has the extra indices $i$'s and $j$'s in addition to 
the dotted and undotted spinor indices. 
From the structure of the right hand side of Eq. (\ref{6.1}), 
it is obvious that the number of upper (lower) extra indices of $\varPsi$ 
is equal to the number of its undotted (dotted) spinor indices.  
From Eq. (\ref{6.1}), we also see the following totally symmetric properties 
with respect to the pairs of the spinor and extra indices: 
\begin{subequations}
\label{6.3}
\begin{align}
\varPsi^{\:\! i_1\ldots i_m \ldots i_n \ldots i_{p}}_{\alpha_1 \ldots \alpha_m \ldots \alpha_n \ldots \alpha_{p} ; \;\! 
 j_1\ldots j_q, \:\! \dot{\alpha}_1 \ldots \dot{\alpha}_q} 
&= \varPsi^{\:\! i_1\ldots i_n \ldots i_m \ldots i_{p}}_{\alpha_1 \ldots \alpha_n \ldots \alpha_m \ldots \alpha_{p} ; \;\! 
 j_1\ldots j_q, \:\! \dot{\alpha}_1 \ldots \dot{\alpha}_q} \,, 
\label{6.3a}
\\[5pt]
\varPsi^{\:\! i_1\ldots i_{p}}_{\alpha_1 \ldots \alpha_{p} ; \;\! 
 j_1\ldots j_a \ldots j_b \ldots j_q, \:\! \dot{\alpha}_1 \ldots \dot{\alpha}_a \ldots \dot{\alpha}_b \ldots \dot{\alpha}_q} 
&= \varPsi^{\:\! i_1\ldots i_{p}}_{\alpha_1 \ldots \alpha_{p} ; \;\! 
 j_1\ldots j_b \ldots j_a \ldots j_q, \:\! \dot{\alpha}_1 \ldots \dot{\alpha}_b \ldots \dot{\alpha}_a \ldots \dot{\alpha}_q} \,.
 \label{6.3b}
\end{align}
\end{subequations}

Substituting Eq. (\ref{5.17}) into Eq. (\ref{6.1}) gives 
\begin{align}
\varPsi^{\:\! i_1\ldots i_{p}}_{\alpha_1 \ldots \alpha_{p} ; \;\! 
 j_1\ldots j_q, \:\! \dot{\alpha}_1 \ldots \dot{\alpha}_q} (z) 
=\sum_{s_{\ast} \in {\textstyle{\frac{1}{2}}} \Bbb{Z}}
\frac{C_{Js_{\ast}}}{(2\pi i)^{4}} \oint_{\varSigma} 
\pi_{j_{1} \dot{\alpha}_{1}} \cdots \pi_{j_{q} \dot{\alpha}_{q}} 
\frac{\partial}{\partial {\omega}{}_{i_{1}}^{\alpha_{1}}} \cdots  
\frac{\partial}{\partial {\omega}{}_{i_{p}}^{\alpha_{p}}} 
F_{Js_{\ast}} \big({Z} \big) \:\! d^4 \pi \,. 
\label{6.4}
\end{align} 
Now we suppose that among the indices $i_1,\ldots, i_p$, 
the number of 1's is $p_{1}$ and the number of 2's is $p_{2}(=p-p_{1})$. 
We also suppose that among the indices $j_1,\ldots, j_q$, the number of 1's is $q_{1}$ and the number of 2's is $q_{2}(=q-q_{1})$. 
Since $F_{Js_{\ast}} \big({Z} \big)$ 
is a homogeneous twistor function of degree $-2s_{\ast}-2$ with respect to each of 
${Z}^{A}_{1}=\big({\omega}{}_{1}^{\alpha}, \pi_{1 \dot{\alpha}} \big)$ 
and ${Z}^{A}_{2}=\big({\omega}{}_{2}^{\alpha}, \pi_{2 \dot{\alpha}} \big)$,  
only one integral in the infinite sum in Eq. (\ref{6.4}) can remain nonvanishing,  
provided that the following two conditions are simultaneously satisfied in this integral:
\begin{subequations}
\label{6.5}
\begin{align}
s_{\ast}&=\frac{1}{2} (q_1 -p_1) \,, 
\label{6.5a}
\\[3pt]
s_{\ast}&=\frac{1}{2} (q_2 -p_2) \,.
\label{6.5b}
\end{align}
\end{subequations}
(In order to obtain a nonvanishing result of Eq. (\ref{6.4}), it is necessary to choose a suitable contour $\varSigma$,  
in addition to imposing Eq. (\ref{6.5}).) 
Equation (\ref{6.4}) eventually reduces to 
\begin{align}
\varPsi^{\:\! i_1\ldots i_{p}}_{\alpha_1 \ldots \alpha_{p} ; \;\! 
 j_1\ldots j_q, \:\! \dot{\alpha}_1 \ldots \dot{\alpha}_q} (z) 
=\frac{C_{Js_{\ast}}}{(2\pi i)^{4}} \oint_{\varSigma} 
\pi_{j_{1} \dot{\alpha}_{1}} \cdots \pi_{j_{q} \dot{\alpha}_{q}} 
\frac{\partial}{\partial {\omega}{}_{i_{1}}^{\alpha_{1}}} \cdots  
\frac{\partial}{\partial {\omega}{}_{i_{p}}^{\alpha_{p}}} 
F_{Js_{\ast}} \big({Z} \big) \:\! d^4 \pi 
\label{6.6}
\end{align} 
with $s_{\ast}$ satisfying Eq. (\ref{6.5}). 
Equations (\ref{6.5a}) and (\ref{6.5b}) yield $q_1 -p_1=q_2 -p_2$. 
Using this, it can be shown that 
\begin{subequations}
\label{6.7}
\begin{align}
N_{+}: &= q+p=2(q_1+p_2)=2(q_2+p_1) \,, 
\label{6.7a}
\\[2pt]
N_{-}: &=q-p=2(q_1 -p_1)=2(q_2 -p_2)=4s_{\ast} \,.
\label{6.7b}
\end{align}
\end{subequations}
Thus we see that the rank of $\varPsi$, denoted by $N_{+}$, is even,  
and that the difference between the numbers of the dotted and undotted spinor indices of $\varPsi$, 
denoted by $N_{-}$, is also even. 
In Appendix B, it is proven that$^{\;\!}$\footnote{~Some readers may think that the proof of Eq. (\ref{6.8}) 
can simply be accomplished within the framework of group theory. 
However, the spin quantum number $J$ at present is a quantum number that 
appears in the context of treating the Pauli-Lubanski pseudovector of a massive particle with rigidity. 
For this reason, the proof of Eq. (\ref{6.8}) must be achieved within the present framework 
by taking into account the origin of $J$.} 
\begin{align}
J=\frac{N_{+}}{2}  \,. 
\label{6.8}
\end{align}
Since $N_{+}$ is non-negative and even,  
it follows from Eq. (\ref{6.8}) that $J$ takes only non-negative integer values: 
$J=0, 1, 2, \ldots\,$.  
Therefore we can conclude that the massive particle with rigidity is allowed to possess only integer spin.   
This result is in agreement with the one obtained by Plyushchay \cite{Ply01, Ply04} (see also Appendix C).

Next, we demonstrate that $\varPsi$ satisfies generalized DFP equations with extra indices. 
To this end, it is useful to exploit that 
\begin{align}
\frac{\partial}{\partial z_{\beta\dot{\beta}}} F_{J} \big({Z} \big)
=\epsilon^{\beta\alpha} \epsilon^{\dot{\beta} \dot{\alpha}} 
\frac{\partial {\omega}^{\gamma}_{k}}{\partial z^{\alpha\dot{\alpha}}} 
\frac{\partial}{\partial {\omega}^{\gamma}_{k}} F_{J} \big({Z} \big) 
=i \pi_{k}^{\dot{\beta}} \epsilon^{\beta \gamma}
\frac{\partial}{\partial {\omega}^{\gamma}_{k}} F_{J} \big({Z} \big)  \,.  
\label{6.9}
\end{align}
The derivative of $\varPsi$ with respect to $z_{\beta\dot{\beta}}$ can be calculated 
by using Eqs. (\ref{6.1}) and (\ref{6.9}) as follows: 
\begin{align}
&\frac{\partial}{\partial z_{\beta\dot{\beta}}} 
\varPsi^{\:\! i_1\ldots i_{p}}_{\alpha_1 \ldots \alpha_{p} ; \;\! 
 j_1\ldots j_q, \:\! \dot{\alpha}_1 \ldots \dot{\alpha}_q} (z) 
\notag
\\
&=\frac{1}{(2\pi i)^{4}} \oint_{\varSigma} 
\pi_{j_{1} \dot{\alpha}_{1}} \cdots \pi_{j_{q} \dot{\alpha}_{q}} 
\frac{\partial}{\partial {\omega}{}_{i_{1}}^{\alpha_{1}}} \cdots  
\frac{\partial}{\partial {\omega}{}_{i_{p}}^{\alpha_{p}}} 
\frac{\partial}{\partial z_{\beta\dot{\beta}}} F_{J} \big({Z} \big) \:\! d^4 \pi 
\notag 
\\
&=\frac{i}{(2\pi i)^{4}} \oint_{\varSigma} 
\pi_{j_{1} \dot{\alpha}_{1}} \pi_{k}^{\dot{\beta}} \pi_{j_{2} \dot{\alpha}_{2}} \cdots \pi_{j_{q} \dot{\alpha}_{q}} 
\frac{\partial}{\partial {\omega}{}_{i_{2}}^{\alpha_{2}}} \cdots  
\frac{\partial}{\partial {\omega}{}_{i_{p}}^{\alpha_{p}}} 
\epsilon^{\beta \gamma} 
\frac{\partial}{\partial {\omega}{}_{i_{1}}^{\alpha_{1}}} 
\frac{\partial}{\partial {\omega}^{\gamma}_{k}} 
F_{J} \big({Z} \big)  \!\: d^4 \pi \,. 
\label{6.10}
\end{align}
Contracting over the indices $\dot{\beta}$ and $\dot{\alpha}_{1}$ in Eq. (\ref{6.10}) and using Eq. (\ref{5.15a}), 
we have 
\begin{align}
& \frac{\partial}{\partial z_{\beta\dot{\beta}}} 
\varPsi^{\:\! i_1\ldots i_{p}}_{\alpha_1 \ldots \alpha_{p} ; \;\! 
 j_1\ldots j_q, \, \dot{\beta} \dot{\alpha}_2 \ldots \dot{\alpha}_q} (z) 
\notag
\\
&= \frac{M_{J}}{\sqrt{2}} \epsilon^{\beta\gamma} \epsilon_{j_{1} k}
\frac{i}{(2\pi i)^{4}} \oint_{\varSigma} 
\pi_{j_{2} \dot{\alpha}_{2}} \cdots \pi_{j_{q} \dot{\alpha}_{q}} 
\frac{\partial}{\partial {\omega}^{\gamma}_{k}} 
\frac{\partial}{\partial {\omega}{}_{i_{1}}^{\alpha_{1}}} \cdots  
\frac{\partial}{\partial {\omega}{}_{i_{p}}^{\alpha_{p}}} 
F_{J} \big({Z} \big) \:\! d^4 \pi 
\notag
\\
&= \frac{iM_{J}}{\sqrt{2}} \epsilon^{\beta\gamma} \epsilon_{j_{1} k}  
\varPsi^{\:\! k i_1\ldots i_{p}}_{\gamma \alpha_1 \ldots \alpha_{p} ; \;\! 
 j_2\ldots j_q, \:\! \dot{\alpha}_2 \ldots \dot{\alpha}_q} (z) \,. 
\label{6.11}
\end{align}
Similarly, contracting over the indices $\beta$ and $\alpha_{1}$ in Eq. (\ref{6.10}) and using Eq. (\ref{5.15b}), 
we have 
\begin{align}
& \frac{\partial}{\partial z_{\beta\dot{\beta}}} 
\varPsi^{\:\! i_1\ldots i_{p}}_{\beta \alpha_2 \ldots \alpha_{p} ; \;\! 
 j_1\ldots j_q, \:\! \dot{\alpha}_1 \ldots \dot{\alpha}_q} (z) 
\notag
\\
&= \frac{M_{J}}{\sqrt{2}} \epsilon^{\dot{\beta} \dot{\gamma}} \epsilon^{i_{1} k}
\frac{i}{(2\pi i)^{4}} \oint_{\varSigma} 
\pi_{k \dot{\gamma}} \pi_{j_{1} \dot{\alpha}_{1}} \cdots \pi_{j_{q} \dot{\alpha}_{q}} 
\frac{\partial}{\partial {\omega}{}_{i_{2}}^{\alpha_{2}}} 
\cdots  \frac{\partial}{\partial {\omega}{}_{i_{p}}^{\alpha_{p}}} 
F_{J} \big({Z} \big) \:\! d^4 \pi 
\notag
\\
&= \frac{iM_{J}}{\sqrt{2}} \epsilon^{\dot{\beta} \dot{\gamma}} \epsilon^{i_{1} k}
\varPsi^{\:\! i_2 \ldots i_{p}}_{\alpha_2 \ldots \alpha_{p} ; \;\! 
k j_1\ldots j_q, \:\! \dot{\gamma} \dot{\alpha}_1 \ldots \dot{\alpha}_q} (z) \,. 
\label{6.12}
\end{align}
Note here that the rank $N_{+}$ does not change 
under the manipulations in Eqs. (\ref{6.11}) and (\ref{6.12}). 
In this way, it turns out that the spinor field $\varPsi$ 
satisfies the generalized DFP equations with the extra indices $i$'s and $j$'s \cite{DegOka} 
\begin{subequations}
\label{6.13}
\begin{align}
& i \sqrt{2}\:\! \frac{\partial}{\partial z_{\beta\dot{\beta}}} 
\varPsi^{\:\! i_1\ldots i_{p}}_{\alpha_1 \ldots \alpha_{p} ; \;\! 
 j_1\ldots j_q, \, \dot{\beta} \dot{\alpha}_2 \ldots \dot{\alpha}_q} 
+M_{J} \epsilon^{\beta\gamma} \epsilon_{j_{1} k}  
\varPsi^{\:\! k i_1\ldots i_{p}}_{\gamma \alpha_1 \ldots \alpha_{p} ; \;\! 
 j_2\ldots j_q, \:\! \dot{\alpha}_2 \ldots \dot{\alpha}_q} =0 \,, 
 \label{6.13a}
 \\[3pt]
& i \sqrt{2}\:\! \frac{\partial}{\partial z_{\beta\dot{\beta}}} 
\varPsi^{\:\! i_1\ldots i_{p}}_{\beta \alpha_2 \ldots \alpha_{p} ; \;\! 
 j_1\ldots j_q, \:\! \dot{\alpha}_1 \ldots \dot{\alpha}_q} 
+M_{J} \epsilon^{\dot{\beta} \dot{\gamma}} \epsilon^{i_{1} k}
\varPsi^{\:\! i_2 \ldots i_{p}}_{\alpha_2 \ldots \alpha_{p} ; \;\! 
k j_1\ldots j_q, \:\! \dot{\gamma} \dot{\alpha}_1 \ldots \dot{\alpha}_q}=0 \,.
\label{6.13b}
\end{align}
\end{subequations}
Using Eqs. (\ref{6.13a}) and (\ref{6.13b}) and noting 
\begin{align}
\frac{\partial}{\partial z^{\alpha\dot{\beta}}} \frac{\partial}{\partial z_{\beta\dot{\beta}}} 
=\frac{1}{2} \delta_{\alpha}^{\;\!\beta} 
\frac{\partial}{\partial z^{\gamma\dot{\gamma}}} \frac{\partial}{\partial z_{\gamma\dot{\gamma}}} \,, 
\label{6.14}
\end{align}
we can derive the Klein-Gordon equation  
\begin{align}
\left( \frac{\partial}{\partial z^{\;\!\beta\dot{\beta}}} \frac{\partial}{\partial z_{\beta\dot{\beta}}} 
+M_{J}^2 \right) 
\varPsi^{\:\! i_1\ldots i_{p}}_{\alpha_1 \ldots \alpha_{p} ; \;\! 
 j_1\ldots j_q, \:\! \dot{\alpha}_1 \ldots \dot{\alpha}_q} =0 \,.  
\label{6.15}
\end{align}
This makes it clear that $\varPsi$ is a field of mass $M_{J}$. 
Thus, taking into account Eq. (\ref{6.8}), we can conclude that 
the spinor field of rank $2J$ with mass $M_{J}$ has been obtained by means of the Penrose transform (\ref{6.1}).

Before closing this section, we make mention of the total symmetrization of $\varPsi$ 
and a bra-ket formalism of the Penrose transform. 

\vspace{3mm}

\noindent
\subsection{Total symmetrization of $\varPsi$}  

\vspace{2mm}

We consider the total symmetrization of $\varPsi$ with respect to all the extra indices, specified by  
\begin{align}
\varPsi^{(\mathrm{S}) \:\! i_1\ldots i_{p+q}}_{\alpha_1 \ldots \alpha_{p} ; \;\! \dot{\alpha}_1 \ldots \dot{\alpha}_q} 
:=\frac{1}{N_{+} !} \sum_{\varsigma} 
\epsilon^{i_{\varsigma\:\! (p+1)} \:\! j_{1}} \cdots \epsilon^{i_{\varsigma\:\! (p+q)} \:\! j_{q}} 
\:\! 
\varPsi^{\:\! i_{\varsigma\:\!(1)} \ldots i_{\varsigma\:\!(p)}}_{\alpha_1 \ldots \alpha_{p} ; \;\! 
 j_1\ldots j_q, \:\! \dot{\alpha}_1 \ldots \dot{\alpha}_q} \,, 
\label{6.16}
\end{align}
where $\varsigma$ denotes a permutation acting on $(i_{1}, \ldots, i_{p+q})$, and 
the sum is taken over all the permutations of $(i_{1}, \ldots, i_{p+q})$ \cite{Okano}. 
(The spinor field 
$\varPsi{}^{(\mathrm{S}) \:\! i_1\ldots i_{p+q}}_{\alpha_1 \ldots \alpha_{p} ; \;\! \dot{\alpha}_1 \ldots \dot{\alpha}_q}$ 
is often abbreviated as $\varPsi^{(\mathrm{S})}$.) 
Using Eq. (\ref{6.3}) and the totally symmetric property of $\varPsi^{(\mathrm{S})}$ with respect to the extra indices, 
we can show that 
\begin{subequations}
\label{6.17}
\begin{align}
\varPsi^{(\mathrm{S}) \:\! i_1\ldots i_{p+q}}_{\alpha_1 \ldots  \alpha_m \ldots \alpha_n \ldots  \alpha_{p} ; \;\! 
\dot{\alpha}_1 \ldots \dot{\alpha}_q} 
&=\varPsi^{(\mathrm{S}) \:\! i_1\ldots i_{p+q}}_{\alpha_1 \ldots  \alpha_n \ldots \alpha_m \ldots  \alpha_{p} ; \;\! 
\dot{\alpha}_1 \ldots \dot{\alpha}_q} \,, 
\label{6.17a}
\\[5pt]
\varPsi^{(\mathrm{S}) \:\! i_1\ldots i_{p+q}}_{\alpha_1 \ldots \alpha_{p} ; \;\! 
\dot{\alpha}_1 \ldots \dot{\alpha}_a \ldots \dot{\alpha}_b \ldots \dot{\alpha}_q} 
&=\varPsi^{(\mathrm{S}) \:\! i_1\ldots i_{p+q}}_{\alpha_1 \ldots \alpha_{p} ; \;\! 
\dot{\alpha}_1 \ldots \dot{\alpha}_b \ldots \dot{\alpha}_a \ldots \dot{\alpha}_q} \,.
\label{6.17b}
\end{align}
\end{subequations}
Hence it follows that $\varPsi^{(\mathrm{S})}$ is totally symmetric with respect to 
each of the dotted and undotted spinor indices. 
The field $\varPsi^{(\mathrm{S})}$ is eventually   
identified as an irreducible spinor field belonging to a spin $J$ representation. 
It can be verified with Eq. (\ref{6.16}) that Eqs. (\ref{6.13a}) and (\ref{6.13b}), respectively, lead to  
\begin{subequations}
\label{6.18}
\begin{align}
& i \sqrt{2}\:\! \frac{\partial}{\partial z_{\beta\dot{\beta}}} 
\varPsi^{(\mathrm{S}) \:\! i_1\ldots i_{p+q}}_{\alpha_1 \ldots \alpha_{p} ; 
\, \dot{\beta} \dot{\alpha}_2 \ldots \dot{\alpha}_q} 
-M_{J} \epsilon^{\beta\gamma} 
\varPsi^{(\mathrm{S}) \:\! i_1\ldots i_{p+q}}_{\gamma \alpha_1 \ldots \alpha_{p} ; 
\;\! \dot{\alpha}_2 \ldots \dot{\alpha}_q} =0 \,, 
 \label{6.18a}
\\[3pt]
& i \sqrt{2}\:\! \frac{\partial}{\partial z_{\beta\dot{\beta}}} 
\varPsi^{(\mathrm{S}) \:\! i_1\ldots i_{p+q}}_{\beta \alpha_2 \ldots \alpha_{p} ; 
\;\! \dot{\alpha}_1 \ldots \dot{\alpha}_q} 
+M_{J} \epsilon^{\dot{\beta} \dot{\gamma}} 
\varPsi^{(\mathrm{S}) \:\! i_1\ldots i_{p+q}}_{\alpha_2 \ldots \alpha_{p} ; 
\;\! \dot{\gamma} \dot{\alpha}_1 \ldots \dot{\alpha}_q} =0 \,. 
\label{6.18b}
\end{align}
\end{subequations}
We thus see that $\varPsi^{(\mathrm{S})}$ satisfies the (ordinary) DFP equations \cite{Dirac1, Fierz, FiePau, IsaPod}.    
This result is consistent with the totally symmetric property of $\varPsi^{(\mathrm{S})}$ 
with respect to the spinor indices. 
It is evident from Eqs. (\ref{6.15}) and (\ref{6.16}) that $\varPsi^{(\mathrm{S})}$ fulfills  
the Klein-Gordon equation with the mass parameter $M_{J}$.  

\vspace{3mm}

\noindent
\subsection{Bra-ket formalism of the Penrose transform} 

\vspace{2mm}

In the same manner as 
the $F_{J} \big(Z, a_{-}, h, \bar{h} \big)=\big\langle Z, a_{-}, h, \bar{h}{}^{\;\!} \big|F_{J} \big\rangle$ 
given under Eq. (\ref{5.13}), we can express the twistor function $F_{J} \big({Z} \big)$ as 
\begin{align}
F_{J} \big({Z} \big)=\big\langle {Z} \:\! \big|F_{J} \big\rangle \,, 
\label{6.19}
\end{align}
where 
\begin{align}
\big\langle {Z} \:\! \big|
& :=\langle 0 | \exp \left( -{Z}^{A}_{i} \Hat{\Bar{Z}}{}^{i}_{A} \right)
=\langle 0 | \exp \left( -{\omega}{}_{i}^{\alpha} \Hat{\Bar{\pi}}{}^{i}_{\alpha} 
-\pi{}_{i \dot{\alpha}} \Hat{\Bar{\omega}}{}^{i \dot{\alpha}} \right) 
\label{6.20}
\end{align}
[$^{\:\!}$see Eq. (\ref{5.7})]. Using Eq. (\ref{6.20}), it is easily shown that 
\begin{subequations}
\label{6.21}
\begin{align}
\big\langle {Z} \:\!\big| \:\! \hat{\pi}_{i \dot{\alpha}}
&=\pi_{i\dot{\alpha}} \big\langle {Z} \:\!\big| \,, 
\label{6.21a}
\\
\big\langle {Z} \:\!\big| \:\! \Hat{\Bar{\pi}}{}^{i}_{\alpha} 
&=-\frac{\partial}{\partial {\omega}{}_{i}^{\alpha}} \big\langle {Z} \:\!\big| \,. 
\label{6.21b}
\end{align}
\end{subequations}
Substituting Eq. (\ref{6.19}) into Eq. (\ref{6.1}) and using Eqs. (\ref{6.21a}) and (\ref{6.21b}) repeatedly, 
we can write the Penrose transform (\ref{6.1}) as 
\begin{align}
\varPsi^{\:\! i_1\ldots i_{p}}_{\alpha_1 \ldots \alpha_{p} ; \;\! 
 j_1\ldots j_q, \:\! \dot{\alpha}_1 \ldots \dot{\alpha}_q} (z) 
=\frac{1}{(2\pi i)^{4}} \oint_{\varSigma} 
\big\langle {Z} \:\! \big| \:\! \hat{\mathcal{P}}^{\:\! i_1\ldots i_{p}}_{\alpha_1 \ldots \alpha_{p} ; \;\! 
 j_1\ldots j_q, \:\! \dot{\alpha}_1 \ldots \dot{\alpha}_q} \big|F_{J} \big\rangle \:\! d^4 \pi 
\label{6.22}
\end{align} 
with the operator 
\begin{align}
\hat{\mathcal{P}}^{\:\! i_1\ldots i_{p}}_{\alpha_1 \ldots \alpha_{p} ; \;\! 
 j_1\ldots j_q, \:\! \dot{\alpha}_1 \ldots \dot{\alpha}_q} 
:=(-1)^{p} \:\! 
\hat{\pi}_{j_{1} \dot{\alpha}_{1}} \cdots \hat{\pi}_{j_{q} \dot{\alpha}_{q}} 
\Hat{\Bar{\pi}}{}^{i_{1}}_{\alpha_{1}} \cdots \Hat{\Bar{\pi}}{}^{i_{p}}_{\alpha_{p}} \,. 
\label{6.23}
\end{align}
Furthermore it is possible to formally express Eq. (\ref{6.22}) as 
\begin{align}
\varPsi^{\:\! i_1\ldots i_{p}}_{\alpha_1 \ldots \alpha_{p} ; \;\! 
 j_1\ldots j_q, \:\! \dot{\alpha}_1 \ldots \dot{\alpha}_q} (z) 
=\big\langle z \:\! \big| \:\! \varPsi^{\:\! i_1\ldots i_{p}}_{\alpha_1 \ldots \alpha_{p} ; \;\! 
 j_1\ldots j_q, \:\! \dot{\alpha}_1 \ldots \dot{\alpha}_q} \big\rangle 
\label{6.24}
\end{align}
with the ket and bra vectors  
\begin{align}
& \big\langle z \;\! \big|:=\frac{1}{(2\pi i)^{4}} \oint_{\varSigma} 
\big\langle {Z} \:\! \big| \:\! d^4 \pi
=\frac{1}{(2\pi i)^{4}} \oint_{\varSigma} 
\big\langle iz^{\alpha\dot{\alpha}} \pi_{i\dot{\alpha}}, \pi_{i \dot{\alpha}} \big| \:\! d^4 \pi \,, 
\label{6.25}
\\[7pt]
& \big| \:\! \varPsi^{\:\! i_1\ldots i_{p}}_{\alpha_1 \ldots \alpha_{p} ; \;\! 
 j_1\ldots j_q, \:\! \dot{\alpha}_1 \ldots \dot{\alpha}_q} \big\rangle 
:=\hat{\mathcal{P}}^{\:\! i_1\ldots i_{p}}_{\alpha_1 \ldots \alpha_{p} ; \;\! 
 j_1\ldots j_q, \:\! \dot{\alpha}_1 \ldots \dot{\alpha}_q}  \big|F_{J} \big\rangle \,. 
\label{6.26}
\end{align}
 
The total symmetrization of $\big| \:\! \varPsi^{\:\! i_1\ldots i_{p}}_{\alpha_1 \ldots \alpha_{p} ; \;\! 
 j_1\ldots j_q, \:\! \dot{\alpha}_1 \ldots \dot{\alpha}_q} \big\rangle$ 
is carried out in accordance with Eq. (\ref{6.16}), yielding 
\begin{align}
\big| \:\! \varPsi^{(\mathrm{S}) \:\! i_1\ldots i_{p+q}}_{\alpha_1 \ldots \alpha_{p} ; \;\! 
\dot{\alpha}_1 \ldots \dot{\alpha}_q} \big\rangle 
:=\hat{\mathcal{P}}^{(\mathrm{S}) \:\! i_1\ldots i_{p+q}}_{\alpha_1 \ldots \alpha_{p} ; \;\! 
\dot{\alpha}_1 \ldots \dot{\alpha}_q}  \big|F_{J} \big\rangle 
\label{6.27}
\end{align}
with 
\begin{align}
\hat{\mathcal{P}}^{(\mathrm{S}) \:\! i_1\ldots i_{p+q}}_{\alpha_1 \ldots \alpha_{p} ; \;\! 
\dot{\alpha}_1 \ldots \dot{\alpha}_q} 
:=\frac{1}{N_{+} !} \sum_{\varsigma} 
\epsilon^{i_{\varsigma\:\! (p+1)} \:\! j_{1}} \cdots \epsilon^{i_{\varsigma\:\! (p+q)} \:\! j_{q}} 
\:\! 
\hat{\mathcal{P}}{}^{\:\! i_{\varsigma\:\! (1)} \ldots i_{\varsigma\:\! (p)}}_{\alpha_1 \ldots \alpha_{p} ; \;\! 
 j_1\ldots j_q, \:\! \dot{\alpha}_1 \ldots \dot{\alpha}_q} \,. 
\label{6.28}
\end{align}
This operator is abbreviated in Appendix B as $\hat{\mathcal{P}}^{(\mathrm{S})}$.

\section{Summary and discussion}

In this paper, we have reformulated the massive rigid particle model, 
defined by the action (\ref{1.1}) with $m\neq 0$, in terms of twistors 
and have investigated both classical and quantum mechanical properties of the twistor model 
established in the reformulation.

We first presented a first-order Lagrangian for a massive particle with rigidity, given in Eq. (\ref{2.3}), 
and verified that it is indeed equivalent to the original Lagrangian (\ref{2.14}). 
We subsequently elaborated a twistor representation of the Lagrangian (\ref{2.14}), 
given in Eq. (\ref{3.26}), via the first-order Lagrangian 
by following the procedure developed in the massless case \cite{DegSuz}. 
It was pointed out that the action with the twistorial Lagrangian (\ref{3.26}) has a form very similar to 
the GGS action for a massive spinning particle \cite{DegOka}. 
The Lagrangian (\ref{3.26}) remains invariant under the $U(1)_{1}\times U(1)_{2}$ transformation.  
Considering this, we carried out a partial gauge-fixing for the $U(1)_{1}\times U(1)_{2}$ symmetry by 
adding the gauge-fixing term $b(a_{1}+a_{2})$ and its associated term $M\zeta\varphi$ 
to the Lagrangian (\ref{3.26}). 
The canonical Hamiltonian formalism of the twistor model was studied on the basis of 
the gauge-fixed Lagrangian (\ref{3.32}). 
We were able to immediately obtain the canonical Dirac bracket relations between twistor variables, 
given in Eq. (\ref{4.30}), by virtue of choosing the appropriate gauge-fixing condition $a_{1}+a_{2}=0$.  
In addition, the associated condition $\varphi=0$ made our formulation simple and clear.\footnote{~Without  
the present gauge-fixing procedure, 
we will have some complicated Dirac bracket relations between twistor variables. 
In this case, we need to define new twistor variables fulfilling the canonical Dirac bracket relations
from the old ones, as carried out in Refs. \cite{FedLuk, DegOka}.}

As a result of this classical mechanical treatment, 
the canonical commutation relations between twistor operators, given in Eq. (\ref{5.2}),  
were obtained, so that the canonical quantization of the twistor model was properly accomplished.  
In the quantization procedure, we found the physical state conditions given in Eq. (\ref{5.4}) 
and saw that they eventually reduce to the algebraic mass-shell condition (\ref{5.13}) 
and the simultaneous differential equations (\ref{5.14d}) and (\ref{5.14e}) for the twistor function $F_{J}$. 
This function seems to correspond to the wave function of a physical state found by Plyushchay \cite{Ply04}.  
In the twistor formulation, however, 
we succeed to derive an arbitrary-rank massive spinor field $\varPsi$ by 
means of the Penrose transform of $F_{J}$ [$^{\:\!}$see Eq. (\ref{6.1})].  
This is an advantage of the twistor formulation developed in this paper. 
Intriguingly, the spinor field $\varPsi$ has extra indices in addition to the usual dotted and undotted 
spinor indices and satisfies the generalized DFP equations (\ref{6.13a}) and (\ref{6.13b}). 
(The extra indices will be related to the particle-antiparticle degrees of freedom \cite{DegOka}.) 
We also verified that the totally symmetrized spinor field $\varPsi^{(\mathrm{S})}$ satisfies 
the (ordinary) DFP equations (\ref{6.18a}) and (\ref{6.18b}). 
It is worth mentioning that the mass of the spinor fields $\varPsi$ and $\varPsi^{(\mathrm{S})}$ 
is determined depending on the spin quantum number $J$. 
More precisely, these spinor fields of rank $2J$ have the mass $M_{J}$ defined in Eq. (\ref{5.6}).

We proved, in the context of the twistor formulation, 
that the spin quantum number $J$ of a massive particle with rigidity can take only non-negative integer values. 
Although the method of proof is completely different from that of Plyushchay \cite{Ply04}, 
both methods led to the same result concerning the allowed values of $J$.  
(Hence the statement of Deriglazov and Nersessian \cite{DerNer} is contradicted.)  
In the twistor formulation, an essential condition for leading to this result is ultimately $\hat{\chi}{}^{(a)-} |F^{\:\!} \rangle =0$, 
given in Eq. (\ref{5.4d}).  
In fact, we derived the allowed values of $J$ by using the conditions (\ref{6.5a}) and (\ref{6.5b}), 
which were found by evaluating the homogeneity degrees of the particular solution $F_{Js_{\ast}}$ 
of the differential equation (\ref{5.14d}) originating from $\hat{\chi}{}^{(a)-} |F^{\:\!} \rangle =0$.

Finally we try to extend the twistor model so that $J$ can take positive half-integer values as well as  
non-negative integer values. For this purpose, we introduce the 1-dimensional $U(1)$ Chern-Simons terms 
\begin{align}
S_{i}:=-2s_{i} \int_{\tau_0}^{\;\!\tau_1} d\tau \:\! a_{i} 
\quad\; (\:\! i=1,2) 
\label{7.1}
\end{align}
with real constants $s_{i}$. 
Since the field $a_{i}$ transforms as Eq. (\ref{3.18}) under the proper reparametrization, 
$S_{i}$ is obviously reparametrization invariant. 
Also, $S_{i}$ remains invariant under the gauge transformation (\ref{3.19}), provided that $\theta_i$ satisfies 
an appropriate boundary condition such as $\theta_{i}(\tau_{1})=\theta_{i}(\tau_{0})$. 
Now we consider the extended twistor model governed by the action 
$\tilde{S}:=\int_{\;\!\tau_0}^{\;\!\tau_1} d\tau^{\,\!} L +S_{1} +S_{2}$, with $L$ being the Lagrangian (\ref{3.32}). 
In quantizing the extended twistor model, the physical state condition 
$\big(^{\:\!} \hat{\chi}{}^{(a)-} -s_{-} \big)|F^{\:\!} \rangle =0$ with $s_{-}:=s_{1}-s_{2}$  
is imposed on $ |F^{\:\!} \rangle$, instead of $\hat{\chi}{}^{(a)-} |F^{\:\!} \rangle =0$. 
Then it can be shown that $s_{-}$ takes only integer or half-integer values. 
In addition, Eq. (\ref{6.7a}) is modified as 
\begin{align}
N_{+}: &= q+p=2(q_1+p_2 -s_{-})=2(q_2+p_1+s_{-}) \,.
\label{7.2}
\end{align}
Using Eqs. (\ref{7.2}) and (\ref{6.8}), and noting $N_{+} \geq 0$, 
we see that $J$ can take both non-negative integer and positive half-integer values. 
In this way, the extended twistor model is established as a model for 
massive particles with integer or half-integer spin. 
However, in the case $s_{-} \neq 0$, 
the extended twistor model cannot be regarded as a twistor representation of the massive rigid particle model.  
In view of this situation, it would be interesting to extend the massive rigid particle model 
so as to correspond to the extended twistor model.

\section*{Acknowledgments}
We would like to thank the anonymous reviewer for helpful comments and suggestions on this paper. 
We are grateful to Shigefumi Naka and Satoshi Okano 
for useful discussions and comments. 
One of us (T.S.) thanks Kenji Yamada, Katsuhito Yamaguchi and Haruki Toyoda 
for their encouragement.



\appendix
\section{Pauli-Lubanski pseudovector and mass eigenvalues}
\label{sec:intro}

In this appendix, we focus, within the present framework,  
on the Pauli-Lubanski pseudovector and related matters. 
We also find specific forms of $M$ and $\gamma$ written in terms of $J$.

We can see that the first-order Lagrangian (\ref{2.3}) remains invariant under the infinitesimal Poincar\'{e} transformation    
\begin{subequations}
\label{A1}
\begin{align}
x^{\mu} & \rightarrow x^{\prime \mu}=x^{\mu} +\varepsilon^{\mu}{}_{\nu} x^{\nu} -\varepsilon^{\mu} , 
\label{A1a}
\\
q^{\mu} & \rightarrow q^{\prime \mu}=q^{\mu} +\varepsilon^{\mu}{}_{\nu} q^{\nu} , 
\label{A1b}
\\
p_{\mu} & \rightarrow p^{\prime}_{\mu}=p_{\mu} +\varepsilon_{\mu}{}^{\nu} p_{\nu} \,,
\label{A1c}
\\
r_{\mu} & \rightarrow r^{\prime}_{\mu}=r_{\mu} +\varepsilon_{\mu}{}^{\nu} r_{\nu} \,.
\label{A1d}
\end{align}
\end{subequations}
Here, $\varepsilon^{\mu\nu}(=-\varepsilon^{\nu\mu})$ are parameters of the infinitesimal Lorentz transformation 
and $\varepsilon^{\mu}$ are parameters of the infinitesimal translation. 
In accordance with Noether's theorem, the conserved quantities corresponding to $\varepsilon^{\mu}$ and 
$\varepsilon^{\mu\nu}$ are found from the Lagrangian (\ref{2.3}) to be $p_{\mu}$ and 
\begin{align}
\mathcal{M}_{\mu\nu}:=x_{\mu} p_{\nu}-x_{\nu} p_{\mu} +q_{\mu} r_{\nu}-q_{\nu} r_{\mu} \,,   
\label{A2}
\end{align}
respectively. Obviously, $p_{\mu}$ is the 4-momentum vector and 
$\mathcal{M}_{\mu\nu}$ is the angular momentum tensor. 
Now we define the Pauli-Lubanski pseudovector $W^{\mu}$ by \cite{PenMac, PenRin2, Gursey}  
\begin{align}
W^{\mu} :=
\frac{1}{2M} \epsilon^{\mu\nu\rho\sigma} p_{\nu} \mathcal{M}_{\rho\sigma} \,, 
\label{A3}
\end{align}
where $\epsilon^{0123}=-1$, and 
$M$ is the physical mass parameter defined in Eq. (\ref{3.8}). 
Since $p_{\mu}$ and $\mathcal{M}_{\mu\nu}$ are conserved quantities, $W^{\mu}$ is also a conserved quantity. 
Substituting Eq. (\ref{A2}) into Eq. (\ref{A3}) leads to 
\begin{align}
W^{\mu} :=
\frac{1}{M} \epsilon^{\mu\nu\rho\sigma} p_{\nu} q_{\rho} r_{\sigma} \,,
\label{A4}
\end{align}
from which we obtain 
\begin{align}
W^{2} &:=W_{\mu} W^{\mu} 
\notag
\\
&\;=\frac{1}{M^2} \left\{ -p^{2} q^{2} r^{2} +p^{2} (q^{\mu} r_{\mu})^{2} +q^{2} (r^{\mu} p_{\mu})^{2} 
+r^{2} (q^{\mu} p_{\mu})^{2} -2q^{\mu} r_{\mu} r^{\nu} p_{\nu} q^{\rho} p_{\rho} \right\} .
\label{A5}
\end{align}
By using Eqs. (\ref{2.5e}), (\ref{2.5f}), (\ref{2.6}), and (\ref{2.7}),  Eq. (\ref{A5}) becomes
\begin{align}
W^{2}=\frac{k^2}{M^2} \left( p^{2}-m^{2} \right) .
\label{A6}
\end{align}
Substitution of Eq. (\ref{3.6a}) into Eq. (\ref{A6}) gives 
\begin{align}
W^{2}=\frac{k^2}{M^2} \left( 2 |\varPi|^{2}-m^{2} \right) .
\label{A7}
\end{align}
In this way, we have an expression of $W^{2}$ based on the first-order Lagrangian.

The Pauli-Lubanski pseudovector derived from the twistorial Lagrangian (\ref{3.32}) takes the following form \cite{DegOka}: 
\begin{align}
W^{\alpha\dot{\alpha}} &=\frac{1}{M} 
\left\{ \left( \bar{\pi}{}^{j}_{\beta} \omega_{i}^{\beta} 
+\bar{\omega}{}^{\:\! j \dot{\beta}} \pi_{i\dot{\beta}} \right) \bar{\pi}{}^{i \alpha} \pi{}_{j}^{\dot{\alpha}} 
-\frac{1}{2} 
\left( \bar{\pi}{}^{i}_{\beta} \omega_{i}^{\beta} 
+\bar{\omega}{}^{i \dot{\beta}} \pi_{i\dot{\beta}} \right) \bar{\pi}{}^{j \alpha} \pi{}_{j}^{\dot{\alpha}} \right\} , 
\label{A8}
\end{align}
or equivalently, 
\begin{align}
W^{\alpha\dot{\alpha}} &=\frac{1}{M} 
\left(\delta_{i}^{l} \delta_{k}^{j} -\frac{1}{2} \delta_{i}^{j} \delta_{k}^{l} \right) 
\bar{Z}{}^{k}_{B} Z_{l}^{B} \bar{\pi}{}^{i \alpha} \pi{}_{j}^{\dot{\alpha}} \,. 
\label{A9}
\end{align}
After substitution of Eqs. (\ref{3.9}) and (\ref{3.10}) into Eq. (\ref{A8}), it reduces to 
Eq. (\ref{A4}) with Eq. (\ref{3.3}). 
By means of the formula 
\begin{align}
\frac{1}{2} \sigma_{ri}{}^{j} \sigma_{rk}{}^{l} 
=\delta_{i}^{l} \delta_{k}^{j} -\frac{1}{2} \delta_{i}^{j} \delta_{k}^{l} \,,
\label{A10}
\end{align}
Eq. (\ref{A9}) can be written as 
\begin{align}
W^{\alpha\dot{\alpha}}=\frac{1}{M} T_{r} \sigma_{ri}{}^{j} \bar{\pi}{}^{i \alpha} \pi{}_{j}^{\dot{\alpha}} ,  
\label{A11}
\end{align}
with
\begin{align}
T_{r} :=\frac{1}{2} \bar{Z}_{B}^{k} \sigma_{rk}{}^{l} Z^{B}_{l} .
\label{A12}
\end{align} 
Here, $\sigma_{ri}{}^{j}$ ($r=1,2,3$) denotes the $(i,j^{\:\!})$ entry of the Pauli matrix $\sigma_{r}$. 
(In Eqs. (\ref{A10}) and (\ref{A11}), summation over $r$ is understood.) 
As can be easily seen, $T_{r}$ is real. 
Using the formulas $\pi_{i\dot{\alpha}} \pi{}_{j}^{\dot{\alpha}} =\epsilon_{ij} \varPi$, 
$\bar{\pi}^{i}_{\alpha} \bar{\pi}{}^{j\alpha} =\epsilon^{ij} \bar{\varPi}$, and 
$\sigma_{2} \sigma_{r} \sigma_{2}=-\sigma_{r}^{\mathrm{T}}$, we obtain 
\begin{align}
W^{2}=-\frac{2}{M^2} \:\! T^{2} |\varPi|^{2}, \qquad T^{2}:= T_{r} T_{r} \,.
\label{A13}
\end{align}

In this paper, as should be emphasized here, 
quantization means the twistor quantization procedure accomplished by setting 
the canonical commutation relations (\ref{5.2a}) and (\ref{5.2b}). 
By using these commutation relations, 
the Weyl ordered operator corresponding to $W^{\alpha\dot{\alpha}}$, which is denoted by $\hat{W}^{\alpha\dot{\alpha}}$, 
can be simplified as 
\begin{align}
\hat{W}^{\alpha\dot{\alpha}}=\frac{1}{M} \hat{T}_{r} \sigma_{ri}{}^{j} 
\Hat{\Bar{\pi}}{}^{i \alpha} \Hat{\pi}{}_{j}^{\dot{\alpha}} ,  
\label{A14}
\end{align}
with
\begin{align}
\hat{T}_{r} :=\frac{1}{2} \Hat{\Bar{Z}}_{B}^{k} \sigma_{rk}{}^{l} \Hat{Z}{}^{B}_{l} .
\label{A15}
\end{align} 
It is easy to show that the operators $\hat{T}_{r}$ fulfill the $\mathit{SU}(2)$ commutation relation 
\begin{align}
\left[\:\! \hat{T}_{r}, \hat{T}_{s} \right]=i\epsilon_{rst} \hat{T}_{t} \,.
\label{A16}
\end{align}
Also, using Eq. (\ref{A16}) and the commutation relation 
\begin{align}
\left[\:\! \hat{T}_{r}, \:\! \sigma_{si}{}^{j} \Hat{\Bar{\pi}}{}^{i \alpha} \Hat{\pi}{}_{j}^{\dot{\alpha}} \right] 
=i\epsilon_{rst} \sigma_{ti}{}^{j} \Hat{\Bar{\pi}}{}^{i \alpha} \Hat{\pi}{}_{j}^{\dot{\alpha}} \,, 
\label{A17}
\end{align}
we can readily show the natural commutation relation
\begin{align}
\left[\:\! \hat{W}^{\alpha\dot{\alpha}}, \hat{T}_{r} \right]=0 \,,  
\label{A18}
\end{align}
which ensures that 
\begin{align}
\left[\:\! \hat{W}^{2}, \hat{T}^{2} \right]=0 \,.  
\label{A19}
\end{align}
Furthermore, using Eq. (\ref{A18}) and the formulas  
%
%
\begin{align}
&i\epsilon_{rst}  \sigma_{si}{}^{j} \sigma_{tk}{}^{l} 
=\sigma_{rk}{}^{j} \delta_{i}^{l} -\sigma_{ri}{}^{l} \delta_{k}^{j} \,, 
\label{A20}
\\
&\epsilon^{\alpha\dot{\alpha}\beta\dot{\beta}\gamma\dot{\gamma}\delta\dot{\delta}}
=i \left( \epsilon^{\alpha\gamma} \epsilon^{\beta\delta} \epsilon^{\dot{\alpha}\dot{\delta}} 
\epsilon^{\dot{\beta}\dot{\gamma}} 
-\epsilon^{\alpha\delta} \epsilon^{\beta\gamma} \epsilon^{\dot{\alpha}\dot{\gamma}} 
\epsilon^{\dot{\beta}\dot{\delta}} \:\! \right)\,, 
\label{A21}
\end{align}
we can prove that 
\begin{align}
\left[\:\! \hat{W}^{\alpha\dot{\alpha}}, \hat{W}^{\beta\dot{\beta}} \:\! \right]
=\frac{i}{M} \epsilon^{\alpha\dot{\alpha}\beta\dot{\beta}\gamma\dot{\gamma}\delta\dot{\delta}}
\hat{p}_{\gamma\dot{\gamma}} \hat{W}_{\delta\dot{\delta}} \,, 
\label{A22}
\end{align}
where 
$\hat{p}_{\gamma\dot{\gamma}}:=\Hat{\Bar{\pi}}{}^{i}_{\gamma} \Hat{\pi}{}_{i \dot{\gamma}}$.  
The operator $\hat{p}_{\alpha\dot{\alpha}}$ is the operator version of the $p_{\alpha\dot{\alpha}}$ 
given in Eq. (\ref{3.3a}) and commutes with $\hat{T}_{r}$ and $\hat{W}^{\alpha\dot{\alpha}}$. 
Equation (\ref{A22}) is a commutation relation that the Pauli-Lubanski pseudovector should satisfy 
\cite{Gursey}. 
It is well known that $\hat{W}^{2}:=\hat{W}_{\alpha\dot{\alpha}} \hat{W}^{\alpha\dot{\alpha}}$ 
is the spin Casimir operator of the Poincar\'{e} algebra.

As can easily be verified, $\hat{W}^{\alpha\dot{\alpha}}$, and hence $\hat{W}^{2}$, 
commutes with the following operators given in Eq. (\ref{5.4}): 
\begin{alignat}{5}
\hat{\phi}{}^{(a)-} &=\hat{P}{}^{(a)-}, 
& \qquad 
\hat{\phi}{}^{(h)} &=\hat{P}{}^{(h)}, 
& \qquad 
\hat{\phi}{}^{(\bar{h})} &=\hat{P}{}^{(\bar{h})},
\notag
\\
\hat{\chi}{}^{(a)-} &=\hat{T}_{3} \,,
& \qquad 
\hat{\chi}{}^{(h)} &=\sqrt{2} \:\! \hat{\varPi} -M , 
& \qquad 
\hat{\chi}{}^{(\bar{h})} &=\sqrt{2} \:\! \Hat{\Bar{\varPi}} -M . 
\label{A23}
\end{alignat}
For this reason, we can put the eigenvalue equation $\hat{W}^{2} |F^{\:\!} \rangle=\varLambda^{\:\!} |F^{\:\!} \rangle$,   
with an eigenvalue $\varLambda$, 
together with the conditions (\ref{5.4a})--(\ref{5.4f}). 
(These conditions can be regarded as eigenvalue equations with the common eigenvector $|F^{\:\!} \rangle$ 
and vanishing eigenvalues.) 
It is straightforward to see that in the particle's rest frame, the commutation relation (\ref{A22}) acting on 
$|F^{\:\!} \rangle$ reduces to $\big[ \hat{W}_{r}, \hat{W}_{s} \big]=i\epsilon_{rst} \hat{W}_{t}$  
by the use of Eqs. (\ref{5.4e}) and (\ref{5.4f}) (see Appendix C). 
Since the operators $\hat{W}_{r}$ fulfill the $\mathit{SU}(2)$ commutation relation, 
the eigenvalue of $\hat{W}_{r} \hat{W}_{r}$ is found to be $J(J+1)$ under a proper condition. 
Here, $J$ denotes the spin quantum number 
that takes non-negative integer or positive half-integer values. 
Noting $\hat{W}^{2}=-\hat{W}_{r} \hat{W}_{r}$, we thus obtain 
\begin{align}
\hat{W}^{2} |F^{\:\!} \rangle=-J(J+1) |F^{\:\!} \rangle\,, \qquad J=0, \frac{1}{2}, 1, \frac{3}{2}, \ldots\,,    
\label{A24}
\end{align}
so that $\varLambda$ is determined to be $-J(J+1)$.

Now, we can expect that the operator version of Eq. (\ref{A7}) holds at the quantum mechanical level.  
Acting this operator on $|F^{\:\!} \rangle$, we have 
\begin{align}
\hat{W}^{2}|F^{\:\!} \rangle
=\frac{k^2}{M^2}\left( 2 \hat{\varPi} \Hat{\Bar{\varPi}} -m^{2} \right) |F^{\:\!} \rangle \,, 
\label{A25}
\end{align}
where relevant commutation relations have been used after applying the Weyl ordering rule. 
With Eqs. (\ref{5.4e}) and (\ref{5.4f}), Eq. (\ref{A25}) becomes 
\begin{align}
\hat{W}^{2}|F^{\:\!} \rangle
=\frac{k^2}{M^2} \left( M^{2} -m^{2} \right) |F^{\:\!} \rangle .
\label{A26}
\end{align}
From Eqs. (\ref{A24}) and (\ref{A26}), $M$ is determined to be 
\begin{align}
M_{J} :=\frac{m}{\sqrt{1+\dfrac{J(J+1)}{k^{2}}}} \,.
\label{A27}
\end{align}
Accordingly, the positive constant $\gamma$ is determined from Eq. (\ref{3.8}) to be 
\begin{align}
\gamma_{J,\pm} :=\left(\sqrt{1+\frac{J(J+1)}{k^{2}}} \pm \sqrt{\frac{J(J+1)}{k^{2}}} \:\right)^{2} .
\label{A28}
\end{align}

We next consider the operator version of Eq. (\ref{A13}) with applying the Weyl ordering rule. 
Acting this operator on $|F^{\:\!} \rangle$ and using relevant commutation relations, we have 
\begin{align}
\hat{W}^{2} |F^{\:\!} \rangle
=-\frac{2}{M^2} \:\! \hat{T}^{2} \hat{\varPi} \Hat{\Bar{\varPi}}^{\:\!}|F^{\:\!} \rangle, \qquad 
\hat{T}^{2}:= \hat{T}_{r} \hat{T}_{r} \,.
\label{A29}
\end{align}
With Eqs. (\ref{5.4e}) and (\ref{5.4f}), Eq. (\ref{A29}) becomes  
\begin{align}
\hat{W}^{2} |F^{\:\!} \rangle=-\hat{T}^{2} |F^{\:\!} \rangle \,.
\label{A30}
\end{align}
We note that $\hat{T}^{2}$ commutes with $\hat{W}^{2}$ and with all the operators in Eq. (\ref{A23}). 
Therefore $|F^{\:\!} \rangle$ can be chosen to be a simultaneous eigenvector of $\hat{T}^{2}$ and 
these other operators. Since $\hat{T}^{2}$ is the Casimir operator of the $\mathit{SU}(2)$ Lie algebra, 
$|F^{\:\!} \rangle$ turns out to satisfy, under a proper condition, the eigenvalue equation 
\begin{align}
\hat{T}^{2} |F^{\:\!} \rangle=I(I+1) |F^{\:\!} \rangle\,, \qquad I=0, \frac{1}{2}, 1, \frac{3}{2}, \ldots\,. 
\label{A31}
\end{align}
Combining Eqs. (\ref{A24}), (\ref{A30}), and (\ref{A31}) leads to the expected result 
\begin{align}
I=J \,.
\label{A32}
\end{align}

\section{A proof of Eq. (\ref{6.8})} 
\label{sec:intro}

This appendix is devoted to prove Eq. (\ref{6.8}) in a twistorial fashion.

Let us recall the commutation relations in Eqs. (\ref{5.2a}) and (\ref{5.2b}). 
In terms of the spinor components of $\hat{Z}{}^{A}_{i}$ and $\Hat{\Bar{Z}}{}_{A}^{i}$, these commutation relations 
are expressed as 
\begin{alignat}{3}
\left[ \:\! \hat{\omega}{}_{i}^{\alpha}, \:\! \Hat{\Bar{\pi}}{}^{j}_{\beta} \:\! \right] 
&=\delta_{i}^{j} \delta_{\beta}^{\alpha} \,, 
&\qquad 
\left[ \:\! \hat{\pi}{}_{i \dot{\alpha}}, \:\! \Hat{\Bar{\omega}}{}^{\:\! j \dot{\beta}}  \:\! \right] 
&= \delta_{i}^{j} \delta_{\dot{\alpha}}^{\dot{\beta}} \,, 
\label{B1} 
\\
\mbox{all others} &=0 \,. 
\notag
\end{alignat}
We now consider the bra-vector 
\begin{align}
\langle \bar{\pi}, \pi |
:=\big\langle \tilde{0} \:\! \big| \exp \!\!\; \left( \bar{\pi}{}^{i}_{\alpha} \hat{\omega}{}_{i}^{\alpha} 
-\pi{}_{i \dot{\alpha}} \Hat{\Bar{\omega}}{}^{i \dot{\alpha}} \right) ,
\label{B2} 
\end{align}
instead of the $\big\langle {Z}^{\:\!} \big|$ defined in Eq. (\ref{6.20}). 
Here, $\big\langle \tilde{0}^{\:\!} \big|$ is a reference bra-vector specified by 
$\big\langle \tilde{0}^{\:\!} \big| \Hat{\Bar{\pi}}{}^{i}_{\alpha} 
=\big\langle \tilde{0}^{\:\!} \big| \hat{\pi}{}_{i \dot{\alpha}} =0$. 
It is easy to show, by using the commutation relations in Eq. (\ref{B1}), that 
\begin{subequations}
\label{B3}
\begin{alignat}{3}
\langle \bar{\pi}, \pi | \Hat{\Bar{\pi}}{}^{i}_{\alpha} 
&=\bar{\pi}{}^{i}_{\alpha} \langle  \bar{\pi}, \pi | \,, 
&\qquad
\langle \bar{\pi}, \pi | \hat{\pi}_{i \dot{\alpha}}
&=\pi_{i\dot{\alpha}} \langle  \bar{\pi}, \pi | \,, 
\label{B3a}
\\
\langle \bar{\pi}, \pi | \:\! \hat{\omega}{}_{i}^{\alpha} 
&=\frac{\partial}{\partial \bar{\pi}{}^{i}_{\alpha}} \langle \bar{\pi}, \pi | \,, 
&\qquad 
\langle \bar{\pi}, \pi | \:\! \Hat{\Bar{\omega}}{}^{i\dot{\alpha}}
&=-\frac{\partial}{\partial \pi_{i\dot{\alpha}}} \langle  \bar{\pi}, \pi | \,. 
\label{B3b}
\end{alignat}
\end{subequations}

\noindent
\paragraph{Proposition 1}
Let $G_{1}$ and $G_{2}$ be piecewise smooth functions of $\bar{\pi}{}^{i}_{\alpha}$ and $\pi_{i\dot{\alpha}}$, 
let $|G_{2} \rangle$ be the ket vector such that $G_{2}(\bar{\pi}, \pi)=\langle \bar{\pi}, \pi |G_{2} \rangle$, 
and let 
$\overrightarrow{T}{}^{2}:=\overrightarrow{T}_{\! r} \overrightarrow{T}_{\! r}$ 
be the differential operator representation of $\hat{T}{}^{2}$, defined from 
\begin{align}
\overrightarrow{T}_{\! r} :=\frac{1}{2} \left(
\bar{\pi}{}^{k}_{\beta} \sigma_{rk}{}^{l} \frac{\partial}{\partial \bar{\pi}{}^{l}_{\beta}}
-\pi_{l\dot{\beta}} \sigma_{rk}{}^{l} \frac{\partial}{\partial \pi_{k\dot{\beta}}} \right). 
\label{B4}
\end{align}
Then 
\begin{align}
\int_{\Bbb{C}^4} G_{1}(\bar{\pi}, \pi) \langle \bar{\pi}, \pi | \:\! \hat{T}{}^{2} |G_{2} \rangle 
\:\! d^{4} \bar{\pi} \wedge d^4 \pi 
=\int_{\Bbb{C}^4} \left\{\;\! \overrightarrow{T}{}^{2} G_{1}(\bar{\pi}, \pi) \right\} G_{2}(\bar{\pi}, \pi) 
\:\! d^{4} \bar{\pi} \wedge d^4 \pi 
\label{B5} 
\end{align}
holds, provided that the integrals on both sides are finite. 
Here, $d^{4} \bar{\pi}:=d\bar{\pi}^{1}_{0} \wedge d\bar{\pi}^{1}_{1} \wedge 
d\bar{\pi}^{2}_{0} \wedge d\bar{\pi}^{2}_{1}$ 
and $d^4 \pi :=d\pi_{1\dot{0}} \wedge d\pi_{1\dot{1}} \wedge d\pi_{2\dot{0}} \wedge d\pi_{2\dot{1}}$. 

\vspace{3mm}

{\it Proof}$\:$: Let us first recall Eq. (\ref{A15}), or equivalently, 
\begin{align}
\hat{T}_{r}:=\frac{1}{2} 
\left( \Hat{\Bar{\pi}}{}^{k}_{\beta} \sigma_{rk}{}^{l} \hat{\omega}{}_{l}^{\beta} 
+\Hat{\Bar{\omega}}{}^{k\dot{\beta}} \sigma_{rk}{}^{l} \hat{\pi}_{l \dot{\beta}} \right) .
\label{B6}
\end{align}
By using Eq. (\ref{B3}) and the traceless property $\sigma_{rk}{}^{k}=0$, it can be shown that 
$\langle \bar{\pi}, \pi |^{\:\!} \hat{T}_{r} =\overrightarrow{T}_{\! r\:\!} \langle \bar{\pi}, \pi|$. 
Applying this twice to the left-hand side of Eq. (\ref{B5}), we have  
\begin{align}
\int_{\Bbb{C}^4} G_{1}(\bar{\pi}, \pi) \langle \bar{\pi}, \pi | \:\! \hat{T}{}^{2} |G_{2} \rangle 
\;\! d^{4} \bar{\pi} \wedge d^4 \pi 
=\int_{\Bbb{C}^4} G_{1}(\bar{\pi}, \pi) 
\:\! \overrightarrow{T}{}^{2} G_{2}(\bar{\pi}, \pi) 
\;\! d^{4} \bar{\pi} \wedge d^4 \pi \,. 
\label{B7} 
\end{align}
Carrying out integration by parts twice on the right-hand side of Eq. (\ref{B7}) leads to 
the right-hand side of Eq. (\ref{B5}):  
\begin{align}
\int_{\Bbb{C}^4} G_{1}(\bar{\pi}, \pi) 
\overrightarrow{T}{}^{2} G_{2}(\bar{\pi}, \pi) 
\;\! d^{4} \bar{\pi} \wedge d^4 \pi 
&=
-\int_{\Bbb{C}^4} \left\{ \;\! \overrightarrow{T}_{\! r} G_{1}(\bar{\pi}, \pi) \right\} 
\overrightarrow{T}_{\! r} G_{2}(\bar{\pi}, \pi) 
\;\! d^{4} \bar{\pi} \wedge d^4 \pi 
\notag
\\
&=
\int_{\Bbb{C}^4} \left\{ \;\! \overrightarrow{T}{}^{2} G_{1}(\bar{\pi}, \pi) \right\} G_{2}(\bar{\pi}, \pi) 
\;\! d^{4} \bar{\pi} \wedge d^4 \pi 
\,.
\label{B8}
\end{align}
Here, in addition to  $\sigma_{rk}{}^{k}=0$, we have used the fact that 
the boundary terms appearing in the integrations by parts vanish, 
because the integrals on both sides of Eq. (\ref{B5}) are assumed to be finite. 
Combining Eqs. (\ref{B7}) and (\ref{B8}) thus gives Eq. (\ref{B5}). 
\hspace{\fill} $\blacksquare$ 

\vspace{5mm}

Now we consider the following function:  
\begin{align}
\varOmega^{\:\! i_1\ldots i_{p+q}}_{\alpha_1 \ldots \alpha_{p} ; \;\! 
\dot{\alpha}_1 \ldots \dot{\alpha}_q}(z)
:=\int_{\Bbb{C}^4} 
\langle \bar{\pi}, \pi | 
\:\! \hat{\mathcal{P}}^{(\mathrm{S}) \:\! i_1\ldots i_{p+q}}_{\alpha_1 \ldots \alpha_{p} ; \;\! 
\dot{\alpha}_1 \ldots \dot{\alpha}_q} \hat{T}{}^{2} |F^{\:\!} \rangle 
E\big(z^{\alpha\dot{\alpha}} \bar{\pi}{}^{i}_{\alpha} \pi_{i\dot{\alpha}} \big)  
\;\! d^{4} \bar{\pi} \wedge d^4 \pi \,, 
\label{B9}
\end{align}
where $E$ is a function of $z^{\alpha\dot{\alpha}} \bar{\pi}{}^{i}_{\alpha} \pi_{i\dot{\alpha}}$, and 
$\hat{\mathcal{P}}^{(\mathrm{S})}$ is the operator defined in Eq. (\ref{6.28}). 
We here employ $z^{\alpha \dot{\alpha}}$ rather than $x^{\alpha \dot{\alpha}}$, 
because $z^{\alpha \dot{\alpha}}$ is useful for defining finite integrals. 
By using the eigenvalue equations in Eq. (\ref{B3a}) repeatedly, Eq. (\ref{B9}) can be written as 
\begin{align}
\varOmega^{\:\! i_1\ldots i_{p+q}}_{\alpha_1 \ldots \alpha_{p} ; \;\! 
\dot{\alpha}_1 \ldots \dot{\alpha}_q}(z)
=\int_{\Bbb{C}^4} 
{\mathcal{P}}^{(\mathrm{S}) \:\! i_1\ldots i_{p+q}}_{\alpha_1 \ldots \alpha_{p} ; \;\! 
\dot{\alpha}_1 \ldots \dot{\alpha}_q} (\bar{\pi}, \pi) 
E\big(z^{\alpha\dot{\alpha}} \bar{\pi}{}^{i}_{\alpha} \pi_{i\dot{\alpha}} \big)  
\langle \bar{\pi}, \pi | \:\! \hat{T}{}^{2} |F^{\:\!} \rangle 
\;\! d^{4} \bar{\pi} \wedge d^4 \pi \,,
\label{B10}
\end{align}
where ${\mathcal{P}}^{(\mathrm{S})}$ is defined by 
\begin{align}
\mathcal{P}^{(\mathrm{S}) \:\! i_1\ldots i_{p+q}}_{\alpha_1 \ldots \alpha_{p} ; \;\! 
\dot{\alpha}_1 \ldots \dot{\alpha}_q} (\bar{\pi}, \pi) 
:=\frac{1}{N_{+} !} \sum_{\varsigma} 
\epsilon^{i_{\varsigma\:\! (p+1)} \:\! j_{1}} \cdots \epsilon^{i_{\varsigma\:\! (p+q)} \:\! j_{q}} 
\:\! 
\mathcal{P}{}^{\:\! i_{\varsigma\:\! (1)} \ldots i_{\varsigma\:\! (p)}}_{\alpha_1 \ldots \alpha_{p} ; \;\! 
 j_1\ldots j_q, \:\! \dot{\alpha}_1 \ldots \dot{\alpha}_q} (\bar{\pi}, \pi) 
\label{B11}
\end{align}
with 
\begin{align}
\mathcal{P}{}^{\:\! i_1\ldots i_{p}}_{\alpha_1 \ldots \alpha_{p} ; \;\! 
 j_1\ldots j_q, \:\! \dot{\alpha}_1 \ldots \dot{\alpha}_q} (\bar{\pi}, \pi) 
:=(-1)^{p} \:\! 
{\pi}_{j_{1} \dot{\alpha}_{1}} \cdots {\pi}_{j_{q} \dot{\alpha}_{q}} 
{\Bar{\pi}}{}^{i_{1}}_{\alpha_{1}} \cdots {\Bar{\pi}}{}^{i_{p}}_{\alpha_{p}} \,. 
\label{B12}
\end{align}
We now apply Proposition 1 to the case of 
$G_{1}={\mathcal{P}}^{(\mathrm{S})} E$ and $|G_{2} \rangle =|F^{\:\!} \rangle$. 
Then Eq. (\ref{B5}) reads 
\begin{align}
\varOmega^{\:\! i_1\ldots i_{p+q}}_{\alpha_1 \ldots \alpha_{p} ; \;\! 
\dot{\alpha}_1 \ldots \dot{\alpha}_q}(z)
=\int_{\Bbb{C}^4} 
\left\{ \;\! \overrightarrow{T}{}^{2\:\!} 
\Big( 
{\mathcal{P}}^{(\mathrm{S}) \:\! i_1\ldots i_{p+q}}_{\alpha_1 \ldots \alpha_{p} ; \;\! 
\dot{\alpha}_1 \ldots \dot{\alpha}_q} (\bar{\pi}, \pi) 
E\big(z^{\alpha\dot{\alpha}} \bar{\pi}{}^{i}_{\alpha} \pi_{i\dot{\alpha}} \big) \Big) \right\}  
\tilde{F}(\bar{\pi}, \pi) 
\;\! d^{4} \bar{\pi} \wedge d^4 \pi \,, 
\label{B13}
\end{align}
where $\tilde{F}(\bar{\pi}, \pi):=\langle \bar{\pi}, \pi |F^{\:\!} \rangle$. 
The functions $E$ and $\tilde{F}$ are assumed to be chosen in such a way that 
they are piecewise smooth and the integrals in Eqs. (\ref{B10}) and (\ref{B13}) are finite. 
By means of the fact that 
$\overrightarrow{T}_{\! r} 
E\big(z^{\alpha\dot{\alpha}} \bar{\pi}{}^{i}_{\alpha} \pi_{i\dot{\alpha}} \big)=0$, 
Eq. (\ref{B13}) becomes 
\begin{align}
\varOmega^{\:\! i_1\ldots i_{p+q}}_{\alpha_1 \ldots \alpha_{p} ; \;\! 
\dot{\alpha}_1 \ldots \dot{\alpha}_q}(z)
=\int_{\Bbb{C}^4} 
\left\{ \;\! \overrightarrow{T}{}^{2} \:\! 
{\mathcal{P}}^{(\mathrm{S}) \:\! i_1\ldots i_{p+q}}_{\alpha_1 \ldots \alpha_{p} ; \;\! 
\dot{\alpha}_1 \ldots \dot{\alpha}_q} (\bar{\pi}, \pi) \right\} 
E\big(z^{\alpha\dot{\alpha}} \bar{\pi}{}^{i}_{\alpha} \pi_{i\dot{\alpha}} \big) 
\tilde{F}(\bar{\pi}, \pi) 
\;\! d^{4} \bar{\pi} \wedge d^4 \pi \,. 
\label{B14}
\end{align}

\noindent
\paragraph{Proposition 2}
The following eigenvalue equation holds: 
\begin{align}
\overrightarrow{T}{}^{2} \:\! 
{\mathcal{P}}^{(\mathrm{S}) \:\! i_1\ldots i_{p+q}}_{\alpha_1 \ldots \alpha_{p} ; \;\! 
\dot{\alpha}_1 \ldots \dot{\alpha}_q} 
=\frac{N_{+}}{2} \left(\frac{N_{+}}{2} +1 \right) 
{\mathcal{P}}^{(\mathrm{S}) \:\! i_1\ldots i_{p+q}}_{\alpha_1 \ldots \alpha_{p} ; \;\! 
\dot{\alpha}_1 \ldots \dot{\alpha}_q} \,, 
\label{B15}
\end{align}
where $N_{+}:=p+q$. 

\vspace{3mm}

{\it Proof}$\:$:
Using Eq. (\ref{B4}) and the formula (\ref{A10}), we can show that
\begin{align}
\overrightarrow{T}{}^{2}
& =\frac{1}{4} \left( \:\! 3\bar{\pi}{}^{i}_{\alpha} \frac{\partial}{\partial \bar{\pi}{}^{i}_{\alpha}} 
+2\bar{\pi}{}^{i}_{\alpha} \bar{\pi}{}^{\:\! j}_{\beta} 
\frac{\partial}{\partial \bar{\pi}{}^{\:\! j}_{\alpha}} \frac{\partial}{\partial \bar{\pi}{}^{i}_{\beta}} 
-\bar{\pi}{}^{i}_{\alpha} \bar{\pi}{}^{\:\! j}_{\beta} \frac{\partial}{\partial \bar{\pi}{}^{i}_{\alpha}} 
\frac{\partial}{\partial \bar{\pi}{}^{\:\!j}_{\beta}} \right.
\notag
\\
& \quad \,
-4\bar{\pi}{}^{i}_{\alpha} \pi_{i\dot{\beta}} \frac{\partial}{\partial \bar{\pi}{}^{\:\! j}_{\alpha}} \frac{\partial}{\partial \pi_{j\dot{\beta}}} 
+2\bar{\pi}{}^{i}_{\alpha} \pi_{j\dot{\beta}} \frac{\partial}{\partial \bar{\pi}{}^{i}_{\alpha}} \frac{\partial}{\partial \pi_{j\dot{\beta}}} 
\notag
\\
& \quad \, \left. 
+3\pi_{i\dot{\alpha}} \frac{\partial}{\partial \pi_{i\dot{\alpha}}} 
+2\pi_{i\dot{\alpha}} \pi_{j\dot{\beta}} \frac{\partial}{\partial \pi_{j\dot{\alpha}}} \frac{\partial}{\partial \pi_{i\dot{\beta}}} 
-\pi_{i\dot{\alpha}} \pi_{j\dot{\beta}} \frac{\partial}{\partial \pi_{i\dot{\alpha}}} \frac{\partial}{\partial \pi_{j\dot{\beta}}} 
\right). 
\label{B16}
\end{align}
It is not difficult to verify the following equalities: 
\begin{subequations}
\label{B17}
\begin{align}
\bar{\pi}{}^{i}_{\alpha} \frac{\partial}{\partial \bar{\pi}{}^{i}_{\alpha}} 
\mathcal{P}^{(\mathrm{S}) \:\! i_1\ldots i_{p+q}}_{\alpha_1 \ldots \alpha_{p} ; \;\! 
\dot{\alpha}_1 \ldots \dot{\alpha}_q} 
&=p\;\! \mathcal{P}^{(\mathrm{S}) \:\! i_1\ldots i_{p+q}}_{\alpha_1 \ldots \alpha_{p} ; \;\! 
\dot{\alpha}_1 \ldots \dot{\alpha}_q} \,, 
\label{B17a}
\\
\bar{\pi}{}^{i}_{\alpha} \bar{\pi}{}^{\:\! j}_{\beta} 
\frac{\partial}{\partial \bar{\pi}{}^{\:\! j}_{\alpha}} \frac{\partial}{\partial \bar{\pi}{}^{i}_{\beta}} 
\mathcal{P}^{(\mathrm{S}) \:\! i_1\ldots i_{p+q}}_{\alpha_1 \ldots \alpha_{p} ; \;\! 
\dot{\alpha}_1 \ldots \dot{\alpha}_q} 
&=p(p-1) \;\! 
\mathcal{P}^{(\mathrm{S}) \:\! i_1\ldots i_{p+q}}_{\alpha_1 \ldots \alpha_{p} ; \;\! 
\dot{\alpha}_1 \ldots \dot{\alpha}_q} \,, 
\label{B17b}
\\
\bar{\pi}{}^{i}_{\alpha} \bar{\pi}{}^{\:\! j}_{\beta} \frac{\partial}{\partial \bar{\pi}{}^{i}_{\alpha}} 
\frac{\partial}{\partial \bar{\pi}{}^{\:\!j}_{\beta}} 
\mathcal{P}^{(\mathrm{S}) \:\! i_1\ldots i_{p+q}}_{\alpha_1 \ldots \alpha_{p} ; \;\! 
\dot{\alpha}_1 \ldots \dot{\alpha}_q} 
&=p(p-1) \;\! 
\mathcal{P}^{(\mathrm{S}) \:\! i_1\ldots i_{p+q}}_{\alpha_1 \ldots \alpha_{p} ; \;\! 
\dot{\alpha}_1 \ldots \dot{\alpha}_q} \,, 
\label{B17c}
\\
\bar{\pi}{}^{i}_{\alpha} \pi_{i\dot{\beta}} \frac{\partial}{\partial \bar{\pi}{}^{\:\! j}_{\alpha}} \frac{\partial}{\partial \pi_{j\dot{\beta}}} 
\mathcal{P}^{(\mathrm{S}) \:\! i_1\ldots i_{p+q}}_{\alpha_1 \ldots \alpha_{p} ; \;\! 
\dot{\alpha}_1 \ldots \dot{\alpha}_q} 
&=0 \,, 
\label{B17d}
\\
\bar{\pi}{}^{i}_{\alpha} \pi_{j\dot{\beta}} \frac{\partial}{\partial \bar{\pi}{}^{i}_{\alpha}} \frac{\partial}{\partial \pi_{j\dot{\beta}}} 
\mathcal{P}^{(\mathrm{S}) \:\! i_1\ldots i_{p+q}}_{\alpha_1 \ldots \alpha_{p} ; \;\! 
\dot{\alpha}_1 \ldots \dot{\alpha}_q} 
&=pq \;\! 
\mathcal{P}^{(\mathrm{S}) \:\! i_1\ldots i_{p+q}}_{\alpha_1 \ldots \alpha_{p} ; \;\! 
\dot{\alpha}_1 \ldots \dot{\alpha}_q} \,,
\label{B17e}
\\
\pi_{i\dot{\alpha}} \frac{\partial}{\partial \pi_{i\dot{\alpha}}} 
\mathcal{P}^{(\mathrm{S}) \:\! i_1\ldots i_{p+q}}_{\alpha_1 \ldots \alpha_{p} ; \;\! 
\dot{\alpha}_1 \ldots \dot{\alpha}_q} 
&=q \;\! 
\mathcal{P}^{(\mathrm{S}) \:\! i_1\ldots i_{p+q}}_{\alpha_1 \ldots \alpha_{p} ; \;\! 
\dot{\alpha}_1 \ldots \dot{\alpha}_q} \,,
\label{B17f}
\\
\pi_{i\dot{\alpha}} \pi_{j\dot{\beta}} \frac{\partial}{\partial \pi_{j\dot{\alpha}}} \frac{\partial}{\partial \pi_{i\dot{\beta}}} 
\mathcal{P}^{(\mathrm{S}) \:\! i_1\ldots i_{p+q}}_{\alpha_1 \ldots \alpha_{p} ; \;\! 
\dot{\alpha}_1 \ldots \dot{\alpha}_q} 
&=q(q-1) \;\! 
\mathcal{P}^{(\mathrm{S}) \:\! i_1\ldots i_{p+q}}_{\alpha_1 \ldots \alpha_{p} ; \;\! 
\dot{\alpha}_1 \ldots \dot{\alpha}_q} \,,
\label{B17g}
\\
\pi_{i\dot{\alpha}} \pi_{j\dot{\beta}} \frac{\partial}{\partial \pi_{i\dot{\alpha}}} \frac{\partial}{\partial \pi_{j\dot{\beta}}} 
\mathcal{P}^{(\mathrm{S}) \:\! i_1\ldots i_{p+q}}_{\alpha_1 \ldots \alpha_{p} ; \;\! 
\dot{\alpha}_1 \ldots \dot{\alpha}_q} 
&=q(q-1) \;\!  
\mathcal{P}^{(\mathrm{S}) \:\! i_1\ldots i_{p+q}}_{\alpha_1 \ldots \alpha_{p} ; \;\! 
\dot{\alpha}_1 \ldots \dot{\alpha}_q} \,.
\label{B17g}
\end{align}
\end{subequations}
Combining Eqs. (\ref{B16}) and (\ref{B17}) leads to Eq. (\ref{B15}). 
\hspace{\fill} $\blacksquare$ 

\vspace{5mm}

Applying Eq. (\ref{A31}) to Eq. (\ref{B10}), and Eq. (\ref{B15}) to Eq. (\ref{B14}), we have 
\begin{align}
&\left\{I(I+1)-\frac{N_{+}}{2} \left(\frac{N_{+}}{2} +1 \right) \right\} 
\notag
\\
&\times \int_{\Bbb{C}^4} 
{\mathcal{P}}^{(\mathrm{S}) \:\! i_1\ldots i_{p+q}}_{\alpha_1 \ldots \alpha_{p} ; \;\! 
\dot{\alpha}_1 \ldots \dot{\alpha}_q} (\bar{\pi}, \pi) 
E\big(z^{\alpha\dot{\alpha}} \bar{\pi}{}^{i}_{\alpha} \pi_{i\dot{\alpha}} \big) 
\tilde{F}(\bar{\pi}, \pi) 
\;\! d^{4} \bar{\pi} \wedge d^4 \pi =0 \,.
\label{B18}
\end{align}
Since the function $E$ is completely arbitrary, as long as it is piecewise smooth and makes the integral in Eq. (\ref{B18}) finite, 
we can conclude that $I=N_{+}/2$. 
Combining this with Eq. (\ref{A32}), we have 
\begin{align}
J=\frac{N_{+}}{2} \,.
\label{B19}
\end{align}
The proof of Eq. (\ref{6.8}) is thus complete.

\section{Plyushchay's noncovariant formulation}
\label{sec:intro}

In this appendix, we extract some part of Plyushchay's noncovariant formulation \cite{Ply01, Ply04} 
from the current twistor formulation.  
It is again shown that the spin quantum number $J$ takes only non-negative integer vales.

In addition to $p_{\alpha\dot{\alpha}}=\bar{\pi}{}^{i}_{\alpha} \pi_{i\dot{\alpha}}$ and 
the Pauli-Lubanski pseudovector $W^{\alpha\dot{\alpha}}$, 
we introduce a new 4-vector 
\begin{align}
n_{\alpha\dot{\alpha}} &:=
\frac{1}{M} \left(\bar{\pi}^{1}_{\alpha} \pi_{1\dot{\alpha}}-\bar{\pi}^{2}_{\alpha} \pi_{2\dot{\alpha}} \right) .
\label{C1}
\end{align}
We can readily show that 
\begin{subequations}
\label{C2}
\begin{align}
p_{\mu} p^{\mu} &=p_{\alpha\dot{\alpha}} p^{\alpha\dot{\alpha}} =2|\varPi|{}^{2} \approx M^{2} >0 \,, 
\label{C2a}
\\
n_{\mu} n^{\mu} &=
n_{\alpha\dot{\alpha}} n^{\alpha\dot{\alpha}} =-\frac{2}{M^{2}} |\varPi|{}^{2} \approx -1 <0 \,, 
\label{C2b}
\\
p_{\mu} W^{\mu} &=p_{\alpha\dot{\alpha}} W^{\alpha\dot{\alpha}} =0 \,, 
\label{C2c}
\\
p_{\mu} n^{\mu} &=p_{\alpha\dot{\alpha}} n^{\alpha\dot{\alpha}} =0 \,, 
\label{C2d}
\\
n_{\mu} W^{\mu} &=n_{\alpha\dot{\alpha}} W^{\alpha\dot{\alpha}}=-\frac{2}{M^2} |\varPi|{}^{2} \chi^{(a)-} 
\approx -\chi^{(a)-} \approx 0\,.
\label{C2e} 
\end{align}
\end{subequations}
In Eqs. (\ref{C2a}) and  (\ref{C2b}), 
the constraints $\chi^{(h)} \approx 0$ and $\chi^{(\bar{h})} \approx 0$ have been used. 
In Eq. (\ref{C2e}), Eq. (\ref{A9}) and 
the constraints $\chi^{(h)} \approx 0$, $\chi^{(\bar{h})} \approx 0$, and $\chi^{(a)-} \approx 0$ 
have been used. 
Equation (\ref{A13}) leads to 
\begin{align}
W_{\mu} W^{\mu}= W_{\alpha\dot{\alpha}} W^{\alpha\dot{\alpha}}
= -\frac{2}{M^2} \:\! T^{2} |\varPi|^{2} \approx -T^{2} <0  \,. 
\label{C3}
\end{align}
We thus see that $p_{\mu}$ is a timelike vector, while $n_{\mu}$ and $W^{\mu}$ are spacelike vectors. 
We also see that the three vectors $p_{\mu}$, $n_{\mu}$, and $W^{\mu}$ are orthogonal to each other.  
Since $p_{\mu}$ is timelike, we can choose the rest frame specified by $p_{\mu}=(p_{0}, \boldsymbol{0})$.

After carrying out the twistor quantization procedure, we have the Weyl ordered operators 
$\hat{p}_{\mu}$, $\hat{n}_{\mu}$, and $\hat{W}^{\mu}$ 
corresponding to $p_{\mu}$, $n_{\mu}$, and $W^{\mu}$, respectively. 
Then Eqs. (\ref{C2c}) and (\ref{C2d}) turn into strong conditions at the operator level: 
\begin{subequations}
\label{C4}
\begin{align}
\hat{p}_{\mu} \hat{W}^{\mu} &=0 \,, 
\label{C4a}
\\
\hat{p}_{\mu} \hat{n}^{\mu} &=0 \,, 
\label{C4b}
\end{align}
\end{subequations}
while Eqs. (\ref{C2a}), (\ref{C2b}), (\ref{C2e}), and (\ref{C3}) 
turn into conditions valid for the physical state vector $|F^{\:\!} \rangle$: 
\begin{subequations}
\label{C5}
\begin{align}
\hat{p}_{\mu} \hat{p}^{\mu} |F^{\:\!} \rangle &=M^2 |F^{\:\!} \rangle \,, 
\label{C5a}
\\
\hat{n}_{\mu} \hat{n}^{\mu} |F^{\:\!} \rangle &=-|F^{\:\!} \rangle \,, 
\label{C5b}
\\
\hat{n}_{\mu} \hat{W}^{\mu} |F^{\:\!} \rangle &=0 \,, 
\label{C5c}
\\
\hat{W}_{\mu} \hat{W}^{\mu} |F^{\:\!} \rangle &=-\hat{T}{}^{2} |F^{\:\!} \rangle \,. 
\label{C5d}
\end{align}
\end{subequations}

The commutation relations among the operators 
$\hat{p}_{\alpha\dot{\alpha}}$, $\hat{n}_{\alpha\dot{\alpha}}$, and $\hat{W}^{\alpha\dot{\alpha}}$ 
can be calculated using 
the canonical commutation relations (\ref{5.2a}) and (\ref{5.2b}). In fact, we obtain 
\begin{subequations}
\label{C6}
\begin{align} 
\left[\:\! \hat{W}^{\alpha\dot{\alpha}}, \hat{n}^{\beta\dot{\beta}} \;\! \right]
&=\frac{i}{M} \epsilon^{\alpha\dot{\alpha}\beta\dot{\beta}\gamma\dot{\gamma}\delta\dot{\delta}}
\hat{p}_{\gamma\dot{\gamma}} \hat{n}_{\delta\dot{\delta}} \,, 
\label{C6a}
\\
\left[\:\! \hat{W}^{\alpha\dot{\alpha}}, \hat{W}^{\beta\dot{\beta}} \;\! \right]
&=\frac{i}{M} \epsilon^{\alpha\dot{\alpha}\beta\dot{\beta}\gamma\dot{\gamma}\delta\dot{\delta}}
\hat{p}_{\gamma\dot{\gamma}} \hat{W}_{\delta\dot{\delta}} 
\label{C6b}
\end{align}
\end{subequations}
[$^{\:\!}$see Eq. (\ref{A22})]. 
The other commutation relations 
$\big[^{\:\!} \hat{W}^{\alpha\dot{\alpha}}, \hat{p}^{\beta\dot{\beta}\;\!} \big]$, 
$\big[^{\:\!} \hat{p}^{\alpha\dot{\alpha}}, \hat{p}^{\beta\dot{\beta}\;\!} \big]$, 
$\big[^{\:\!} \hat{p}^{\alpha\dot{\alpha}}, \hat{n}^{\beta\dot{\beta}\;\!} \big]$, and 
$\big[^{\:\!} \hat{n}^{\alpha\dot{\alpha}}, \hat{n}^{\beta\dot{\beta}\;\!} \big]$ 
vanish.  
In 4-vector notation, Eqs. (\ref{C6a}) and (\ref{C6b}) can be expressed as 
\begin{subequations}
\label{C7}
\begin{align} 
\left[\:\! \hat{W}^{\mu}, \hat{n}^{\nu} \:\! \right]
&=\frac{i}{M} \epsilon^{\mu\nu\rho\sigma}
\hat{p}_{\rho} \hat{n}_{\sigma} \,, 
\label{C7a}
\\
\left[\:\! \hat{W}^{\mu}, \hat{W}^{\nu} \:\! \right]
&=\frac{i}{M} \epsilon^{\mu\nu\rho\sigma}
\hat{p}_{\rho} \hat{W}_{\sigma} \,. 
\label{C7b}
\end{align}
\end{subequations}

In the rest frame, the operators $\hat{p}_{\mu}$ take the specific form 
$\hat{p}_{\mu}=(\hat{p}_{0}, \boldsymbol{0})$. 
Accordingly, it follows from Eq. (\ref{C4a}) that 
$\hat{W}^{\mu}=(0, \hat{\boldsymbol{W}})=(0, \hat{W}_{r})$. 
Likewise, Eq. (\ref{C4b}) gives $\hat{n}^{\mu}=(0, \hat{\boldsymbol{n}})=(0, \hat{n}_{r})$. 
As a result, Eqs. (\ref{C5a}), (\ref{C5b}), (\ref{C5c}), and (\ref{C5d}) become, respectively,  
\begin{subequations}
\label{C8}
\begin{align}
\hat{p}_{0}^{2} \:\! |F^{\:\!} \rangle &=M^2 |F^{\:\!} \rangle \,, 
\label{C8a}
\\
\hat{n}_{r} \hat{n}_{r} |F^{\:\!} \rangle &=|F^{\:\!} \rangle \,, 
\label{C8b}
\\
\hat{n}_{r} \hat{W}_{r} |F^{\:\!} \rangle &=0 \,, 
\label{C8c}
\\
\hat{W}_{r} \hat{W}_{r} |F^{\:\!} \rangle &=\hat{T}{}^{2} |F^{\:\!} \rangle \,. 
\label{C8d}
\end{align}
\end{subequations}
Classical mechanical analogs of Eqs. (\ref{C8b}) and (\ref{C8c}), namely,   
$n_{r} n_{r}=1$ and $n_{r} W_{r}=0$, have indeed been treated by Plyushchay \cite{Ply01, Ply04}. 
Taking into account Eq. (\ref{C8a}), we assume that $\hat{p}_{0} |F^{\:\!} \rangle =M |F^{\:\!} \rangle$. 
Then, upon acting on $|F^{\:\!} \rangle$,  
the commutation relations (\ref{C7a}) and (\ref{C7b}) reduce to 
\begin{subequations}
\label{C9}
\begin{align} 
\left[\:\! \hat{W}_{r}, \hat{n}_{s} \:\! \right] &=i\epsilon_{rst} \hat{n}_{t} \,, 
\label{C9a}
\\
\left[\:\! \hat{W}_{r}, \hat{W}_{s} \:\! \right] &=i\epsilon_{rst} \hat{W}_{t} \,. 
\label{C9b}
\end{align}
\end{subequations}

Using the commutation relation (\ref{C9a}), we can show that 
\begin{align}
\hat{\boldsymbol{n}} \:\! \mathcal{W} \:\! |F^{\:\!} \rangle 
=\mathcal{W} (\hat{\boldsymbol{n}} +\boldsymbol{\varepsilon} \times \hat{\boldsymbol{n}} ) |F^{\:\!} \rangle \,, 
\label{C10}
\end{align}
where $\mathcal{W}:=\exp\! \big(^{\!} -i\boldsymbol{\varepsilon} {\!\; \cdot \!\;} \hat{\boldsymbol{W}} \big)$, 
and $\boldsymbol{\varepsilon}$ is an infinitesimal 3-dimensional vector. 
Multiplying both sides of Eq. (\ref{C10}) by 
an eigenvector $\langle \boldsymbol{n}|$ of the eigenvalue equation 
$\langle \boldsymbol{n}| \hat{\boldsymbol{n}}=\boldsymbol{n} \langle \boldsymbol{n}|$ on the left, 
we obtain 
\begin{align}
\langle \boldsymbol{n}|\mathcal{W} \:\! \hat{\boldsymbol{n}} |F^{\:\!} \rangle 
=(\boldsymbol{n}-\boldsymbol{\varepsilon} \times \boldsymbol{n}) 
\langle \boldsymbol{n}|\mathcal{W} |F^{\:\!} \rangle \,.
\label{C11}
\end{align}
Here, the approximate expression 
$\mathcal{W} (\boldsymbol{\varepsilon} \times \hat{\boldsymbol{n}}) \simeq 
(\boldsymbol{\varepsilon} \times \hat{\boldsymbol{n}}) \mathcal{W}$ has been used. 
From (\ref{C11}), we can see that $\langle \boldsymbol{n}|\mathcal{W}$ is the eigenvector of 
$\hat{\boldsymbol{n}}$ corresponding to the eigenvalue 
$\boldsymbol{n}-\boldsymbol{\varepsilon} \times \boldsymbol{n}$. 
Hence it follows that 
$\langle \boldsymbol{n}|\mathcal{W}
=\langle \boldsymbol{n}-\boldsymbol{\varepsilon} \times \boldsymbol{n}|$ holds at least in the physical subspace 
in which $|F^{\:\!} \rangle$ lives. In this way, we have 
$\langle \boldsymbol{n}|\mathcal{W}|F^{\:\!} \rangle
=\langle \boldsymbol{n}-\boldsymbol{\varepsilon} \times \boldsymbol{n}|F^{\:\!} \rangle$, 
or equivalently, 
\begin{align}
\langle \boldsymbol{n}|\mathcal{W}| F^{\:\!} \rangle
=F(\boldsymbol{n}-\boldsymbol{\varepsilon} \times \boldsymbol{n}) \,,
\label{C12}
\end{align}
where $F(\boldsymbol{n})$ is defined by $F(\boldsymbol{n}):=\langle \boldsymbol{n} | F^{\:\!} \rangle$. 
Expanding both sides of Eq. (\ref{C12}) with respect to $\boldsymbol{\varepsilon}$ and 
equating terms of order $\boldsymbol{\varepsilon}$ yield 
\begin{align}
\langle \boldsymbol{n}|\hat{\boldsymbol{W}}| F^{\:\!} \rangle
=\boldsymbol{L} F(\boldsymbol{n}) \,, 
\qquad \boldsymbol{L}:=-i \boldsymbol{n} \times \partial_{\boldsymbol{n}} \,.  
\label{C13}
\end{align}
The operator $\boldsymbol{L}$ is identified as 
the orbital angular-momentum operator defined in the internal space parametrized by $\boldsymbol{n}$. 
By using the commutation relations 
$\big[^{\:\!} \hat{p}_{0}, \hat{W}_{r} \big]=0$, 
(\ref{C9a}), (\ref{C9b}), and (\ref{A18}), it can be shown that $\hat{W}_r | F^{\:\!} \rangle$ is also 
a physical state vector satisfying the conditions (\ref{C8a})--(\ref{C8d}). 
Then we can use Eq. (\ref{C13}) twice to obtain 
\begin{align}
\langle \boldsymbol{n}|\hat{\boldsymbol{W}}{}^{2} | F \:\! \rangle 
=\boldsymbol{L}^{2} F(\boldsymbol{n}) \,, 
\qquad 
\hat{\boldsymbol{W}}{}^{2} :=\hat{W}_{r} \hat{W}_{r} \,. 
\label{C14}
\end{align}
Combining Eq. (\ref{C14}) with the eigenvalue equation (\ref{A24}) leads to 
\begin{align}
\boldsymbol{L}^{2}  F(\boldsymbol{n})=J(J+1) F(\boldsymbol{n}) \,. 
\label{C15}
\end{align}
This is precisely the eigenvalue equation for the {\it SO}(3) Casimir operator $\boldsymbol{L}^{2}$.  
Equation (\ref{C15}) can be solved as usual in terms of the spherical harmonics. 
Correspondingly, the allowed values of the spin quantum number $J$ are restricted to 
non-negative integer values. In this way, we can reproduce the procedure given by Plyushchay. 
As we have seen, the existence of $\hat{n}^{\mu}$ satisfying the commutation relation (\ref{C9a}) 
is essential for the argument in this appendix.

Taking the 3-dimensional inner product of Eq. (\ref{C13}) with $\boldsymbol{n}$,  
and using the eigenvalue equation 
$\langle \boldsymbol{n}| \hat{\boldsymbol{n}}=\boldsymbol{n} \langle \boldsymbol{n}|$, 
we have 
\begin{align}
\langle \boldsymbol{n}|^{\:\!} \hat{n}_{r} \hat{W}_{r} | F^{\:\!} \rangle=0 \,.  
\label{C16}
\end{align}
This is consistent with the condition (\ref{C8c}). 
Equation (\ref{C16}) actually reduces to Eq. (\ref{C8c}) under the natural assumption that 
the set of $\langle \boldsymbol{n}|$'s constitutes a complete set in the relevant Hilbert space. 
We thus see that Eq. (\ref{C13}) can yield Eq. (\ref{C8c}) and eventually leads to Eq. (\ref{5.4d}), i.e. 
$\hat{\chi}{}^{(a)-} |F^{\:\!} \rangle =0$   
via the operator version of Eq. (\ref{C2e}). 
Since both of the allowed values of $J$ and the condition (\ref{5.4d}) can be obtained by exploiting Eq. (\ref{C13}), 
we can expect that the condition (\ref{5.4d}) is related to deriving the allowed values of $J$. 
In fact, as demonstrated in Sec. 6, the allowed values of $J$, namely the non-negative integer values, 
can be derived on the basis of Eq. (\ref{5.4d}).


\begin{thebibliography}{00}

\bibitem{Pisarski} 
R.~D.~Pisarski,  
^^ ^^ Field theory of paths with a curvature-dependent term," 
Phys. Rev. D {\bf 34} (1986) 670. 

\bibitem{Polyakov}
A.~Polyakov, 
^^ ^^ Fine structure of strings," 
Nucl. Phys. B {\bf 268} (1986) 406. 

\bibitem{Ply01}
M.~S.~Plyushchay,  
^^ ^^ Canonical quantization and mass spectrum of relativistic particle: 
analog of relativistic string with rigidity," 
Mod. Phys. Lett. A {\bf 3} (1988) 1299. 

\bibitem{Ply02}
M.~S.~Plyushchay,  
^^ ^^ Massless point particle with rigidity," 
Mod. Phys. Lett. A {\bf 4} (1989) 837. 

\bibitem{Ply03}
M.~S.~Plyushchay,  
^^ ^^ Massless particle with rigidity as a model for the description of bosons and fermions,"  
Phys. Lett. B {\bf 243} (1990) 383. 

\bibitem{Ply04}
M.~S.~Plyushchay,  
^^ ^^ Massive relativistic point particle with rigidity," 
Int. J. Mod. Phys. A {\bf 4} (1989) 3851. 

\bibitem{BGPR}
C.~Batlle, J.~Gomis, J.~M.~Pons and N.~Rom\'{a}n-Roy, 
^^ ^^ Lagrangian and Hamiltonian constraints for second-order
singular Lagrangians," 
J. Phys. A: Math. Gen. {\bf 21} (1988) 2693. 

\bibitem{Ram01}
E.~Ramos and J.~Roca
^^ ^^ W-symmetry and the rigid particle,"  
Nucl. Phys. B {\bf 436} (1995) 529, arXiv:hep-th/9408019. 

\bibitem{Ban01}
R.~Banerjee, P.~Mukherjee and B.~Paul, 
^^ ^^ Gauge symmetry and W-algebra in higher derivative systems,"  
J. High Energy Phys. 08 (2011) 085, arXiv:1012.2969 [hep-th]. 

\bibitem{Ban02}
R.~Banerjee, B.~Paul and S.~Upadhyay 
^^ ^^ BRST symmetry and W-algebra in higher derivative models,"  
Phys. Rev. D {\bf 88} (2013) 065019, arXiv:1306.0744 [hep-th]. 

\bibitem{Pav01}
M.~Pav\v{s}i\v{c},  
^^ ^^ Classical motion of membranes, strings and point particles with extrinsic curvature,"  
Phys. Lett. B {\bf 205} (1988) 231. 

\bibitem{Pav02}
M.~Pav\v{s}i\v{c},  
^^ ^^ The quantization of a point particle with extrinsic curvature leads to the Dirac equation,"  
Phys. Lett. B {\bf 221} (1989) 264. 

\bibitem{Der01}
T.~Dereli, D.~H.~Hartley, M.~Onder and R.~W.~Tucker, 
^^ ^^ Relativistic elastica,"  
Phys. Lett. B {\bf 252} (1990) 601. 

\bibitem{Ply05}
M.~S.~Plyushchay,  
^^ ^^ Does the quantization of a particle with curvature lead to Dirac equation?,"  
Phys. Lett. B {\bf 253} (1991) 50. 

\bibitem{Nes01}
V.~V.~Nesterenko, A.~Feoli and G.~Scarpetta, 
^^ ^^ Dynamics of relativistic particle with Lagrangian dependent on acceleration,"  
J. Math. Phys. {\bf 36} (1995) 5552, arXiv:hep-th/9408071. 

\bibitem{Pav03}
M.~Pav\v{s}i\v{c},  
^^ ^^ Rigid particle and its spin revisited,"  
Found. Phys. {\bf 37} (2007) 40, arXiv:hep-th/0412324. 

\bibitem{DerNer}
A.~Deriglazov and A.~Nersessian,  
^^ ^^ Rigid particle revisited: extrinsic curvature yields the Dirac equation,"  
Phys. Lett. A {\bf 378} (2014) 1224, arXiv:1303.0483 [hep-th]. 

\bibitem{DegSuz}
S.~Deguchi and T.~Suzuki, 
^^ ^^ Spinor and twistor formulations of massless particles with rigidity," 
Phys. Lett. B {\bf 731} (2014) 337, arXiv:1401.1901 [hep-th]. 

\bibitem{BarPic}
I.~Bars and M.~Pic\'{o}n, 
^^ ^^ Single twistor description of massless, massive, AdS, and other interacting particles," 
Phys. Rev. D {\bf 73} (2006) 064002, arXiv:hep-th/0512091. 

\bibitem{DEN}
S.~Deguchi, T.~Egami, and J.~Note,
^^ ^^ Spinor and twistor formulations of tensionless bosonic strings in four dimensions,'' 
Prog. Theor. Phys. {\bf 124} (2010) 969, arXiv:1006.2438 [hep-th].  

\bibitem{DNOS}
S.~Deguchi, S.~Negishi, S.~Okano, and T.~Suzuki, 
^^ ^^ Canonical formalism and quantization of a massless spinning bosonic particle in four dimensions," 
Int. J. Mod. Phys. A {\bf 29} (2014) 1450044, arXiv:1309.4169 [hep-th]. 

\bibitem{Shirafuji}
T. Shirafuji, 
^^ ^^ Lagrangian mechanics of massless particles with spin," 
Prog. Theor. Phys. {\bf 70} (1983) 18. 

\bibitem{PenMac}
R.~Penrose and M.~A.~H.~MacCallum, 
^^ ^^ Twistor theory: An approach to the quantisation of fields and space-time,"  
Phys. Rep. {\bf 6} (1973) 241. 

\bibitem{PenRin2}
R.~Penrose and W.~Rindler, 
\textit{Spinors and Space-Time}, Vol.~2:  
Spinor and Twistor Methods in Space-Time Geometry, 
Cambridge Monographs on Mathematical Physics,  
(Cambridge University Press, Cambridge, 1986).  

\bibitem{HugTod}
S.~A.~Huggett and K.~P.~Tod, 
\textit{An Introduction to Twistor Theory}, 2nd ed.,  
London Mathematical Society, Student Texts Vol. 4 
(Cambridge University Press, Cambridge, England, 1994). 

\bibitem{Penrose}
R.~Penrose, 
``The twistor programme," 
Rep. Math. Phys. {\bf 12} (1977) 65 . 

\bibitem{Perjes1} 
Z.~Perj\'{e}s, 
``Twistor variables of relativistic mechanics," 
Phys. Rev. D {\bf 11} (1975) 2031. 

\bibitem{Perjes2} 
Z.~Perj\'{e}s, 
``Unitary space of particle internal states," 
Phys. Rev. D {\bf 20} (1979) 1857. 

\bibitem{Perjes3} 
Z.~Perj\'{e}s, 
``Perspectives of Penrose theory in particle physics," 
Rep. Math. Phys. {\bf 12} (1977) 193.  

\bibitem{Perjes4} 
Z. Perj\'{e}s, 
``Internal symmetries in twistor theory," 
Czech. J. Phys. B {\bf 32} (1982) 540.  

\bibitem{Hughston}
L.~P.~Hughston, 
\textit{Twistors and Particles}, Lecture Notes in Physics Vol. 97 
(Springer-Verlag, Berlin, 1979). 


\bibitem{FedZim}
S.~Fedoruk and V.~G.~Zima, 
^^ ^^ Bitwistor formulation of massive spinning particle," 
J. Kharkiv Univ. {\bf 585} (2003) 39, arXiv:hep-th/0308154. 

\bibitem{BALE}
A.~Bette, J.~A.~de Azc\'{a}rraga, J.~Lukierski, and  C.~Miquel-Espanya, 
^^ ^^ Massive relativistic free fields with Lorentz spins and electric charges,"  
Phys. Lett. B {\bf 595} (2004) 491, arXiv:hep-th/0405166. 

\bibitem{AFLM}
J.~A.~de Azc\'{a}rraga, A. Frydryszak, J.~Lukierski, and C.~Miquel-Espanya, 
^^ ^^ Massive relativistic particle model with spin from free two-twistor dynamics and its quantization," 
Phys. Rev. D {\bf 73} (2006) 105011, arXiv:hep-th/0510161. 

\bibitem{FFLM} 
S.~Fedoruk, A.~Frydryszak, J.~Lukierski, and C.~Miquel-Espanya, 
^^ ^^ Extension of the Shirafuji model for massive particles with spin,"  
Int. J. Mod. Phys. A {\bf 21} (2006) 4137, arXiv:hep-th/0510266. 

\bibitem{AIL}
J.~A.~de Azc\'{a}rraga, J.~M.~Izquierdo, and J.~Lukierski, 
^^ ^^ Supertwistors, massive superparticles and $\kappa$-symmetry," 
J. High Energy Phys. {01} (2009) 041, arXiv:0808.2155 [hep-th]. 

\bibitem{MRT} 
L.~Mezincescu, A.~J.~Routh, and P.~K.~Townsend, 
^^ ^^ Supertwistors and massive particles," 
Ann. Phys. (Amsterdam) {\bf 346} (2014) 66, arXiv:1312.2768 [hep-th]. 

\bibitem{FedLuk} 
S.~Fedoruk and J.~Lukierski, 
^^ ^^ Massive twistor particle with spin generated by Souriau-Wess-Zumino term and its quantization," 
Phys. Lett. B {\bf 733} (2014) 309, arXiv:1403.4127 [hep-th]. 

\bibitem{AFIL}
J.~A.~de Azc\'{a}rraga, S.~Fedoruk, J.~M.~Izquierdo, and J.~Lukierski, 
^^ ^^ Two-twistor particle models and free massive higher spin fields," 
J. High Energy Phys. {04} (2015) 010, arXiv:1409.7169 [hep-th]. 

\bibitem{MRT2} 
L.~Mezincescu, A.~J.~Routh, and P.~K.~Townsend, 
^^ ^^ Twistors and the massive spinning particle," J. Phys. A {\bf 49} (2016) 025401, 
arXiv:1508.05350 [hep-th]. 

\bibitem{RouTow} 
A.~J.~Routh and P.~K.~Townsend, 
^^ ^^ Twistor form of massive 6D superparticle," J. Phys. A {\bf 49} (2016) 025402, 
arXiv:1507.05218 [hep-th]. 

\bibitem{DegOka}
S.~Deguchi and S.~Okano, 
^^ ^^ Gauged twistor formulation of a massive spinning particle in four dimensions," 
Phys. Rev. D {\bf 93} (2016) 045016 [Erratum-ibid. D {\bf 93} (2016) 089906(E)], 
arXiv:1512.07740 [hep-th]. 

\bibitem{OkaDeg}
S.~Okano and S.~Deguchi, 
``A no-go theorem for the $n$-twistor description of a massive particle," 
J. Math. Phys. {\bf 58} (2017) 031701, arXiv:1606.01339 [hep-th]. 

\bibitem{PenRin1}
R.~Penrose and W.~Rindler, 
\textit{Spinors and Space-Time}, Vol.~1:  
Two-Spinor Calculus and Relativistic Fields, 
Cambridge Monographs on Mathematical Physics,  
(Cambridge University Press, Cambridge, 1984). 

\bibitem{Dirac2} 
P.~A.~M.~Dirac, \textit{Lectures on Quantum Mechanics} 
(Belfer Graduate School of Science, Yeshiva University, New York 1964). 

\bibitem{HRT} 
A.~J.~Hanson, T.~Regge, and C.~Teitelboim, 
\textit{Constrained Hamiltonian Systems} 
(Accademia Nazionale dei Lincei, Rome, 1976). 

\bibitem{HenTei} 
M.~Henneaux and C.~Teitelboim, 
\textit{Quantization of Gauge Systems} 
(Princeton University Press, Princeton, NJ, 1992). 


\bibitem{Dirac1}
P.~A.~M.~Dirac, 
^^ ^^ Relativistic wave equations," 
Proc. R. Soc. A {\bf 155} (1936) 447. 

\bibitem{Fierz} 
M.~Fierz, 
^^ ^^ \"{U}ber die relativistische theorie Kr\"{a}ftefreier teilchen mit beliebigem spin," 
Helv. Phys. Acta {\bf 12} (1939) 3. 

\bibitem{FiePau}
M.~Fierz and W.~Pauli, 
^^ ^^ On relativistic wave equations for particles of arbitrary spin in an electromagnetic field," 
Proc. R. Soc. A {\bf 173} (1939) 211. 

\bibitem{IsaPod} 
A.~P.~Isaev and M.~A.~Podoinitsyn, 
^^ ^^ Two-spinor description of massive particles and relativistic spin projection operators," 
Nucl. Phys. B {\bf 929} (2018) 452,  arXiv:1712.00833 [hep-th]. 

\bibitem{CRZ}
P.~Claus, J.~Rahmfeld, and Y. Zunger, 
``A simple particle action from a twistor parametrization of $\mathrm{AdS}_5$,"
Phys. Lett. B {\bf 466} (1999) 181, arXiv:hep-th/9906118. 

\bibitem{Okano}
S. Okano, 
``Twistor formulation of a massive spinning particle,"  
Ph.D. thesis, Nihon University (2016).  

\bibitem{Gursey}
E. G\"{u}rsey, in {\em High Energy Physics}, edited by C. DeWitt and M. Jacob 
(Gordon and Breach Science Publishers, New York, 1965), p. 53. @

\end{thebibliography}
\end{document}